\documentclass[twocolumn,twocolappendix]{aastex63}
\usepackage{verbatim}
\usepackage{longtable}
\usepackage{multirow}
\usepackage{tabularx}
\usepackage[flushleft]{threeparttable}

\usepackage{enumitem}
\setlist[enumerate]{align=left}

\newcommand{\hr}{H{$\gamma$}}

\newcommand {\hi}{\ifmmode \left{\rm H}\,{\textsc i}\right \else H\,{\sc i}\fi}
\newcommand{\kms}{$\rm km~s^{-1}$}

\newcommand {\oii}{\ifmmode \left[{\rm O}\,{\textsc ii}\right] \else [O\,{\sc ii}]\fi}
\newcommand {\SII}{\ifmmode \left[{\rm S}\,{\textsc ii}\right] \else [S\,{\sc ii}]\fi}
\newcommand {\ciii}{C\,{\sc iii]}}
\newcommand {\cii}{C\,{\sc ii]}}
\newcommand {\CII}{C\,{\sc ii}}
\newcommand {\SiIII}{Si\,{\sc iii]}}
\newcommand {\AlIII}{Al\,{\sc iii}}
\newcommand {\SiII}{[Si\,{\sc ii]}}
\newcommand {\SiIV}{Si\,{\sc iv}}
\newcommand {\NIII}{[N\,{\sc iii]}}
\newcommand {\NIV}{[N\,{\sc iv]}}
\newcommand {\civ}{C\,{\sc iv}}
\newcommand {\oiv}{O\,{\sc iv]}}
\newcommand {\oi}{O\,{\sc i}}
\newcommand {\HeII}{He\,{\sc ii}}
\newcommand {\OIII}{O\,{\sc iii}}
\newcommand {\FeII}{Fe\,{\sc ii}}

\newcommand {\heii}{\ifmmode {\rm He}\, {\sc II\lambda4686}\ \else He\,{\sc II$\lambda$4686}\fi}

\newcommand {\hb}{\ifmmode {\rm H}\beta \else H$\beta$\fi}
\newcommand {\ha}{\ifmmode {\rm H}\alpha \else H$\alpha$\fi}

\newcommand {\hei}{\ifmmode {\rm He} {textsc{i}} \else He\,{\sc i}\fi}
\newcommand {\mgii}{\ifmmode {\rm Mg}{\textsc{ii}} \else Mg\,{\sc ii}\fi}
\newcommand {\mgiii}{\ifmmode {\rm Mg}{\textsc{iii}} \else Mg\,{\sc iii}\fi}
\newcommand {\lya}{\ifmmode {\rm Ly} {textsc{$\alpha$}} \else Ly\,{\sc $\alpha$}\fi}

\newcommand{\erg}{${\rm erg \ s^{-1}}$ }

\newcommand{\ergcms}{\ifmmode {\rm ergs\,cm}^{-2}\,{\rm s}^{-1} \else ergs\,cm$^{-2}$\,s$^{-1}$\fi}
\newcommand{\ergcmsA}{\ifmmode{\rm ergs}\, {\rm cm}^{-2}\,{\rm s}^{-1}\,{\rm\AA}^{-1} \else ergs\, cm$^{-2}$\, s$^{-1}$\, \AA$^{-1}$\fi}

\usepackage[normalem]{ulem}
\usepackage{CJK}
\usepackage{xcolor,color}
\usepackage{hyperref}
\hypersetup{
   colorlinks,
   linkcolor={blue!88!black!80},
   citecolor={blue!88!black!80},
   urlcolor={blue!88!black!80}
}

\shorttitle{High-redshhift EVQs }
\shortauthors{Guo et al.}
 
\begin{document}
\begin{CJK}{UTF8}{gbsn}

\title{High-redshift Extreme Variability Quasars from Sloan Digital Sky Survey Multi-Epoch Spectroscopy}

\author[0000-0001-8416-7059]{Hengxiao Guo (郭恒潇)}
\affiliation{Department of Astronomy, University of Illinois at Urbana-Champaign, Urbana, IL 61801, USA}
\affiliation{National Center for Supercomputing Applications, University of Illinois at Urbana-Champaign, Urbana, IL 61801, USA}
\affiliation{Department of Physics and Astronomy, 4129 Frederick Reines Hall, University of California, Irvine, CA, 92697-4575, USA}

\author{Jiacheng Peng}
\affiliation{Department of Physics, Hebei Normal University, No. 20 East of South 2nd Ring Road, Shijiazhuang 050024, China}

\author{Kaiwen Zhang}
\affiliation{Department of Physics, University of Illinois at Urbana-Champaign, 1110 West Green Street, Urbana, IL 61801, USA}

\author[0000-0001-9947-6911]{Colin J. Burke}
\affiliation{Department of Astronomy, University of Illinois at Urbana-Champaign, Urbana, IL 61801, USA}
\affiliation{National Center for Supercomputing Applications, University of Illinois at Urbana-Champaign, Urbana, IL 61801, USA}

\author[0000-0003-0049-5210]{Xin Liu}
\affiliation{Department of Astronomy, University of Illinois at Urbana-Champaign, Urbana, IL 61801, USA}
\affiliation{National Center for Supercomputing Applications, University of Illinois at Urbana-Champaign, Urbana, IL 61801, USA}

\author[0000-0002-0771-2153]{Mouyuan Sun}
\affiliation{Department of Astronomy, Xiamen University, Xiamen, Fujian 361005, China}

\author{Shu Wang}
\affiliation{Kavli Institute for Astronomy and Astrophysics, Peking University, Beijing 100871, China}
\affiliation{Department of Astronomy, School of Physics, Peking University, Beijing 100871, China}

\author{Minzhi Kong}
\affiliation{Department of Physics, Hebei Normal University, No. 20 East of South 2nd Ring Road, Shijiazhuang 050024, China}
\affiliation{Department of Astronomy, University of Illinois at Urbana-Champaign, Urbana, IL 61801, USA}

\author[0000-0001-6938-8670]{Zhenfeng Sheng}
\affiliation{CAS Key Laboratory for Researches in Galaxies and Cosmology, University of Sciences and Technology of China, Hefei, Anhui 230026, China}
\affiliation{School of Astronomy and Space Science, University of Science and Technology of China, Hefei 230026, China}

\author[0000-0002-1517-6792]{Tinggui Wang}
\affiliation{CAS Key Laboratory for Researches in Galaxies and Cosmology, University of Sciences and Technology of China, Hefei, Anhui 230026, China}
\affiliation{School of Astronomy and Space Science, University of Science and Technology of China, Hefei 230026, China}

\author{Zhicheng He}
\affiliation{CAS Key Laboratory for Researches in Galaxies and Cosmology, University of Sciences and Technology of China, Hefei, Anhui 230026, China}
\affiliation{School of Astronomy and Space Science, University of Science and Technology of China, Hefei 230026, China}

\author[0000-0002-4455-6946]{Minfeng Gu}
\affiliation{Key Laboratory for Research in Galaxies and Cosmology, Shanghai Astronomical Observatory, Chinese Academy of Sciences,
80 Nandan Road, Shanghai 200030, China}

\email{hengxiaoguo@gmail.com (HXG), confucious\_76@163.com (MZK)}

\begin{abstract}
We perform a systematic search for high-redshift ($z >$ 1.5) extreme variability quasars (EVQs) using repeat spectra from the Sixteenth Data Release of Sloan Digital Sky Survey, which provides a baseline spanning up to $\sim$18 yrs in the observed frame. We compile a sample of 348 EVQs with a maximum continuum variability at rest frame 1450 \AA\ of more than 100\% (i.e., $\delta$V $\equiv$ (Max$-$Min)/Mean $>$1). The EVQs show a range of emission line variability, including 23 where at least one line in our redshift range disappears below detectability, which can then be seen as analogous to low-redshift changing-look quasars (CLQs)". Importantly, spurious CLQs caused by SDSS problematic spectral flux calibration, e.g., fiber drop issue, have been rejected. The similar properties (e.g., continuum/line, difference-composite spectra and Eddington ratio) of normal EVQs and CLQs, implies that they are basically the same physical population with analogous intrinsic variability mechanisms, as a tail of a continuous distribution of normal quasar properties. In addition, we find no reliable evidence ($\lesssim$ 1$\sigma$) to support that the CLQs are a subset of EVQs with less efficient accretion. Finally, we also confirm the anti-breathing of \civ\ (i.e., line width increases as luminosity increases) in EVQs, and find that in addition to $\sim$ 0.4 dex systematic uncertainty in single-epoch \civ\ virial black hole mass estimates, an extra scatter of $\sim$ 0.3 dex will be introduced by extreme variability.

\end{abstract}




\section{Introduction}\label{sec:intro}
The canonical unification scheme of active galactic nuclei (AGNs)\footnote{Throughout this paper, the terms of AGN and quasar are used interchangeably regardless of the luminosity difference.} dictates that broad-line (Type 1) and narrow-line (Type 2) objects are the same underlying population viewed at different orientations \citep{Antonucci93,Urry95}. The discovery of a rare Changing-Look (CL; or ``changing-state'') phenomenon where an AGN changes type and exhibits flux variations of more than a factor of a few with significantly enhanced (or reduced) broad emission line flux over months to decades \citep[e.g.,][]{Tohline76,Cohen86,Aretxaga99,Shappee14,Denney14,LaMassa15,Macleod16,Ruan16b,Runco16, Runnoe16,Yang18,Guo19a,Trakhtenbrot19,Ai20,Sheng20} challenges this simple unification picture.  

Dust obscuration, whereby broad-line region clouds move in and out of the line of sight, should be a good explanation for the CL behavior given their comparable timescales. Indeed, the dust obscuration scenario has been invoked in some earlier cases \citep[e.g., NGC 7603, Mrk 993,][]{Goodrich89,Tran92}. However, measurements of the low polarization ($<$1\%) of CLQs \citep{Hutsemekers17,Hutsemekers19} disfavor the obscuration picture. In addition, \citet{Sheng17} observed large mid-infrared variability in 10 CLQs that echoes their optical variability with a time lag expected from dust reprocessing, also supporting the idea that the CL phenomenon is due to physical changes in the accretion disk or accretion rate. Meanwhile, \cite{MacLeod19} suggest that the CL phenomenon in general is probably not due to tidal disruption events or microlensing by foreground stars, unless these events are strongly preferred in quasars with lower Eddington ratios.

To date, over 100 CLQs have been discovered with photometric and spectroscopic methods \citep[e.g.,][]{MacLeod19,Yang18}. Most of those observations only reveal the appearance/disappearance of broad Balmer lines (e.g., \hb\ or \ha), while the transition of broad \mgii\ is rarely observed even in the dim state \citep{Homan20}. \cite{Roig14} noticed some unusual \mgii-emitters which show strong and broad \mgii\ lines, but very weak or unrecognizable emission in other normal indicators of AGN activity, such as \ha, \hb, and near-UV power-law continuum. Thanks to the discovery of \mgii-emitters, \cite{Guo19a} discovered the first \mgii\ CLQ based on the repeat observations of 361 \mgii-emitters in SDSS DR 14. However, the CL phenomenon at high redshift (e.g., $z>$ 1.5) is barely explored \citep{Ross19}, and the transition behaviors of different UV emission lines are still unclear.

The primary approaches to search for CLQs are photometric and spectroscopic methods. However, their behavior is not always associated. On one hand, previous studies have used their extremely variable nature to search CLQs \citep{Graham20,Yang20}. For instance \cite{Macleod16} found that most of their $\sim$1000 photometrically selected objects with variations of $|\Delta g| >$ 1 mag are EVQs rather than CLQs (only 10 confirmed CLQs). On the other hand, 15 CLQs selected from repeat SDSS (SDSS/LAMOST) spectra show dispersed variabilities ranging from 0.03 to 1.59 mag in the $g$-band with a median of 0.3 mag \citep{Yang18}. Furthermore, J1525$+$2920, the \mgii\ CLQ in \cite{Guo19a}, exhibits small variability of only 0.1 mag in its $g$-band light curve. This clearly indicates that large photometric variability does not guarantee CL behavior, and the real CL phenomenon does occur with inconspicuous variability. Therefore, searching for CLQs with repeat spectra is more straightforward given the advent of powerful modern spectroscopic surveys such as the Time-Domain Spectroscopic Survey \citep[TDSS;][]{Ruan16a}, SDSS-V \citep{Kollmeier17}, the Large Sky Area Multi-Object Fiber Spectroscopic Telescope \citep[LAMOST;][]{Cui12} and the forthcoming Dark Energy Spectroscopic Instrument \citep[DESI;][]{DESICollaboration16}. In addition, the delayed follow-up of photometrically selected candidates may miss the best opportunity to capture the faintest state.

According to the expectation of the photoionization model, if the continuum luminosity decreases continuously, all the broad emission lines (e.g., \civ, \ciii\ and \mgii) will gradually fade away, and finally be swamped by spectral noise as a CLQ \citep{Guo20}. Therefore, the purpose of this paper is as follows:
\begin{enumerate}
\item Compile a sample of high-redshift EVQs and identify new CLQs with transitions in different emission lines (e.g., \civ\ and \ciii ).
\item Explore the prerequisites for CL behaviors (e.g., critical Eddington ratios).
\item Check the relationship between CLQs and EVQs in AGN family and try to understand their intrinsic variability mechanism.  
\end{enumerate}

The paper is organized as follows. In \S \ref{sec:data}, we describe the data set of photometries and spectroscopies used to identify EVQs/CLQs. In \S \ref{sec:sample}, we illustrate the sample selections of the high-redshift EVQs/CLQs, and reject spurious EVQs/CLQs caused by SDSS fiber dropping. Then we compare the properties of the non-CL EVQs and CLQs in our total EVQ sample to explore the prerequisites of CL behaviors in \S \ref{sec:result}. In \S \ref{sec:diss}, we discuss the CL behavior, the intrinsic variability mechanism of EVQs/CLQs, and the single-epoch black hole (BH) mass estimated from \civ\ in EVQs. Finally, we draw our conclusions in \S \ref{sec:con}. A concordance $\Lambda$CDM cosmology with $\Omega_m = 0.3$, $\Omega_{\Lambda} = 0.7$, and $H_{0}=70$ km s$^{-1}$ Mpc$^{-1}$ is assumed throughout.

\section{Data}\label{sec:data}
Repeat SDSS spectra with a baseline up to 18 yrs is an ideal tool to identify CL behavior, while photometric light curves are expected to verify the concurrent spectral variability. There is no current single survey that has sufficient sky coverage, baseline, depth, and cadence to support a verification of CLQs (targets of opportunity). Therefore the best strategy is to combine different surveys (e.g., CRTS and PTF/iPTF/ZTF) to extend the baseline and gain as high cadence as possible to trace the transition in CLQs with a typical timescale of months to decades.

\subsection{SDSS Spectrum}
All the spectra in this work are obtained from the public Sixteenth Data Release of Sloan Digital Sky Survey (SDSS DR16) database \citep{Ahumada19}, which covers 14,555 deg$^2$, mostly in the northern sky. DR16 is also notable as the final data release from the Extended Baryon Oscillation Spectroscopic Survey \citep[eBOSS,][]{Dawson16}. Benefiting from its cumulative data archive from 2000 to the present, many objects have been observed more than once. This is quite suitable for investigating AGN spectral variability and searching for CLQs. The repeat spectroscopic observations are mainly from three parts: 1) the overlapped survey areas between adjacent plates; 2) dedicated programs, e.g., TDSS \citep{Ruan16a} and SDSS reverberation mapping \citep[SDSS-RM,][]{Shen15}; 3) plates re-observed due to insufficient Signal-to-Noise Ratio (SNR). The spectral wavelength coverage for SDSS I\&II (SDSS III\&IV) is 3800 -- 9200 (3600 -- 10400) \AA\ with spectral resolution $R \sim$ 1850 -- 2200.

The asserted accuracy of absolute spectral flux calibration of stars is about 6\%\footnote{\url{https://www.sdss.org/dr16/algorithms/spectrophotometry/}} after SDSS III with a 2\arcsec\ fiber size. Compared to the calibration algorithm in DR 14, DR 16 used a new set of stellar templates to fit absorption lines of standard stars, which reduces residuals by a factor of 2 in the blue spectrograph (3600 to 6000 \AA) relative to previous releases through improved modeling of spectral lines in F-stars. However, comparison between repeat observations indicates the precision of spectral flux calibration is significantly underestimated in high-redshift quasars (see details in Appendix \ref{sec:calibration}).

\subsection{CRTS}
The Catalina Real-Time Transient Survey \citep[CRTS][]{Drake09} is a $\sim$33,000 deg$^2$ survey designed to discover rare and interesting transients (e.g., supernovae, TDEs, and CLQs). It was conducted with three 1 m-class telescopes in both Northern and Southern hemispheres---the 0.7 m Catalina Sky Survey (CSS) Schmidt, the 1.5 m Mount Lemmon Survey (MLS) telescopes in Arizona and the 0.5 m Siding Springs Survey (SSS) Schmidt in Australia. CRTS covers up to $\sim$2500 deg$^2$ per night, with four exposures per visit, separated by 10 min. It contains time series for approximately 500 million sources with a limiting magnitude of $V\sim$ 21 mag (Vega system) from 2003 to 2016. All the data are automatically processed in real-time to report potential transients. All photometries are aperture-based and broadly calibrated to Johnson V \citep{Drake13}.

\subsection{PTF, iPTF and ZTF}
The Palomar Transient Factory (PTF) is a fully-automated, wide-field, time-domain survey designed to explore the transient and variable sky conducted from 2009 to 2012 \citep{Rau09,Law09}. The observations are made at Palomar Observatory with the 48 inch (P48, 1.2 m) Samuel Oschin Schmidt telescope equipped with the CHF12K camera. It covers $\sim$3000 deg$^2$ of the northern sky with a 5$\sigma$ limiting AB magnitude of $\sim$20.6 in Mould$-R$ and $\sim$21.3 in SDSS$-g$ bands with an average 5-day cadence. The intermediate Palomar Transient Factory (iPTF) ran from 2013 to 2017 as the successor to the PTF on P48 telescope with a relatively higher cadence. All the PTF/iPTF light curves are stored in NASA/IPAC infrared science archive\footnote{\url{https://irsa.ipac.caltech.edu/Missions/ptf.html}}

ZTF\footnote{\url{https://irsa.ipac.caltech.edu/Missions/ztf.html}} is a new robotic time-domain survey on P48 telescope mounted a new 600 megapixel camera with a 47 deg$^2$ field of view \citep{Bellm19}. It covers the entire visible northern sky (Dec $> -30^{\circ}$) from 2018 to present. ZTF's extremely wide field and fast readout electronics enables a survey that scans more than 3750 deg$^2$ per hour, to a 5$\sigma$ detection limit of $g\sim$20.8 and $r\sim$ 20.6 mag (AB system) with a 30s exposure during new moon \citep{Masci19}. The ZTF data are aperture-based photometry with a typical aperture diameter of 2\arcsec. The data can be accessed via ZTF commands\footnote{{\url{https://www.ztf.caltech.edu/page/dr2\#12b.iii}}}.

\subsection{PanSTARRS}
The Panoramic Survey Telescope \& Rapid Response System \citep[Pan-STARRS or PS1,][]{Chambers16} covers 30,000 deg$^2$ of the sky north of Dec $> -30^{\circ}$ from 2010 to 2016 using a 1.8 meter telescope in Hawaii, with typically $\sim$12 epochs for each filter ($grizy$). The mean 5$\sigma$ point source limiting sensitivities in single-epoch exposure of $grizy$ are 22.0, 21.8, 21.5, 20.9, 19.7 AB magnitude, respectively. All the detections are obtained via the Pan-STARRS Catalog Search interface\footnote{\url{https://catalogs.mast.stsci.edu/panstarrs/}}.

\subsection{DECaLS}
The Dark Energy Camera Legacy Survey (DECaLS) aims to provide the optical imaging for target selection in the DESI survey, covering $\sim$ 9000 deg$^2$ both the North Galactic Cap region at Dec $\le$ 32$^{\circ}$ and the South Galactic Cap region at Dec $\le$ 34$^{\circ}$ \citep{Dey19}. The 5$\sigma$ depths in final stacked $grz-$band images are 24.0, 23.4 and 22.5 AB magnitude. The single epoch exposure is from the forced phototmetry with different aperture sizes. All the data are obtained from the data lab in NOAO\footnote{\url{https://datalab.noao.edu/query.php}} combining three tables (i.e., ls\_dr8.tractor, ls\_dr8.ccds\_annotated and ls\_dr8.forced).

\subsection{Photometric Data Calibration}
To verify the CL behavior with photometric evidence from different surveys, we compile light curve for each candidate in the $r-$band, which usually has the best filter transmission and/or SNR with sufficient photometries. We convert all magnitudes to AB system. Then we correct the differences between various filter systems to calibrate the magnitudes to the same scale, though the deviations are very small (corrections to SDSS$-r$ for CRTS$-V$, PTF/iPTF$-R$, ZTF$-r$, PS1$-r$, DECaLs$-r$ are $-0.115, -0.102, -0.034, -0.002, -0.035$ AB mag, respectively). We have adopted the aperture-based ($\sim$2\arcsec) photometries from all surveys. The synthetic magnitudes of the SDSS spectra are obtain via convolving the spectrum with SDSS$-r$ filter with {\tt PySynphot}\footnote{\url{https://pysynphot.readthedocs.io/en/latest/}} \citep{STScIDevelopmentTeam13}.

\begin{figure*}
\centering
\includegraphics[width=18.cm]{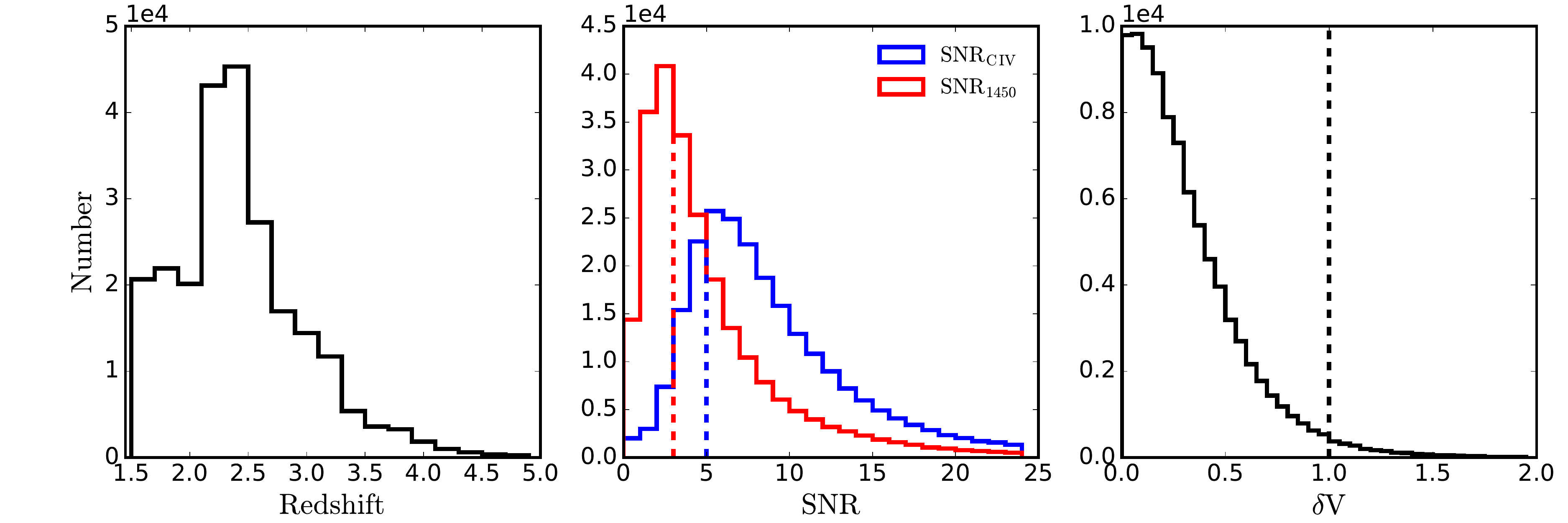}
\caption{Left: the redshift distribution for 237,958 objects (92,422 objects) ranging from $z = 1.5$ to 4.9 with a peak around $z = 2.2$. Middle: distributions of the median SNR per pixel for the continuum at 1450\AA\ (red) and \civ\ line (blue). The red/blue dashed lines present the different SNR cuts (i.e., $\rm SNR_{\rm 1450} > 3$ and $\rm SNR_{\rm CIV} >5$). Right: distribution of continuum variability at 1450\AA\ for 92,422 objects. Criterion 4 selects EVQ objects with $\delta$V $\equiv$ $\rm \frac{Max-Min}{0.5(Max+Min)} >$ 1. }
\label{fig:sn}
\end{figure*}

\begin{deluxetable*}{lrr}[htbp]
\caption{EVQs selection from SDSS DR16 repeat spectra} \label{tab:selection}
\tablewidth{0pt}
\tablehead{
\colhead{Selection} &
\colhead{Spectra} &
\colhead{Objects}
}
\startdata
Parent sample  &5,789,200 & \dots\\
1) Class = ``QSO" or ``GALAXY" & \dots &\dots\\
2) Nepoch $\ge$ 2 \& zWarning = 0 & 748,769 & 313,877 \\
3) Redshift:  1.5 $<z<$4.9  & 237,958 &92,422  \\
4) $\rm SNR_{\rm 1450} >3$ and $\rm SNR_{\rm C\,IV} >5$  & 129,877 & 61,037 \\
5) $\rm Flux_{\rm 1450}$: (MAX-MIN)/MEAN $>$ 1 & 5,474& 1,039\\
6) Reject SDSS-RM objects   & 1818(+4041 recycled) &951\\
7) Require spectral change or variability confirmed by light curves & 638(+451 recycled) &348\\
8) CLQ: any prominent broad-UV-line flux is consistent with zero  & 27(+32 recycled) & 23\\
\enddata
\begin{tablenotes}
\small
\item Note. the recycled objects denote low-SNR EVQs removed by Criterion 4. In (8), the broad UV line refer to any of \lya, \SiIV, \civ, \ciii, \mgii. 
\end{tablenotes}
\end{deluxetable*}

\section{Sample selection}\label{sec:sample}
\subsection{Preliminary Screening}
Our strategy is using repeat spectra in SDSS to search for high-redshift EVQs/CLQs. We first select the EVQ/CLQ candidates from the whole SDSS spectroscopic database via Criteria 1 to 6 in Table \ref{tab:selection}. Then we remove spurious EVQs/CLQs by visual inspection due to problematic flux calibration of SDSS spectra with assessments from the concurrent multi-survey photometries and spectral variation (Criterion 7, see the details in \S\ref{sec:vis}). Finally, we separate CLQs and normal EVQs with Criterion 8. In this section, we describe this selection procedure in detail.

We start with all spectra (5.8 million) in the SDSS DR16 database \citep{Ahumada19}, which was released in December 2019. Criteria 1 \& 2 are applied to guarantee that repeat quiescent/active galaxy observations are not subject to the problematic redshifts. The target classification and redshift measurements are obtained via the software {\tt REDMONSTER}\footnote{\url{https://github.com/timahutchinson/redmonster}} \citep{Hutchinson16}. This yields a repeat spectral catalog archived on Zenodo [10.5281/zenodo.3892020], which contains 313,877 multi-epoch quiescent/active galaxies. This catalog includes basic continuum and line measurements based on the SDSS pipeline \citep{Bolton12}. In addition to CLQ searches, this repeat catalog is also useful for other scientific goals, e.g., post-TDE candidates with strong variability of their broad nitrogen line \citep{Liu18b,Jiang08}, studies of continuum and emission-line variability \citep{Wilhite05,Ruan14,Guo14b,Guo16b}, identifying supermassive BH binary candidates with the radial velocity shift of the broad emission lines \citep{Eracleous12,Liu14,Runnoe17,Guo19a}.

In order to discover CLQs with prominent UV lines (e.g., \civ) and estimate the BH mass at high redshift, Criterion 3 is set to ensure the \civ\ line is in the SDSS wavelength coverage. The redshift distribution is shown in Figure \ref{fig:sn} (left panel), whose peak is around 2.2 (a feature designed by the baryon oscillation spectroscopic survey to map the large-scale structure traced by the \lya\ forest). We measure the SNR of the \civ\ line in [1500,1600] \AA\ and a line-free continuum window [1450, 1460] \AA\ for those spectra that satisfy criteria 1-3. Figure \ref{fig:sn} (middle panel) presents the SNR distributions of continuum around 1450\AA\ and \civ\ line. We cut the $\rm SNR_{\rm C\,IV} >$5 and $\rm SNR_{\rm 1450} >$3 for emission-line and continuum SNRs (Criterion 4). 

Previous investigations \citep[e.g.,][]{Yang18} indicate that CLQs presenting appearance/disappearance of broad emission lines are always accompanied with continuum variations, although sometimes the variation is tiny. Criterion 5 requires the continuum variability at 1450\AA, $\delta$V $\equiv$ $\rm \frac{Max-Min}{Mean} >1 $, where Mean = $\rm \frac{1}{2}(Max+Min)$. This is equivalent to Max/Min $>$ 3, approximately consistent with previous EVQ/CLQ selection ($|\Delta g| $ $>1 $ mag) in other works \citep{Macleod16,Rumbaugh18} considering the median redshift is around 2.2 in our sample. Note that the average number of repeat observations for each object after this criterion rapidly increases to 5 from 2, indicating that many other EVQs were not selected, probably due to the lack of frequent observations. The distribution of continuum variability is also plotted in Figure \ref{fig:basic_para} (right panel), which rejects $\sim$98\% objects with little/mild variability. This variability cut may miss some CLQs with tiny continuum variation \citep[e.g., the \mgii\ CLQ with variability of 0.1 mag in $r-$band,][]{Guo19a}.  Criterion 6 removes all the objects in the SDSS-RM program (centered at RA = 213.704, DEC = +53.083 with a field of view 7 deg$^2$) since only a low-redshift CLQ were discovered in the SDSS-RM sample \citep{Wang18,Dexter19b}, and more careful analysis could be conducted for SDSS-RM EVQs with extensive spectroscopies and simultaneous photometries over 6 yrs \citep{Shen15}.

Finally, we are left with 951 EVQ objects with 1818 spectra (also see Table \ref{tab:selection}). For these 951 objects,  we recycle every epoch rejected in Criterion 4 due to the low spectral SNR, since these low-SNR epochs are likely to be the faint-states of real CLQs experiencing large continuum variations. The SNR cut ensures that at least one epoch in a object meets the SNR criterion for BH mass measurement.

\subsection{Spectral Flux Calibration Problem}
In Appendix \ref{sec:scatter}, we demonstrate that SDSS high-redshift quasar spectra are subject to an intrinsic flux calibration scatter of $\sim$ 20\%. The variability criterion ($\delta$V $>$1) is much larger than the intrinsic calibration scatter, which should have less influence on our EVQ/CLQ selection. We also describe the fiber-drop issue in Appendix \ref{sec:fiberdrop}, which leads to a significant reduction of the spectral flux. Epochs suffering from fiber-drop with huge flux reduction can mimic EVQs/CLQs (see Figure \ref{fig:fake_CLQ}). Therefore, cross-identification, e.g., with concurrent photometries, is crucial for robust EVQ/CLQ selection.

\begin{figure*}
\hspace*{-2cm} 
\centering
\includegraphics[width=\paperwidth]{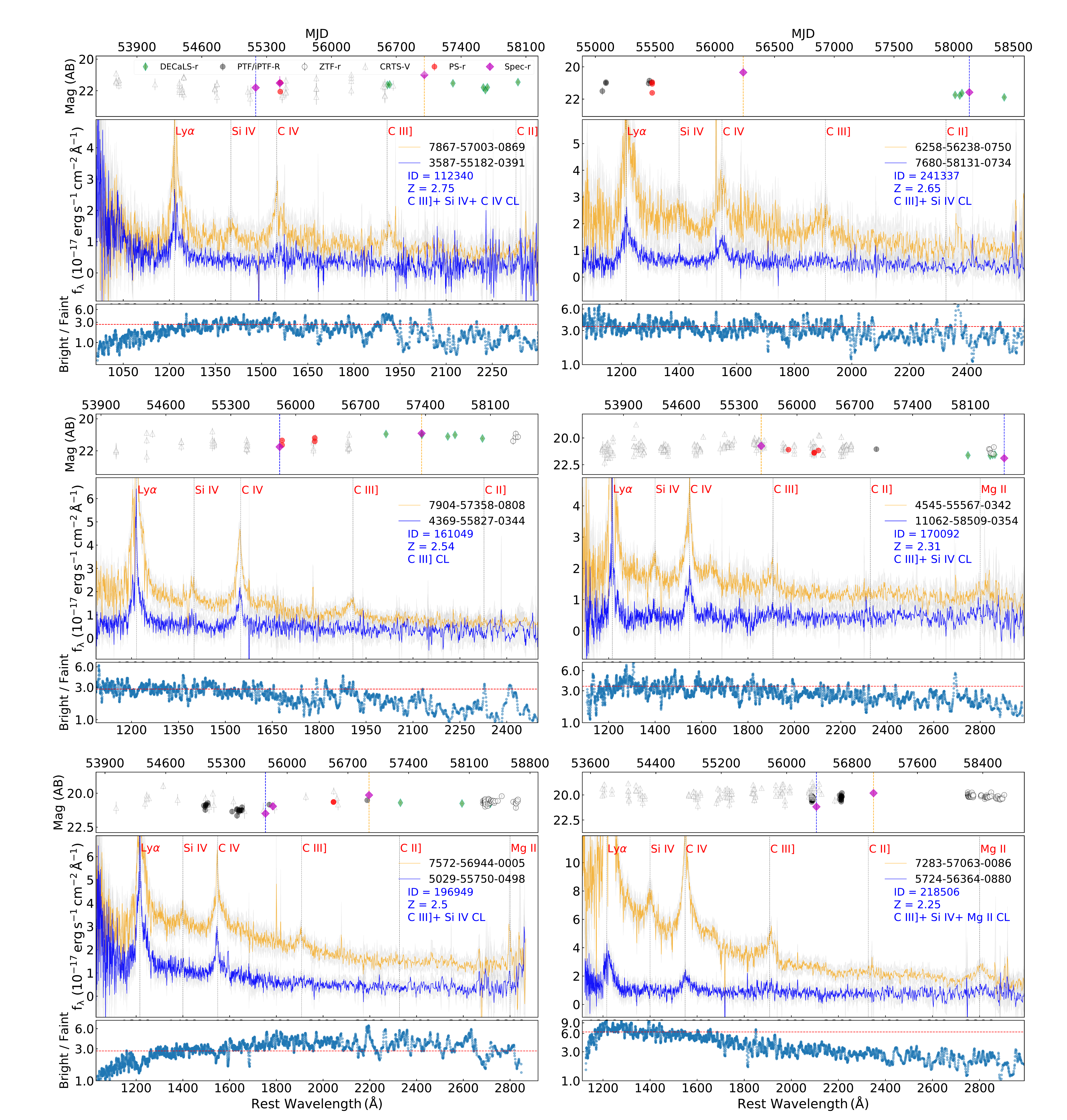}
\caption{Examples of CLQs. We demonstrate six out of 23 CLQs showing transitions in different lines (listed in blue text). The rest are shown in Appendix \ref{app:rest}. Each panel presents the brightest and faintest epochs (middle), $r-$band multi-survey light curve corrected for the difference among photometric systems (top), and the spectral ratio between bright and faint epochs in each object (bottom). All the spectra are smoothed with box-car of 5 pixels for clarity except the original SDSS spectra (light grey). The red line in bottom panel is the flux ratio at 1450\ \AA\ to guide the eye.}
\label{fig:example}
\end{figure*}

\subsection{Visual Inspection for EVQs/CLQs} \label{sec:vis}
We then visually inspect the brightest and faintest epochs to screen for real EVQs/CLQs following the rules: 1) objects with significant spectral changes, e.g., spectral slope or emission/absorption line profiles; 2) or the spectral synthetic magnitude of the faintest epoch is consistent with the nearest multi-survey photometric/synthetic magnitudes (e.g., within a window of $\pm$ 1 yr). The brightest spectrum in each object is usually consistent with the concurrent photometries and barely affected by the fiber-drop problem. Also, we checked that multi-survey photometries are still reliable even when they exceed the 5$\sigma$ limiting magnitude (e.g., CRTS photometries fainter than 21 mag are still broadly consistent with SDSS spectrophotometry). This selection may miss some real EVQs/CLQs with neither concurrent photometries nor obvious spectral variation (uncertain epochs). However, our selection is not pursuing the completeness of EVQs/CLQs at high redshift, but rather a pure sample to investigate their physical properties. Totally, these two rules effectively reject $\sim$ 10\% fiber-loose objects and $\sim$ 50\% uncertain epochs. In addition, we also eliminate $<5\%$ problematic spectra due to lack of data or strong telluric-line residuals.


This leaves 348 significant EVQs. Then we use the spectral decomposition technique to measure the continuum/emission-line properties (see \S \ref{sec:fit}). As suggested, CLQs are likely to be a part of the EVQ tail of normal quasars and EVQs are good candidates to search for CLQs \citep{Rumbaugh18}. However, the definition of a CLQ is pretty ambiguous and arbitrary, as well as strongly dependent on the spectral SNR \citep[][also see \S \ref{sec:SNR}]{Guo20}. Therefore, we here adopt a temporary definition \footnote{Because we eventually find that CLQs and EVQs are the same population, and we recommend to use EVQ to refer to the whole extremely variable quasar population.} of CLQ, analogous to what has been done at low redshift, to explore the properties of high-redshift CLQs. We quantitatively define CLQs as those whose faint-epoch line flux (at least one prominent broad UV line, i.e., \lya, \SiIV, \civ, \ciii, \mgii) is consistent with zero within its uncertainty ($flux < 3\sigma$). This yields 23 CLQs showing broad-line appearance/disappearance, or are very close to completing this transition as shown in Figure \ref{fig:example} and \ref{fig:example2}-\ref{fig:example1}. In total, our sample includes 1 \civ, 19 \ciii, 16 \SiIV\ and 4 \mgii\ CLQs with some individual objects showing CL behavior in more than one line. Note that we also occasionally discover an obvious \cii\ CLQ in Figure \ref{fig:example2}.

The broad line component in CLQs usually gradually transitions from a strong broad emission line, to a weak one easily swamped by noise, and finally to an absorption line with the continuum decreasing continuously. Figure \ref{fig:example} exhibits six randomly selected examples of high-redshift CLQs with changes in different lines. In each panel, we present the brightest and faintest epochs (middle panel), spectral ratio (bottom panel), and $r-$band light curve (top panel). Most objects present the typical bluer-when-brighter (BWB) trend \citep[i.e., quasar continuum becomes bluer when it gets brighter,][]{Sun14,Guo16b,Cai19}, accompanied with at least one transitioning line. The CL behavior is confirmed either by spectral change (see spectral ratio) or concurrent photometries in $g-$band light curve. The weak UV emission lines (e.g., 19 \ciii, 16 \SiIV, and 4 \mgii\ CLQs) are usually easier to change than the strong lines, for instance, \civ\ (only ID = 112340 in Figure \ref{fig:example} if not very clean) and \lya\ (no \lya\ CLQ). During the transitions, the line core component is usually more variable than the line wings, e.g., line core of \ciii\ in ID = 23445 in Figure \ref{fig:example2} has transferred into an absorption line while the line wing is still persistent, which is consistent with the photoionization model.

\subsection{Spectral Fitting}\label{sec:fit}
To determine the variability properties of the broad emission lines and to measure their profiles for virial BH mass estimates, we fit spectral models following \cite{Shen19} using a software \texttt{PyQSOFit} \citep{Guo18}. The model is a linear combination of a power-law continuum, a 3rd-order polynomial (to account for reddening), a pseudo continuum constructed from UV/optical \FeII\ emission templates, and single or multiple Gaussians for the emission lines. We do not include a host galaxy component since such a component is negligible for quasars at $z>$ 1.5. Balmer continuum is also not included due to its major contribution is in $\sim$ 3000 to 3600 \AA, which exceeds our wavelength coverage. 

We adopt a global fit to model the relatively emission-line free region to quantify the continuum. We then fit multiple Gaussian models to the continuum-subtracted spectrum on individual line complexes. Table \ref{tab:linefit} lists the detailed information of line complexes and the fitting parameters. In each line complex, we simultaneously fit a set of Gaussians to individual lines with a boundary of 1200 \kms\ to separate broad and narrow components. The measurement uncertainties of the spectral properties (e.g., continuum luminosity, line width, etc.) are estimated with a Monte Carlo approach by repeatedly fitting the spectrum for 50 trials perturbed with a zero-mean random Gaussian noise whose $\sigma$ is the original error in each pixel. All the related measurements are saved in our EVQ catalog (see \S \ref{sec:catalog} and Table \ref{tab:format}).

Since high-redshift quasars are potentially subject to broad/narrow absorption lines that may bias the continuum and emission-line fits, we perform following steps to amend this effect: 1) mask pixels that are 3$\sigma$ below the SDSS built-in model, which is constructed by principal component analysis using quasar/galaxy templates with {\tt REDMONSTER}; 2) perform one iteration for the continuum fit to reject pixels that fall 3$\sigma$ below the previous fit; 3) reject pixels in emission-line regions fall 3$\sigma$ below the continuum model for the emission line fitting. The combination of these criteria significantly alleviate measurement bias from narrow/broad absorption lines.

\begin{table}
\caption{Line Fitting Parameters}\label{tab:linefit}
\centering
\begin{tabular}{cccc}
\hline\hline
Line Complex & Fitting Range [\AA] & Line &  $n_{\rm gauss}$ \\
(1) & (2) & (3) & (4)  \\
\hline
\SiIV\ & 1290-1450 & Broad  \SiIV/\oiv  & 2\\
& & \CII\ 1335 & 1\\
& & \oi\ 1304 & 1\\
\civ\ & 1500-1700 & Broad \civ\ & 3 \\
 & & Broad \HeII\,1640 & 1 \\
 & & Narrow \HeII\,1640 & 1 \\
 & & Broad \OIII\,1663 & 1 \\
 & & Narrow \OIII\,1663 & 1 \\
\ciii\ & 1700-1970 & Broad \ciii & 2 \\
 & & \SiIII\,1892 & 1 \\
 & & \AlIII\,1857 & 1 \\
 & & \SiII\,1816 & 1 \\
 & & \NIII\,1750 & 1 \\
 & & \NIV\,1718 & 1 \\
 \mgii & 2700-2900 & Broad \mgii\ & 2 \\
  & & Narrow \mgii\ & 1 \\
\hline
\hline
\end{tabular}
\end{table}

\subsection{The EVQ Catalog}\label{sec:catalog}
We have tabulated the measured quantities of 348 EVQs including 23 CLQs from the spectral fitting in the online catalog of this paper, and the catalog format is listed in Table \ref{tab:format}. Below we describe the specifics of the cataloged quantities.\\

1. Object ID: the repeat catalog index of SDSS DR 16.\\
2-4. Spectroscopic plate ID, Modified Julian Date (MJD) and fiber ID: the combination of plate-MJD-fiber locates a particular spectroscopic observation in SDSS.\\
5-7. RA, Dec (in J2000.0) and redshift. All the spectra are flagged with zWarning = 0, indicating a good quality of the spectroscopic redshift measurement.\\
8. Nepoch: number of repeat spectroscopic observations. \\
9. FIRST detection: If there is a source in the FIRST radio catalog (version December 2014) within 2.0\arcsec\ of the quasar position. 1 if detected in FIRST; 0 if not; $-1$ if the quasar lies outside of the FIRST footprint.\\
10. FIRST peak flux density: observed flux in 20 cm; $-1$ = not in FIRST footprint; 0 = FIRST undetected.\\
11. Flag: 1 = brightest epoch; $-1$ = faintest epoch.\\
12. $\delta$V: the continuum variability at 1450\AA; $\delta$V $\equiv$ $\rm \frac{Max-Min}{0.5(Max+Min)}$. \\
13. $\rm f_{1450}$: observed flux in rest frame at 1450\AA\ without corrections of the intrinsic extinction/reddening.\\
14-15. $\rm SNR_{1450}$ and $\rm SNR_{\rm CIV}$: the median SN per pixel in continuum region [1425,1475] \AA\ and \civ\ emission line region [1500,1600] \AA. \\
16-17. Continuum lunimosity and uncertainties at 1350\AA. \\
18-21. Line FWHM, rest frame EW, and their uncertainties for \civ; $-1$ if not available.\\
22-23. Virial BH masses and uncertainties using calibrations of \civ\ (VP06); $-1$ if not available. \\
24-25. Average virial BH masses and uncertainties of the brightest and faintest epochs. If only brightest epoch is available, we use $M_{\rm BH,Mean}$ = $\rm M_{\rm BH,Bright} - 0.15$ dex, otherwise, $M_{\rm BH,Mean}$ = $M_{\rm BH,Faint} + 0.15$ dex to mitigate the BH mass discrepancy of 0.3 dex in bright/faint states.\\
26. Eddington ratio: based on the mean virial BH mass.\\
27. The reduced $\chi^2_{\nu}$ for the \civ\ fitting in [1500,1700] \AA.\\
28. Type: 0 = CLQ; 1 = EVQ. \\

\section{EVQ/CLQ Properties} \label{sec:result}
\cite{Rumbaugh18} and \cite{luo20} have extensively explored the distinctions between EVQs and normal quasars with light curves over $\sim$ 16 yrs and suggested that EVQs seem to be in the tail of a continuous distribution of quasar properties, rather than standing out as a distinct population. One of our main purposes of this work is to further examine the possible difference between the EVQs and CLQs, although with a relatively smaller (but currently the largest) sample size of high-redshift CLQs. Note that we use EVQ to denote the non-CL EVQs in the following sections.

\subsection{Basic Properties}
\begin{figure*}
\hspace{1mm}
\centering
\includegraphics[width=18.cm]{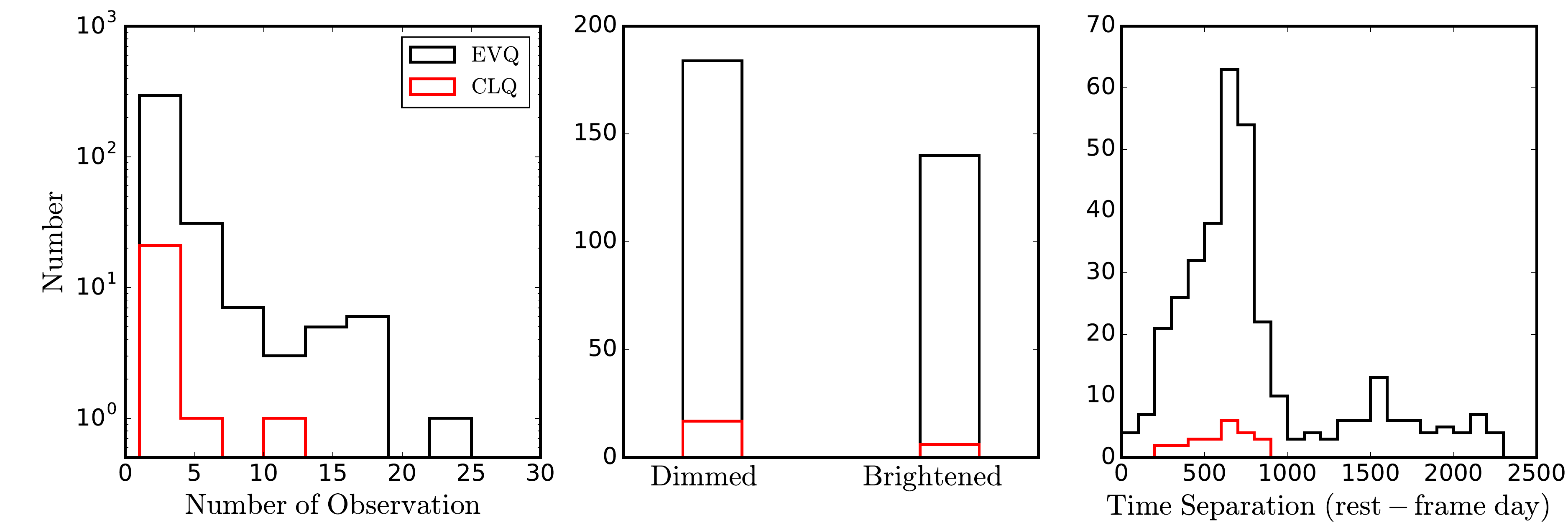}
\caption{Distributions of number of observation, brightened and dimmed objects and maximum time separation in the rest frame for EVQs/CLQs. Black and red lines represent EVQ and CLQ, respectively. }
\label{fig:basic_para}
\end{figure*}

A sample of 348 EVQs are selected through repeat observations in SDSS with $\delta V > 1$. As shown in the first two panels of Figure \ref{fig:basic_para}, most EVQs/CLQs concentrate upon number of epoch less than 5. We found 140 (6) EVQs (CLQs) brightened and 185 (17) EVQs (CLQs) dimmed. The last panel shows the distribution of the rest frame time separations between the brightest and faintest epochs of each EVQ/CLQ. The short-term large variability is intrinsically rare, such that only a few objects with time separation less than 100 days are observed. The drop of objects beyond $\sim$1000 days ($\sim$3000 days in observed frame considering a median redshift of 2.2) is largely due to the quasar selection bias before DR9 (2010), which primarily selected quasars with z $<$ 2, rather than a real rarity of EVQs at these timescales. All the CLQs are distributed in a range of $\sim$100 to 900 days in the rest frame.

319 out of 348 EVQs are within the Faint Images of the Radio Sky at Twenty-centimeters (FIRST) coverage, and 9 of them are radio detected, resulting a radio fraction of $\sim$ 3\%, which is consistent with previous results that the radio fraction of quasars will gradually decrease (e.g., $<$5\% at $z >$ 1 ) with increasing redshift and decreasing luminosity \citep{Jiang07}.

\subsection{Emission Line Properties}
The spectral fitting technique is used to decompose the brightest and faintest spectra of 348 EVQs/CLQs, as described above in \S \ref{sec:fit}. Measurements of \SiIV, \ciii\ and \mgii\ are not as accurate as \civ\ due to their weaker intensity or even disappeared in some CLQs. In addition, \mgii\ always very closely approaches or exceeds the long-wavelength edge, leading to only about half of our objects having reliable \mgii\ spectral measurements. Therefore, we focus on the measurements based on \civ\ in the following sections, and related measurements are listed in the EVQ catalog (Table \ref{tab:format}).

\subsubsection{$\delta$V, $L_{\rm1350\mbox{}}$, EW and FWHM}\label{sec:L}
\begin{figure*}
\centering
\includegraphics[width=12.cm]{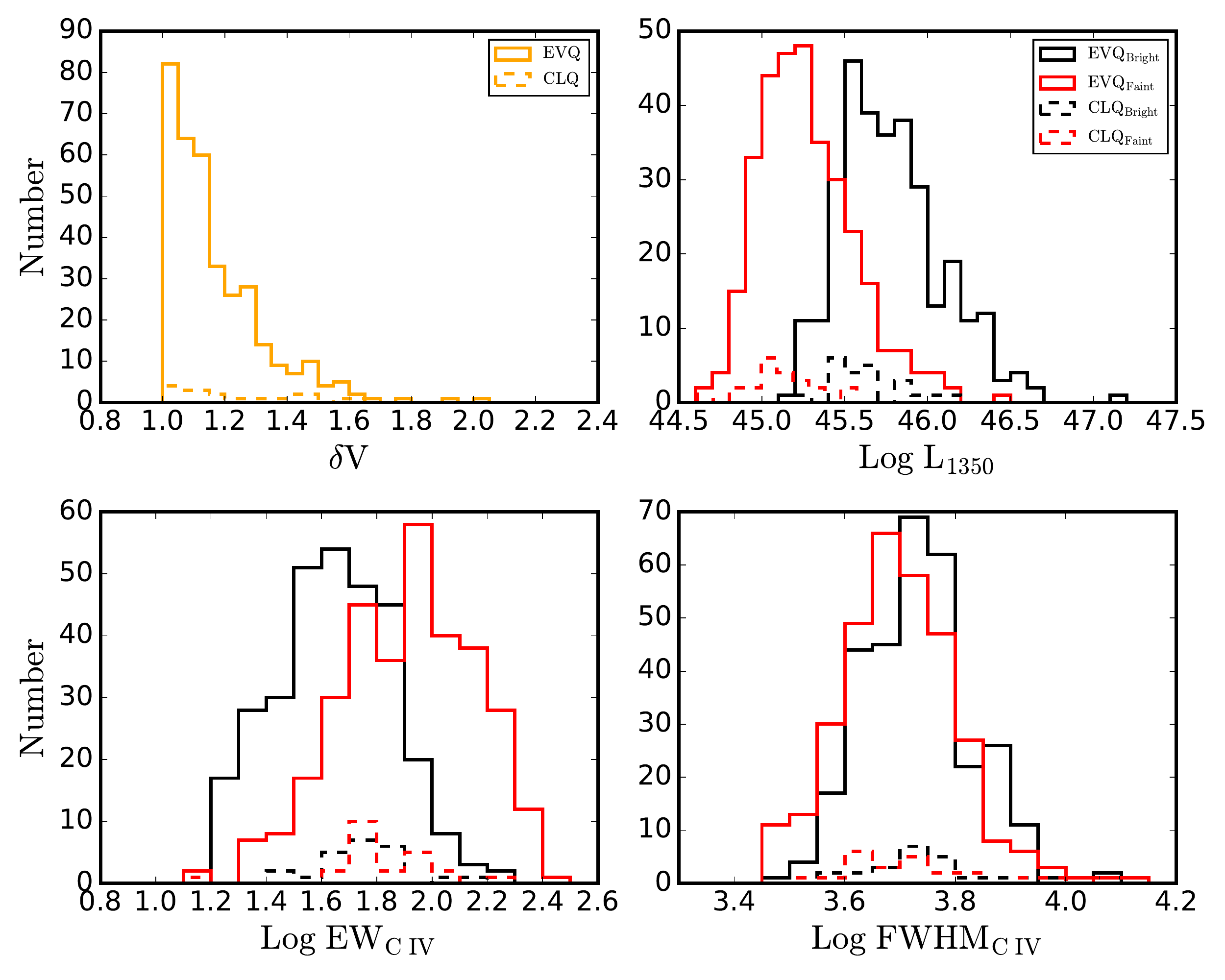}
\caption{Distributions of continuum luminosity at 1350\AA, EW and FWHM of \civ. Black and red solid (dashed) lines denote the bright and faint states of EVQs (CLQs). Their mean values and the same-state K-S test results are listed in Table \ref{tab:EVQ_para}, respectively.}
\label{fig:EVQ_para}
\end{figure*}

Figure \ref{fig:EVQ_para} presents the distributions of continuum variability at rest frame 1450\AA, continuum luminosity at rest frame 1350\AA, Equivalent Width (EW) and Full Width at Half Maximum (FWHM) of the broad component of \civ. The corresponding means and uncertainties are listed in Table \ref{tab:EVQ_para}. The continuum variabilities $\delta$V of EVQ/CLQ show no significant difference considering the errors, and the CLQs are not biased to the largest variability end. As expected, the average $L_{\rm 1350}$ (i.e., $10^{45.78}$ and $10^{45.27}$ \erg\ for the bright and faint states) are much fainter than that of the normal high-redshift quasars (i.e., $\sim$46.15 \erg) in DR 7 \citep{Shen11}, given the anti-correlation between the Eddington ratio and variability \citep[e.g.,][]{Rumbaugh18,Guo16b}. The average difference of 0.5 dex between bright and faint states is consistent with our variability criterion (i.e., continuum flux at 1450\AA: Max/Min $>$ 3). The largest variability in our sample reaches $\sim$1 dex in SDSS baseline over $\sim$18 yrs. The \civ\ EW distributions of bright and faint states of EVQs are also well separated by $\sim$0.3 dex. According to the well-known Baldwin effect \citep{Baldwin77}, $L_{\rm con} \propto 1/EW$, fainter EVQs exhibit higher emission line EWs, consistent with the \civ\ result in the second panel. The third panel shows that the FWHM distribution for the faint EVQs is on average slightly lower than that of the bright EVQs. This is the opposite of what the breathing model predicts: $L_{\rm con}$ $\propto$ 1/FWHM, discovered in Balmer lines \citep[e.g.,][]{Cackett06,Denney09,Barth15,Wang20}. However, the anti-breathing of \civ\ is not newly discovered and has been reported in previous investigations \citep[e.g.,][]{Wills93,Wilhite06,Shen13,Denney12,Guo16b,Wang20}. The origin of this anti-breathing phenomenon is likely due to the existing of a non-reverberating component in \civ, which may originate from the intermediate line region (between the broad and narrow emitting regions with FWHM $\sim$ 1200 to 2000 \kms) or an outflow (see details in \S \ref{sec:mass} \& \ref{sec:civ_mass}).

\begin{table*}
\caption{EVQ and CLQ properties}\label{tab:EVQ_para}
\centering
\begin{tabular}{lccccccc}
\hline\hline
&EVQ & CLQ & EVQ & CLQ & \multicolumn{2}{c}{KS test between EVQ and CLQ}\\
& \multicolumn{2}{c}{Bright state} & \multicolumn{2}{c}{Faint state}  & Bright state & Faint state \\
\hline
$\delta V$ & 1.17$\pm$0.16& 1.27$\pm$0.24&1.17$\pm$0.16 &1.27$\pm$0.24 & $6.4 \times 10^{-2}$ & $6.4 \times 10^{-2}$\\ 
$\rm LogL_{\rm 1350}$ (\erg) & 45.78$\pm$0.31  & 45.62$\pm$0.21 & 45.27$\pm$0.29& 45.08$\pm$0.22 & $1.5 \times 10^{-2}$ &$5.3 \times 10^{-3}$\\
$\rm EW_{\rm CIV}$ (\AA) & 50$\pm$26 & 62$\pm$26 & 94$\pm$50 &73$\pm$32 & $5.5 \times 10^{-3}$ &$5.2 \times 10^{-2}$\\
$\rm FWHM_{\rm CIV} $ (\kms)& 5522$\pm$1322 & 5813$\pm$1591 & 5144$\pm$1349 &5614$\pm$1975&$5.2 \times 10^{-2}$&$8.3 \times 10^{-1}$\\
$\rm Log\lambda_{Edd}$ & $-$0.64$\pm$0.23 & $-$0.76$\pm$0.23 & $-$1.11$\pm$0.24 & $-$1.30$\pm$0.25 & $1.2 \times 10^{-2}$&$3.6 \times 10^{-2}$\\

\hline\hline
\end{tabular}
\end{table*}

\subsubsection{A non-reverberating component in \civ}
\begin{figure}
\centering
\includegraphics[width=9.cm]{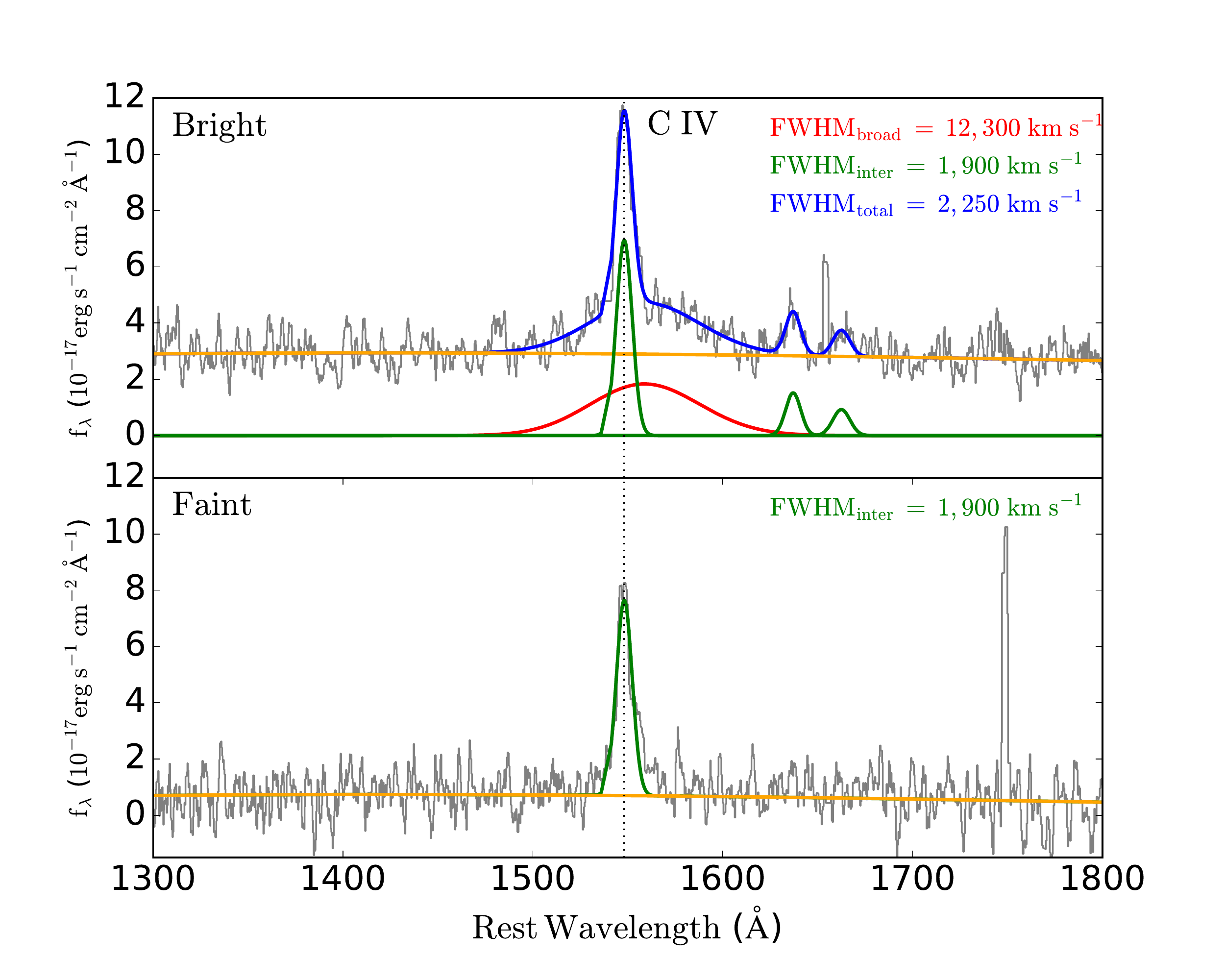}
\caption{Example of an anti-breathing \civ\ (SDSS J225519.57$-$010009.4). The red broad component (FWHM = 12,300 \kms) has disappeared as its continuum decreased within 2.1 years in the rest frame, indicating a positive correlation between $L_{\rm con}$ and $FWHM_{\rm C\ IV}$ (i.e., anti-breathing). The bright state is best fitted with four Gaussian profiles (two for \civ, and one each for \HeII\ 1640\AA\ and \OIII\ 1663\AA), while the width of all the green Gaussian profiles in both the bright and faint states are tied together. }
\label{fig:IL}
\end{figure}

Figure \ref{fig:IL} presents a clear example of an EVQ with a non-reverberating component, leading to the \civ\ anti-breathing phenomenon \citep[e.g.,][]{Wang20}. \civ\ consists of a very broad component (red; FWHM = 12,300 \kms) and an Intermediate Broad Component (IBC, green; FWHM = 1900 \kms) in the bright state, while the very broad component has disappeared in the faint state, resulting in the total FWHM being significantly reduced (i.e., 2250 \kms\ to 1900 \kms; anti-breathing). Note that the green broad component in the faint epoch is tied to that in the bright state assuming this IBC is constant. The goodness of fit with $\chi^2_{\nu}$ = 2.6 in the faint spectrum indicates the IBC indeed does not change too much even as an EVQ. If one considers the IBC as a ``narrow" component in \civ, it could be classified as a generalized CLQ according to the traditional definition.    

\subsubsection{Same-state comparison between CLQs and EVQs} \label{sec:ks}
Comparing the same-state property distributions of EVQs and CLQs in Figure \ref{fig:EVQ_para} with Kolmogorov-Smirnov (K-S) test, we find that most of p values of luminosity, EW and FWHM are not small than 0.01 in Table \ref{tab:EVQ_para}; that is, at significance level of 0.01, we cannot reject the null hypothesis that EVQs and CLQs are drawn from the same continuous distribution. On the other hand, the same-state CLQs on average present similar but slightly lower continuum luminosity and broader FWHM in Table \ref{tab:EVQ_para}, hint of possible lower Eddington ratios for CLQs. However, those slight differences are still much smaller than the uncertainties. We therefore suggest that there is no strong evidence that EVQs and CLQs are different in these various properties, supporting that CLQs are likely to be a subset of EVQs.

\subsection{BH Mass}\label{sec:mass}
We estimate the AGN BH mass using the single-epoch estimator assuming virialized motion in the broad-line region (BLR) clouds \citep{Shen13}. With the continuum luminosity as a proxy for the BLR radius and the broad emission line width, characterized by the FHWM, as an indicator of the virial velocity, the virial mass estimate can be expressed as,
\begin{equation}\label{eq:virialmass}
    \log \bigg(\frac{M_{\rm BH}}{M_{\odot}} \bigg) = a + b \log \bigg(\frac{\lambda L_{\lambda}}{10^{44}\,{\rm erg\,s^{-1}}}\bigg) + 2 \log \bigg(\frac{\rm{FWHM}}{{\rm km\,s^{-1}}}\bigg),
\end{equation}
where $L_{\lambda}=L_{{\rm 1350}}$ for \civ. The coefficients $a$ and $b$ are empirically calibrated based on the local reverberation mapped AGNs and scaling relations between UV and optical lines. We adopt the calibrations of \cite{Vestergaard06} (VP06) with a = 0.66 and b =0.53 for \civ\ line. 

According to the photoionization model \citep{Goad93, Baldwin95,korista00,Guo20}, the observed broad-line flux is actually dominated by the BLR clouds in a relatively narrow ring assuming a disk-like BLR\citep{Gravity18}, where the clouds are best able to reprocess the incident continuum. The average cloud distance to the center will increase when the continuum luminosity increases, yielding a slower rotational velocity of the clouds (FWHM), namely the broad-line breathing model. However, the \civ\ FWHM increases when the quasar brightens in Figure \ref{fig:EVQ_para}, which presents an opposite behavior. 

\begin{figure}
\centering
\includegraphics[width=9.cm]{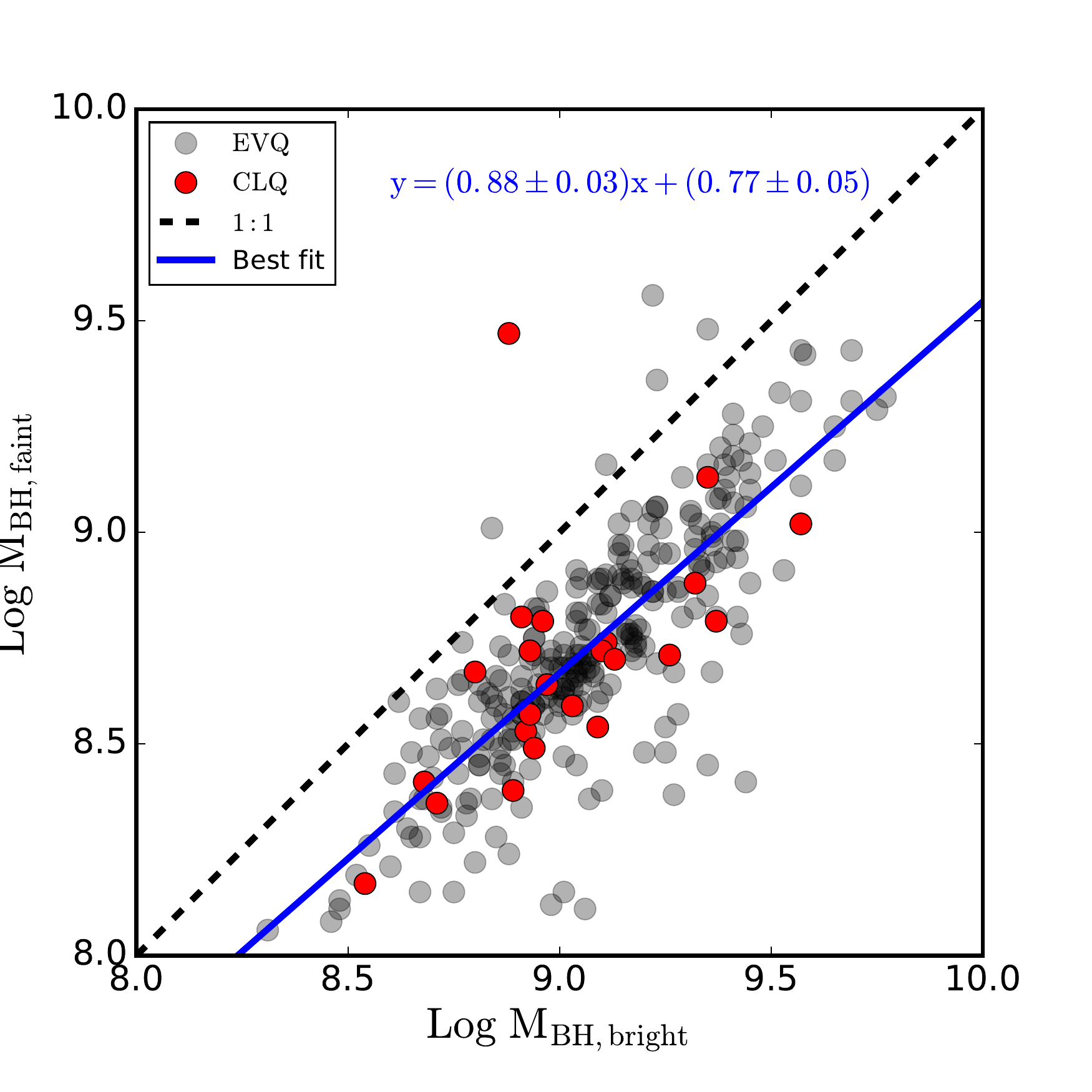}
\caption{Discrepancy of BH masses in bright and faint states. The blue line is the best linear fit for EVQ+CLQ sample and the coefficient errors are estimated with the bootstrap method. The anti-breathing behavior in \civ\ results in a $\sim$0.3 dex offset relative to the 1:1 BH mass ratio. The red outlier with the highest faint-state BH mass is a \civ\ CLQ (ID = 112340).}
\label{fig:BH}
\end{figure}

This anti-breathing behavior of \civ\ in Figure \ref{fig:EVQ_para} will result in a serious problem: luminosity-dependent BH mass \citep[i.e., higher BH mass with increasing continuum luminosity,][]{Shen13}. Figure \ref{fig:BH} presents the BH masses measured in the brightest/faintest states for each EVQ/CLQ. The best linear fit of the EVQ sample is significantly offset from a 1-to-1 relation, with an average difference of $\sim$ 0.3 dex between the bright and faint BH masses. CLQs are randomly distributed in EVQs without any obvious difference, except the outlier with the highest faint-state BH mass, which is a \civ\ CLQ (ID = 112340). We simply adopt the average as our fiducial BH mass, since there is still no ideal solution for this issue to date (see discussion in Section \S \ref{sec:civ_mass}). However, we caution that this BH mass will suffer an extra scatter (0.3 dex) introduced by quasar variability.

\subsection{Eddington Ratio}\label{sec:edd}
With the continuum luminosities at 1350\AA\ derived from the spectral fitting for the bright and faint states, we can estimate the bolometric luminosity with $L_{\rm bol}$ = 3.81$L_{\rm 1350}$ according to \cite{Richards06}, hence the Eddington ratio $\lambda_{\rm Edd}=L_{{\rm bol}}/L_{{\rm Edd}}$, where $L_{\rm Edd}= 1.26\times 10^{38} M_{{\rm BH,Mean}}/M_{\odot}$ $\rm erg\, s^{-1}$ based on the averaged BH mass.

\begin{figure*}
\centering
\includegraphics[width=19.cm]{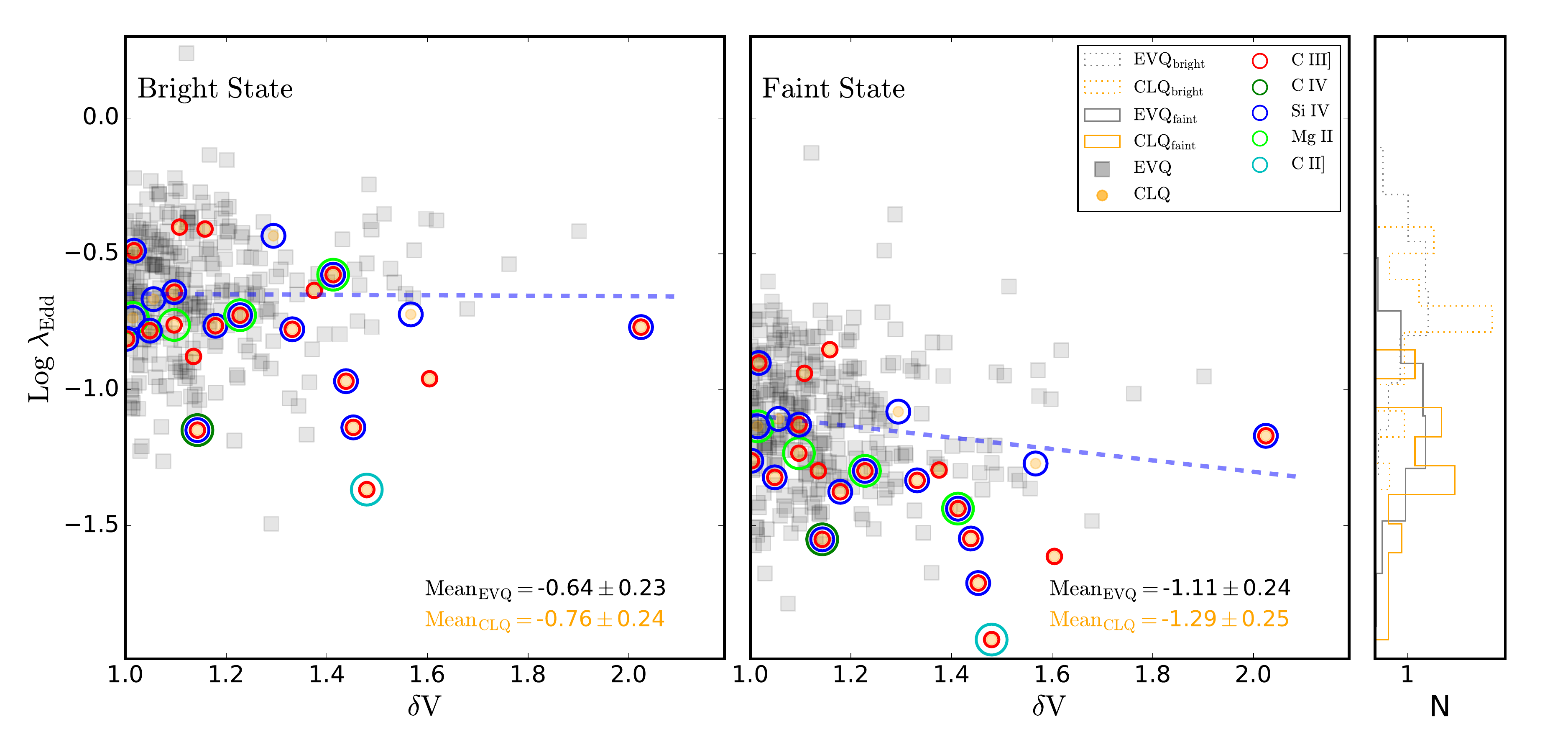}
\caption{Eddington ratio versus spectral variation at 1450 \AA. The CLQs show a relatively lower average Eddington ratio both in bright and faint states compared to normal EVQs (also see the normalized histgram). The Eddington ratio is computed with the average BH mass of the bright/faint states, and the mean values are listed in the lower right corner and Table \ref{tab:EVQ_para}. The whole EVQ sample shows an anti-correlation between the continuum variation and Eddington ratio with slopes of $-0.01\pm0.03$ (bright) and $-0.21\pm0.05$ (faint), indicated by the blue dashed lines. Different CLQs are marked with circles in different colours. }
\label{fig:Edd}
\end{figure*}

Extensive investigations have demonstrated an anti-correlation between variability and Eddington ratio in normal quasars \citep[e.g.,][]{Wilhite08,MacLeod10,Ai10,Guo14b}, and \cite{Rumbaugh18} exhibit this negative relation further extending to EVQs with $\sim$1000 photometrically selected quasars. Figure \ref{fig:Edd} also confirms this relation between the $M_{\rm BH,Mean}$-based Eddington ratio and maximum spectral variation at 1450 \AA , especially in the faint state. The most striking feature is that the average Eddington ratio of CLQs is lower than that of EVQs, particularly in the faint state ($\sim$ 0.2 dex) but only with a confidence level of $\lesssim$ 1$\sigma$. Actually, this feature is implicitly inferred by their slightly lower luminosity and broader FWHM relative to the normal EVQs in Table \ref{tab:EVQ_para} and Figure \ref{fig:EVQ_para}. However, the sample size of CLQs is still not large enough, and the bolometric correction may be different when the spectral energy distribution changes. Furthermore, the uncertainty of the Eddington ratio for individual objects is quite large due to the intrinsic BH mass uncertainty \citep[$\sim$0.4 dex,][]{Shen13} and the extra variability-induced scatter ($\sim$0.3 dex) in \S\ref{sec:mass}. Therefore, we suggest that the evidence here is not conclusive enough to prove that CLQs is a subset of EVQs with less efficient accretion.             

Table \ref{tab:EVQ_para} shows the same-state comparison between CLQ and EVQ with K-S test. P values of maximum variability and Eddington ratio are all not smaller than 0.01, which indicates that we cannot reject the null hypothesise at 0.01 significance level. This is generally consistent with the K-S results of continuum luminosity, EW and FWHM, further supporting that CLQ and EVQ are likely to be the same population.

We also distinguish different CLQs with different circles in Figure \ref{fig:Edd}. The majority of CL behaviors are \ciii\ and \SiIV\ CLQs primarily due to their relatively weak line intensity (see the details in \S \ref{sec:CLQ}). In addition, the critical Eddington ratio for state transition of UV lines seems to be around 0.01 to 0.1, and the critical $\lambda_{\rm Edd}$ for \ciii\ and \SiIV\ is definitely higher than \civ\ and \lya, since the broad \civ\ and \lya\ are still persistent in almost all EVQs. \cii\ CL behavior is not common in our sample and most \cii\ lines have already disappeared in bright state of \ciii\ CLQs due to its low luminosity, easily swamped by spectral noise. Therefore, we suggest that the critical Eddington ratio of CL behavior in different lines basically follows $\lambda_{\rm Edd,\ C\ II]} > \lambda_{\rm Edd,\ Si\ IV,\ C\ III],\ Mg\ II} > \lambda_{\rm Edd,\ C\ IV} > \lambda_{\rm Edd,\ Ly\alpha}$.

\subsection{Composite Spectra}
\begin{figure*}
\centering
\includegraphics[width=18.cm]{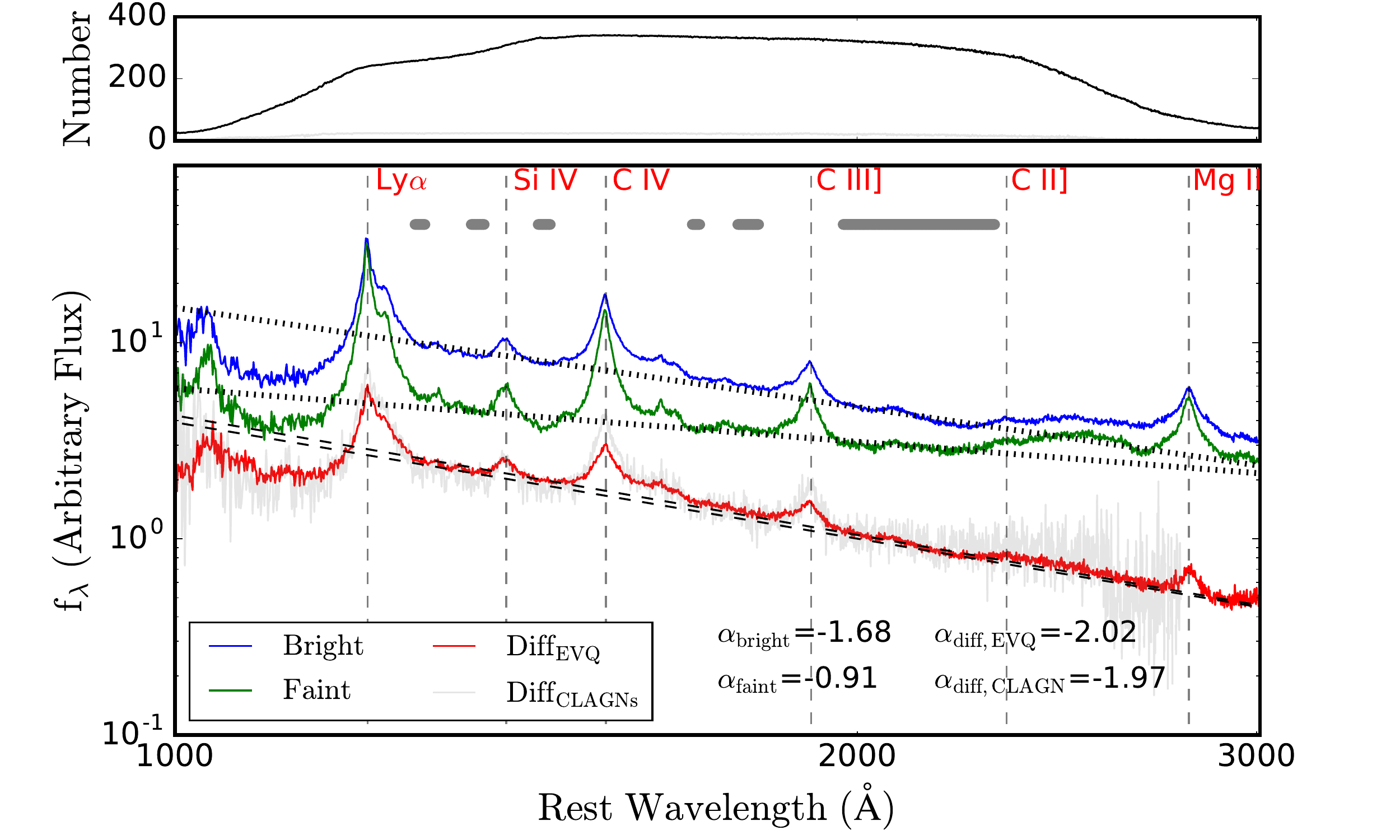}
\caption{Geometric mean composite spectra for bright and faint states of EVQs, represented in blue and green colors in the lower panel, respectively. The red (grey) curve is the composite difference spectrum for EVQs (CLQs). All the composite spectra are shifted for clarity. The black dashed/dotted lines are power-law fits to the line-free continuum windows (grey bars). The upper small panel shows the number of objects (black for EVQ and gray for CLQ) in each wavelength bin (1 \AA) when constructing the corresponding composite spectra. Note that noisy pixels on the edges are removed during constructing the composites.
}
\label{fig:composite}
\end{figure*}

To further study the spectral evolution in EVQs, we construct composite spectra separately for bright and faint epochs. In this work, we use the geometric mean spectrum in order to preserve the global continuum shape, instead of the arithmetic mean spectrum, which preserves the relative fluxes of emission features. The geometric mean spectrum is generated following the procedure in \cite{VandenBerk01}, including correcting galactic extinction, rebinning the individual spectra to the source rest frame, scaling the spectra, and finally stacking the spectra into the composite with $<f_{\lambda}> = (\prod_{i=0}^{n} f_{\lambda,i})^{1/n}$, where $f_{\lambda,i}$ is the flux of each spectrum at wavelength $\lambda$ and n is the total number of spectra in spectral bins. The composite difference spectrum is also derived in a similar way. We fit composite spectra with a power-law to several line free regions between 1275 and 2400\AA\ (see the gray bars in  Figure \ref{fig:composite}), which are the most reliable regions. 

Figure \ref{fig:composite} presents the bright, faint and difference composite spectra of EVQs and CLQs. The slope ($\alpha_{\rm bright}$ = $-1.68\pm0.02$ ) of the bright EVQ composite is similar to the typical QSO spectral slope of $-1.56$ \citep{VandenBerk01}. The faint spectrum is very flat with $\alpha_{\rm faint}$ = $-0.91\pm0.03$, further confirming the well-known BWB trend in quasars. Previous studies \citep[e.g.,][]{ Wilhite05,Ruan14,Guo16b} have suggested that the difference spectrum slope ($\alpha_{\rm diff} = -2$) of normal QSOs is slightly shallower than the prediction of standard disk model \citep[e.g., $\alpha_{\rm \lambda}=-2.33$,][]{Shakura73}, likely due to the local extinction in quasar host \citep{Xie16}. Slopes of both difference spectra are $\sim-2$ in EVQ and CLQ sample at least indicating that the variability mechanism should be very similar among normal quasars, EVQs and CLQs.

\subsection{Detection Rate}
Among the total 61,037 objects of high-redshift repeat quasars, we find 1039 objects of EVQ, uncorrected for the selection incompleteness, selection bias and rare spurious EVQs, yielding an EVQ detection rate of $\sim$2\%. This is broadly consistent with the result ($\sim$4\%, 1011 EVQs of 25484 repeat spectra in SDSS with $\rm |\Delta g| > 1$ mag) from \cite{Macleod16}. However, \cite{Rumbaugh18} reported that the EVQ fraction is about 10\% with $\rm |\Delta g| > 1$ mag based on the Stripe 82 SDSS-DES \citep[dark energy survey,][]{Flaugher15} light curves. Accounting for selection effects, they further suggest an intrinsic EVQ fraction of $\sim$ 30 $-$ 50\% among quasars brighter than 22 mag in $g-$band over a baseline of 15 yrs. We suggest that the higher fraction of EVQ in SDSS-DES search is due to the selection bias that DES photometry is much deeper than SDSS single-epoch spectroscopic exposure.

The CLQ fraction in EVQs is about 7\% (23/348), yielding a detection rate of $\sim$0.1\% in all high-redshift repeat quasars in SDSS. This is about one order of magnitude higher than the fraction of the low-redshift CLQ searching with the same SDSS repeat spectra, e.g., $\sim$0.01\% in \cite{Yang18} and $0.04$\% in \cite{Macleod16}. This is most likely due to the stronger variability in shorter wavelengths for high-redshift quasars.

\section{Discussions}\label{sec:diss}
\subsection{Decisive Roles for the CL Behavior}\label{sec:CLQ}
\cite{Guo20} have demonstrated that the CLQs showing appearance/disappearance of broad emission line with continuum increasing/decreasing is a natural phenomenon under the photoionization model. Whether the transition occurs primarily depends on several factors: 1) Eddington ratio (perhaps, although we only find a marginal clue to support it); 2) the spectral SNR of the faint state; 3) host galaxy contamination for the low redshift CLQs. 

A quantitative photoionization calculation is shown in Figure 8 of \cite{Guo20}. While the central continuum gradually drops, the broad emission line responds to the continuum variation with reduction of the line strength ($L_{\rm con} \propto L_{\rm line}$) and shrinkage of the line emitting region, leading to a smaller emitting ring and faster rotational velocity of broad-line-emitting clouds (i.e., breathing model). When the continuum luminosity (or the Eddington ratio) is low enough assuming a constant BH mass, the broad emission line will be too weak (or too broad) to be detected, resulting in CL behavior. This simple theoretical model is in agreement with the statistical result in Figure \ref{fig:Edd}, which presents CLQ Eddington ratios are on average lower than that of normal EVQs, although the significance of the difference is less than 1$\sigma$ level. Differently, the observed \civ\ shows an anti-breathing mode due to the non-reverberating component, we argue that the IBC may just need more time to show the breathing at further distances \citep{Denney12}.       
      
\begin{figure*}
\centering
\includegraphics[width=18.cm]{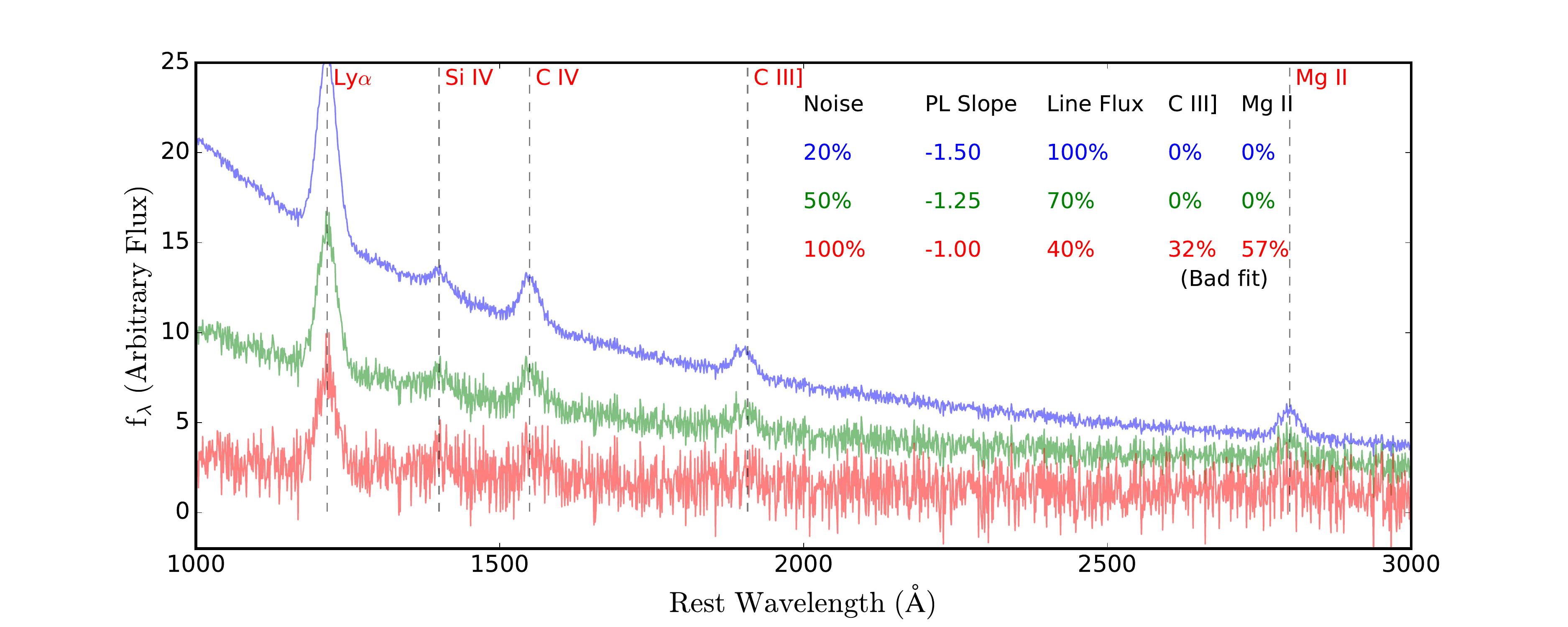}
\caption{Simulated CLQs with increasing noise level, flattening slopes and fading line luminosities. Each spectrum consists of a continuum power-law, multi-Gaussian profiles and random Gaussian noise. With decreasing continuum emission, the broad component of each line becomes more and more illegible. The line (i.e., \ciii\ and \mgii) fitting results to 1000 simulated spectra are listed, indicating the CL phenomenon strongly depends on the noise level with fading line flux.}
\label{fig:SN}
\end{figure*}

Moreover, the spectral noise will dilute the weak broad component, as shown in Figure \ref{fig:SN}. We conduct a simple simulation assuming increasing Gaussian noise levels, flattening power-law slopes and decreasing line fluxes with fading continuum at three levels. The emission lines are modeled by single-Gaussian profiles with a fixed FWHM of 5000 \kms. Each level is simulated for 1000 times with random Gaussian noise ($\mu = 0$ and $\sigma$ up to 1). Then we use \texttt{PyQSOFit} to fit each spectrum with a continuum power-law and two Gaussian profiles for \ciii\ and \mgii\ with a maximum line width of 10000 \kms. In the faintest state, we find 32\% and 57\% of the modeled profiles hit the boundary, which means the automatic code is almost unable to correctly recognize the broad component, although the line flux is not zero. Consequently, we may classify those objects as a \ciii\ or \mgii\ CLQ, whose weak broad components are actually swamped by spectral noise. 

This raises an important question that what if some EVQs classified as CLQs due to the SNR problem in our work. We stress that all previous discovered CLQs are subject to SNR issue, and our previous analysis are based on the frame of SDSS resolution. It is possible to resolve hidden broad components with better telescopes, which is exactly the reason why we need to push to the limit to inspect the smoothed SDSS spectra for CLQ searches (e.g., in Figure \ref{fig:example}). Because of the changeable classification of CLQ under different scenarios, we suggest that using a strict definition to separate the CLQ and EVQ may not a good choice.

Finally, the broad emission lines (e.g., \ha\ and \hb) at low redshift are easier to be swamped in their bright host galaxy. However, the host contamination in our high-redshift example is negligible.

\begin{table}
\caption{Line relative flux}\label{tab:lineflux}
\centering
\begin{tabular}{ccccc}
\hline\hline
Name& Vacuum $\lambda$ & Rel.Flux & Sim. Flux & Decay rate \\
& (\AA) &\multicolumn{2}{c}{100$\times$F/F(\lya)}& (dex)\\
\hline
\lya & 1215.67 & 100 & 100 & 0.40\\
\SiIV & 1396.76 & 8.92 & 6.31 & 0.50\\
\civ & 1549.06 &25.29 & 22.38 & 0.45\\
\ciii & 1908.73 & 15.94 & 10.00 & 0.40\\
\mgii & 2798.75 & 14.73 & 11.22 & 0.30\\
\hline\hline
\multicolumn{5}{c}{Note. the relative flux is derived from the quasar composite}\\
\multicolumn{5}{c}{in \cite{VandenBerk01}, and the simulated flux and}\\
\multicolumn{5}{c}{decay rate are calculated from the photoionization model} \\
\multicolumn{5}{c}{ranging from 43.8 to 43.3 \erg\ following \cite{Guo20}}\\
\multicolumn{5}{c}{(also see Figure \ref{fig:loc})}\\
\end{tabular}
\end{table}

\subsection{Line Transition Order in CLQs}

We suggest that the intrinsic line intensity mostly determines the transition order of different lines, while the other factors (e.g., spectral SNR) are second-order effects, which may slightly exchange the order of some lines with similar intensity (e.g., \ciii\ and \mgii\, also see Table \ref{tab:lineflux}). In low-redshift CLQs \citep[e.g.,][]{Macleod16,Yang18}, the weakest high-order broad Balmer lines (e.g., \hr) usually disappears first, then relatively stronger \hb, finally to the strongest \ha. In addition, the broad \mgii\ is usually more persistent than \hb\, but seems less persistent than \ha\ \citep[e.g., see the \mgii\ CLQ in][]{Guo19a}. In high-redshift CLQs, the UV lines (see Figure \ref{fig:example} \& \ref{fig:example2}-\ref{fig:example1}) also basically follow this rule, .e.g., \cii $>$ \SiIV\ \& \ciii\ \& \mgii $>$ \civ $>$ \lya\ from first to last disappears. Despite the line intensity of \SiIV\ is sightly smaller than \mgii\ and \ciii, it seems not disappear significantly earlier than \ciii\ and \mgii\ in most of our CLQs.       
 
\begin{figure}
\centering
\includegraphics[width=9.cm]{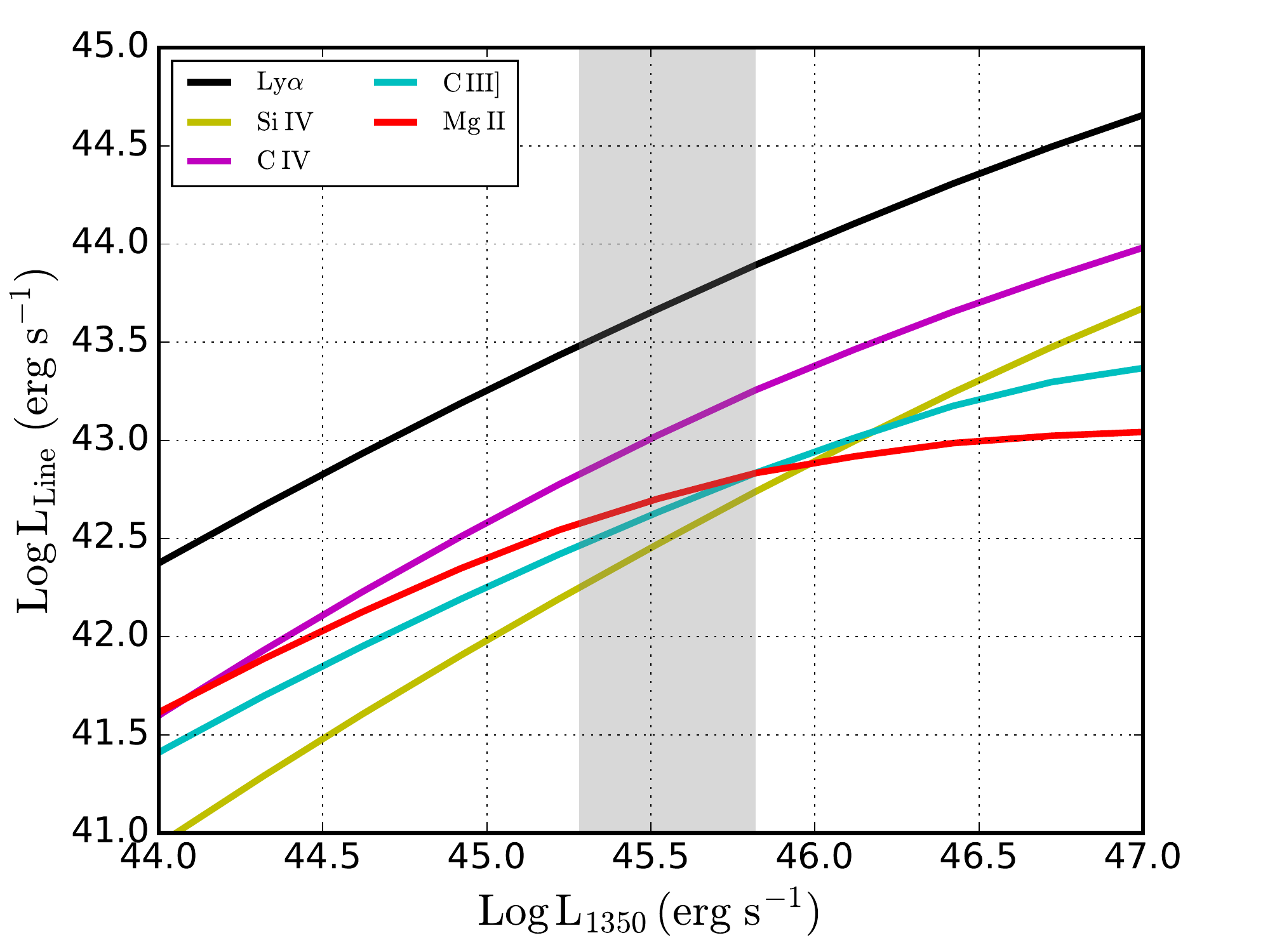}
\caption{The simulated emission line decay rate in photoionization frame. The gray region of [45.3, 45.8] \erg is our fiducial luminosity coverage according to the medians of the bright/faint continuum luminosity distribution in Figure \ref{fig:EVQ_para}. The line ratios relative to \lya\ at 45.5 \erg\ are listed in Table \ref{tab:lineflux}.}
\label{fig:loc}
\end{figure}

The decay rate of different lines is another factor which might have an influence on the transition order of CLQs. We mimic a quasar emission-line flux changing with its decreasing continuum within the photoionzation frame following \cite{Guo20}. The basic idea is that with decreasing inputs of hydrogen-ionizing photons assuming a locally optimally emitting cloud (LOC) model \citep{Baldwin95}, we can track the variation of the emission lines. The details of the model setup is specified in \cite{Guo20}. Figure \ref{fig:loc} illustrates the different decay rates with a fading continuum. The grey shadow brackets the average continuum variation range in the bright and faint states of EVQs (see Figure \ref{fig:EVQ_para}). However, within the limited variability range (e.g., only 0.5 dex in $x$-axis), the discrepancy is negligible (the quantities are listed in Table \ref{tab:lineflux}). Note that the less variable \mgii\ compared to other UV lines is also recovered \citep{Goad93}.

\subsection{Relation between EVQ and CLQ}\label{sec:SNR}
We have demonstrated that CLQs generally have similar properties (Figures \ref{fig:EVQ_para} \& \ref{fig:Edd}) compared to normal EVQs, but perhaps with slightly lower continuum luminosities and broader line widths, and hence lower Eddington ratios. Moreover, the difference-composite slopes of EVQs/CLQs in Figures \ref{fig:composite} are both identical to that of normal quasars, indicating they are dominated by a similar variability mechanism \citep[also see][]{Noda18}. The K-S test in Table \ref{tab:EVQ_para} also suggests that EVQs and CLQs likely belong to the same population. Therefore, we conclude that CLQs and EVQs are basically the same population. Whether the broad component fully disappeared or is just too weak to be detectable also strongly depends on the spectral SNR, as shown in Figure \ref{fig:SN}. Even if the line core of a strong broad emission line transitions into an absorption line, the broad component (line wing) may still exist (e.g., \ciii\ in ID = 234455 in Figure \ref{fig:example1}), but is possibly too weak to be decomposed from the host component. Thus the CLQs are just an observational class, and we recommend to use EVQ instead to refer the whole extreme variable quasar population since it is more physical.

\subsection{Variability Mechanism}
The AGN variability mechanism is still poorly understood. Our discovery of a bunch of high-redshift CLQs further challenges the standard disk model, which requires a variability timescale of thousands of years.  It was suggested that the observed optical light curve is a mixture of various timescale variabilities originated from the accretion disk \citep[e.g.,][]{Cai16,Lawrence18}. The standard thin disk model is expected to control the very long-term variation with a viscous timescale of $10^5$ yr, while the short-term variation (days to years) is dominated by other mechanisms e.g., the X-ray reprocessing \citep[e.g.,][]{Krolik91}, local temperature fluctuation in disk \citep{Dexter11,Cai16} and corona-heated accretion-disk reprocessing \citep{Sun20}. 

Through an investigation of the CLQ Mrk 1018, \cite{Noda18} suggest that the CLQ is more likely triggered by rapid mass accretion rate drop, accompanied with state transition \citep[also see ][]{Ruan19}. They point out that the sound speed could be much faster when considering a radiation-pressure-dominated disk and the magnetic pressure to help stabilize the disk \citep{Jiang13}, which could alleviate timescale problem. However, \cite{Lawrence18} indicate that the standard viscous accretion disk models are close to the edge, and simply crank up the viscosity parameter would not solve the timescale problem. The extreme reprocessing where an erratically variable central quasi-point source is entirely responsible for heating an otherwise cold and passive low-viscosity disk, may be a better route forward.

\subsection{\civ\ Line Profile} \label{sec:civ_mass}
AGN broad emission lines are usually dominated by the photoionization from the accretion continuum. As mentioned, the virialized broad-line-emitting clouds will respond to the increasing of the central continuum luminosity by increasing average distance of the clouds accompanied and a decreased averaged line width, known as the AGN breathing model \citep[e.g.,][]{Guo20}. This validates some empirical single-epoch BH estimators assuming the virial motion of the clouds, in particular the \hb\ for low redshift AGNs.    

\civ\ is the only broad emission line widely used to estimate the BH mass at high redshift (i.e., $z>2.3$). However, the anti-breathing of \civ\ results in a luminosity-independent BH mass with a discrepancy of 0.3 dex in bright and faint states as shown in \S \ref{sec:mass}, challenging the widely used empirical mass estimator \citep{Vestergaard06}. The anti-breathing of \civ\ is usually explained by the presence of a non-varying core component in addition to a reverberating broad-base component \citep{Denney12}, consistent with the behavior in Figure \ref{fig:IL}. Previous studies speculate that the non-reverberating core component is more likely associated with a disk wind \citep[e.g.,][]{Proga00} or originates from an intermediate line region \citep[e.g.,][]{Wills93}, commonly accompanied with a blueshift relative to the systemic velocity of the quasar \citep{Richards02,Sun18b}. 

Among our 348 objects, 244 show the anti-breathing behavior, and 104 present the expected breathing mode, indicating the anti-breathing dominates the \civ\ variability, but normal breathing \civ\ still exists. To date, there is still no ideal solution to fix the BH mass discrepancy in bright and faint states, although previous work tried to recalibrate the \civ-based BH mass to that from \hb, e.g., using the an ultraviolet (or Eigenvector 1) indicator: line peak flux ratio of \SiIV\ and \civ\ \citep{Runnoe13}. Therefore we use the mean value as our fiducial BH mass from \civ. Note that the variability introduces the extra scatter of 0.3 dex for \civ-based BH mass, which is already non-negligible compared with the systematic scatter of $\sim$0.4 dex uncertainty in single-epoch mass estimates \citep{Shen13}.    

\cite{Wang20} have suggested that \hb\ (and to a lesser extent, \ha) displays the most consistent normal breathing expected from the virial relation, \mgii\ mostly shows no breathing \citep[also see ][]{Shen13,Yang20} and \civ\ (and similarly C III]) mostly shows anti-breathing \citep{Wills93,Denney12}. The BH mass based on \mgii\ should be more precise than \civ\ BH mass \citep{Shen08,Shen12}. However, the \mgii\ lines in most EVQs are too noisy due to the faintness of our sample and approaching/exceeding the edge of wavelength coverage of the SDSS spectra.

\section{Conclusions}\label{sec:con}
We have compiled a sample of 348 spectroscopically selected EVQs at redshift $>$ 1.5 with repeat observations (Nepoch $\ge$ 2) from SDSS DR16. The continuum variability at 1450\AA\ between the brightest and faintest epochs in each object is larger than 100\%, i.e., $\delta$V $\equiv$ (Max-Min)/Mean $>1$. Among those 348 EVQ objects,  we have discovered 23 high-redshift CLQs (Figure \ref{fig:example} and \ref{fig:example2}--\ref{fig:example1}) with transitions in distinctive broad emission lines (e.g., \ciii, \mgii), yielding a detection rate of $\sim$ 7\% in EVQs. We caution that the problematic spectral flux calibration of SDSS spectrum (i.e., fiber-drop epoch) may mimic the faint state of an CLQ (Figure \ref{fig:fake_CLQ}). We explore the properties of EVQs and CLQs; the main findings are as follows: 

\begin{enumerate}
    \item Through the comparison of various properties, i.e., continuum/line properties (Figure \ref{fig:EVQ_para}), composites (Figure \ref{fig:composite}) and Eddington ratio (Figure \ref{fig:Edd}), we conclude that EVQs and CLQs are basically the same population, as a tail of a continuous distribution of normal quasar properties \citep{Rumbaugh18}.  Further dividing the observed CLQ and EVQ with a strict and clean definition is not easy due to spectral noise (Figure \ref{fig:SN}).
    
    \item We originally expect that the CLQs is a subset of EVQs with less efficient accretion. However, our Eddington ratio distributions (Figure \ref{fig:Edd}) show no reliable evidence to support that, with a confidence level of $\lesssim$ 1$\sigma$, regardless of other potential bias.  
    
    \item The disappearance order of different UV emission lines in high-redshift CLQs generally dominated by line intensity (Table \ref{tab:lineflux}), basically following \cii $>$ \SiIV\ \& \ciii\ \& \mgii $>$ \civ $>$ \lya\ (from first to last disappears). The line decay rate (Figure \ref{fig:loc}) and host contamination have little influence on the transition order. 
    
    \item The anti-breathing behavior (i.e., $L_{\rm con} \propto$ FWHM) of \civ\ is confirmed in our EVQ sample, which is caused by a relatively narrow non-reverberating component probably originated from a disk wind \citep{Proga00} or an intermediate line region \citep{Wills93}. We caution that due to the anti-breathing of \civ, extreme AGN variability will introduce an extra BH mass scatter of $\sim$ 0.3 dex at $L_{\rm 1350} \sim 45.5$ \erg, which is comparable to the systematic scatter of $\sim$0.4 dex with the empirical single-epoch estimator. 
\end{enumerate}

The variability mechanism of CLQs/EVQs will continue to be a primary issue in the near future. Multi-wavelength data (e.g., UV/optical and X-ray) and large sample of CLQs (e.g., even faint CLQs of \civ\ and \lya\ ) will be helpful to understand quasar variability more generally. In addition, theoretical calculations based on the LOC model by connecting concurrent X-ray and UV/optical data may further unveil the mystery of quasar variability.

\acknowledgments
We thank the referee, Yue Shen and Zhenyi Cai for constructive comments to improve the paper. H.X.G. acknowledges support from the NSF grant AST-1907290. C.J.B. acknowledges support from the Illinois graduate survey science fellowship. M.Y.S. acknowledges support from the National Natural Science Foundation of China (NSFC-11973002).  Z.C.H. is supported by NSFC-11903031 and USTC Research Funds of the Double First-Class Initiative. T.G.W. is supported by NSFC-11833007. M.F.G. is supported by NSFC-11873073. M.Z.K. is supported by the Joint Research Fund in Astronomy (No. U1831126) under cooperative agreement between the NSFC and CAS and Natural Science Foundation of Hebei Province No. A2019205100.

SDSS-IV is managed by the Astrophysical Research Consortium for the Participating Institutions of the SDSS Collaboration including the Brazilian Participation Group, the Carnegie Institution for Science, Carnegie Mellon University, the Chilean Participation Group, the French Participation Group, Harvard-Smithsonian Center for Astrophysics, Instituto de Astrof\'isica de Canarias, The Johns Hopkins University, Kavli Institute for the Physics and Mathematics of the Universe (IPMU) / University of Tokyo, Lawrence Berkeley National Laboratory, Leibniz Institut f\"ur Astrophysik Potsdam (AIP),  Max-Planck-Institut f\"ur Astronomie (MPIA Heidelberg), Max-Planck-Institut f\"ur Astrophysik (MPA Garching), Max-Planck-Institut f\"ur Extraterrestrische Physik (MPE), National Astronomical Observatories of China, New Mexico State University, New York University, University of Notre Dame, Observat\'ario Nacional / MCTI, The Ohio State University, Pennsylvania State University, Shanghai Astronomical Observatory, United Kingdom Participation Group,Universidad Nacional Aut\'onoma de M\'exico, University of Arizona, University of Colorado Boulder, University of Oxford, University of Portsmouth, University of Utah, University of Virginia, University of Washington, University of Wisconsin, Vanderbilt University, and Yale University.

\software{Astropy \citep{AstropyCollaboration13}, Matplotlib \citep{Hunter07}, Numpy \& Scipy \citep{Numpy}}, Pysynphot\citep{STScIDevelopmentTeam13}, PyQSOFit \citep{Guo18}

\clearpage



\bibliography{ref}

\begin{thebibliography}{}
\expandafter\ifx\csname natexlab\endcsname\relax\def\natexlab#1{#1}\fi
\providecommand{\url}[1]{\href{#1}{#1}}
\providecommand{\dodoi}[1]{doi:~\href{http://doi.org/#1}{\nolinkurl{#1}}}
\providecommand{\doeprint}[1]{\href{http://ascl.net/#1}{\nolinkurl{http://ascl.net/#1}}}
\providecommand{\doarXiv}[1]{\href{https://arxiv.org/abs/#1}{\nolinkurl{https://arxiv.org/abs/#1}}}

\bibitem[{{Ahumada} {et~al.}(2019){Ahumada}, {Allende Prieto}, {Almeida},
  {Anders}, {Anderson}, {Andrews}, {Anguiano}, {Arcodia}, {Armengaud},
  {Aubert}, {Avila}, {Avila-Reese}, {Badenes}, {Balland}, {Barger},
  {Barrera-Ballesteros}, {Basu}, {Bautista}, {Beaton}, {Beers}, {Benavides},
  {Bender}, {Bernardi}, {Bershady}, {Beutler}, {Moni Bidin}, {Bird}, {Bizyaev},
  {Blanc}, {Blanton}, {Boquien}, {Borissova}, {Bovy}, {Brandt}, {Brinkmann},
  {Brownstein}, {Bundy}, {Bureau}, {Burgasser}, {Burtin}, {Cano-Diaz},
  {Capasso}, {Cappellari}, {Carrera}, {Chabanier}, {Chaplin}, {Chapman},
  {Cherinka}, {Chiappini}, {Choi}, {Chojnowski}, {Chung}, {Clerc}, {Coffey},
  {Comerford}, {Comparat}, {da Costa}, {Cousinou}, {Covey}, {Crane}, {Cunha},
  {da Silva Ilha}, {Dai}, {Damsted}, {Darling}, {Davidson}, {Davies}, {Dawson},
  {De}, {de la Macorra}, {De Lee}, {Queiroz}, {Deconto Machado}, {de la Torre},
  {Dell'Agli}, {du Mas des Bourboux}, {Diamond-Stanic}, {Dillon}, {Donor},
  {Drory}, {Duckworth}, {Dwelly}, {Ebelke}, {Eftekharzadeh}, {Davis Eigenbrot},
  {Elsworth}, {Eracleous}, {Erfanianfar}, {Escoffier}, {Fan}, {Farr},
  {Fernandez-Trincado}, {Feuillet}, {Finoguenov}, {Fofie}, {Fraser-McKelvie},
  {Frinchaboy}, {Fromenteau}, {Fu}, {Galbany}, {Garcia}, {Garcia-Hernandez},
  {Garma Oehmichen}, {Ge}, {Geimba Maia}, {Geisler}, {Gelfand }, {Goddy}, {Le
  Goff}, {Gonzalez-Perez}, {Grabowski}, {Green}, {Grier}, {Guo}, {Guy},
  {Harding}, {Hasselquist}, {Hawken}, {Hayes}, {Hearty}, {Hekker}, {Hogg},
  {Holtzman}, {Horta}, {Hou}, {Hsieh}, {Huber}, {Hunt}, {Ider Chitham}, {Imig},
  {Jaber}, {Jimenez Angel}, {Johnson}, {Jones}, {Jonsson}, {Jullo}, {Kim},
  {Kinemuchi}, {Kirkpatrick}, {Kite}, {Klaene}, {Kneib}, {Kollmeier}, {Kong},
  {Kounkel}, {Krishnarao}, {Lacerna}, {Lan}, {Lane}, {Law}, {Leung}, {Lewis},
  {Li}, {Lian}, {Lin}, {Long}, {Longa-Pena}, {Lundgren}, {Lyke}, {Mackereth},
  {MacLeod}, {Majewski}, {Manchado}, {Maraston}, {Martini}, {Masseron},
  {Masters}, {Mathur}, {McDermid}, {Merloni}, {Merrifield}, {Meszaros},
  {Miglio}, {Minniti}, {Minsley}, {Miyaji}, {Gohar Mohammad}, {Mosser},
  {Mueller}, {Muna}, {Munoz-Gutierrez}, {Myers}, {Nadathur}, {Nair}, {Nandra},
  {Correa do Nascimento}, {Nevin}, {Newman}, {Nidever}, {Nitschelm},
  {Noterdaeme}, {O'Connell}, {Olmstead}, {Oravetz}, {Oravetz}, {Osorio},
  {Pace}, {Padilla}, {Palanque-Delabrouille}, {Palicio}, {Pan}, {Pan},
  {Parker}, {Paviot}, {Peirani}, {Pena Ramrez}, {Penny}, {Percival},
  {Perez-Fournon}, {Perez-Rafols}, {Petitjean}, {Pieri}, {Pinsonneault},
  {Poovelil}, {Povick}, {Prakash}, {Price-Whelan}, {Raddick}, {Raichoor},
  {Ray}, {Barboza Rembold}, {Rezaie}, {Riffel}, {Riffel}, {Rix}, {Robin},
  {Roman-Lopes}, {Roman-Zuniga}, {Rose}, {Ross}, {Rossi}, {Rowlands}, {Rubin},
  {Salvato}, {Sanchez}, {Sanchez-Menguiano}, {Sanchez-Gallego}, {Sayres},
  {Schaefer}, {Schiavon}, {Schimoia}, {Schlafly}, {Schlegel}, {Schneider},
  {Schultheis}, {Schwope}, {Seo}, {Serenelli}, {Shafieloo}, {Shamsi}, {Shao},
  {Shen}, {Shetrone}, {Shirley}, {Silva Aguirre}, {Simon}, {Skrutskie},
  {Slosar}, {Smethurst}, {Sobeck}, {Cervantes Sodi}, {Souto}, {Stark},
  {Stassun}, {Steinmetz}, {Stello}, {Stermer}, {Storchi-Bergmann},
  {Streblyanska}, {Stringfellow}, {Stutz}, {Suarez}, {Sun}, {Taghizadeh-Popp},
  {Talbot}, {Tayar}, {Thakar}, {Theriault}, {Thomas}, {Thomas}, {Tinker},
  {Tojeiro}, {Hernandez Toledo}, {Tremonti}, {Troup}, {Tuttle}, {Unda-Sanzana},
  {Valentini}, {Vargas-Gonzalez}, {Vargas-Magana}, {Vazquez-Mata}, {Vivek},
  {Wake}, {Wang}, {Weaver}, {Weijmans}, {Wild}, {Wilson}, {Wilson}, {Wolthuis},
  {Wood-Vasey}, {Yan}, {Yang}, {Yeche}, {Zamora}, {Zarrouk}, {Zasowski},
  {Zhang}, {Zhao}, {Zhao}, {Zheng}, {Zheng}, {Zhu}, \& {Zou}}]{Ahumada19}
{Ahumada}, R., {Allende Prieto}, C., {Almeida}, A., {et~al.} 2019, arXiv
  e-prints, arXiv:1912.02905.
\newblock \doarXiv{1912.02905}

\bibitem[{{Ai} {et~al.}(2020){Ai}, {Dou}, {Yang}, {Sun}, {Xie}, {Yao}, {Wu},
  {Wang}, {Shu}, \& {Jiang}}]{Ai20}
{Ai}, Y., {Dou}, L., {Yang}, C., {et~al.} 2020, \apjl, 890, L29,
  \dodoi{10.3847/2041-8213/ab7306}

\bibitem[{{Ai} {et~al.}(2010){Ai}, {Yuan}, {Zhou}, {Wang}, {Dong}, {Wang}, \&
  {Lu}}]{Ai10}
{Ai}, Y.~L., {Yuan}, W., {Zhou}, H.~Y., {et~al.} 2010, \apjl, 716, L31,
  \dodoi{10.1088/2041-8205/716/1/L31}

\bibitem[{{Antonucci}(1993)}]{Antonucci93}
{Antonucci}, R. 1993, \araa, 31, 473,
  \dodoi{10.1146/annurev.aa.31.090193.002353}

\bibitem[{{Aretxaga} {et~al.}(1999){Aretxaga}, {Joguet}, {Kunth}, {Melnick}, \&
  {Terlevich}}]{Aretxaga99}
{Aretxaga}, I., {Joguet}, B., {Kunth}, D., {Melnick}, J., \& {Terlevich}, R.~J.
  1999, \apjl, 519, L123, \dodoi{10.1086/312114}

\bibitem[{{Astropy Collaboration} {et~al.}(2013){Astropy Collaboration},
  {Robitaille}, {Tollerud}, {Greenfield}, {Droettboom}, {Bray}, {Aldcroft},
  {Davis}, {Ginsburg}, {Price-Whelan}, {Kerzendorf}, {Conley}, {Crighton},
  {Barbary}, {Muna}, {Ferguson}, {Grollier}, {Parikh}, {Nair}, {Unther},
  {Deil}, {Woillez}, {Conseil}, {Kramer}, {Turner}, {Singer}, {Fox}, {Weaver},
  {Zabalza}, {Edwards}, {Azalee Bostroem}, {Burke}, {Casey}, {Crawford},
  {Dencheva}, {Ely}, {Jenness}, {Labrie}, {Lim}, {Pierfederici}, {Pontzen},
  {Ptak}, {Refsdal}, {Servillat}, \& {Streicher}}]{AstropyCollaboration13}
{Astropy Collaboration}, {Robitaille}, T.~P., {Tollerud}, E.~J., {et~al.} 2013,
  \aap, 558, A33, \dodoi{10.1051/0004-6361/201322068}

\bibitem[{{Baldwin} {et~al.}(1995){Baldwin}, {Ferland}, {Korista}, \&
  {Verner}}]{Baldwin95}
{Baldwin}, J., {Ferland}, G., {Korista}, K., \& {Verner}, D. 1995, \apjl, 455,
  L119, \dodoi{10.1086/309827}

\bibitem[{{Baldwin}(1977)}]{Baldwin77}
{Baldwin}, J.~A. 1977, \apj, 214, 679, \dodoi{10.1086/155294}

\bibitem[{{Barth} {et~al.}(2015){Barth}, {Bennert}, {Canalizo}, {Filippenko},
  {Gates}, {Greene}, {Li}, {Malkan}, {Pancoast}, {Sand }, {Stern}, {Treu},
  {Woo}, {Assef}, {Bae}, {Brewer}, {Cenko}, {Clubb}, {Cooper},
  {Diamond-Stanic}, {Hiner}, {H{\"o}nig}, {Hsiao}, {Kand rashoff}, {Lazarova},
  {Nierenberg}, {Rex}, {Silverman}, {Tollerud}, \& {Walsh}}]{Barth15}
{Barth}, A.~J., {Bennert}, V.~N., {Canalizo}, G., {et~al.} 2015, \apjs, 217,
  26, \dodoi{10.1088/0067-0049/217/2/26}

\bibitem[{{Bellm} {et~al.}(2019){Bellm}, {Kulkarni}, {Graham}, {Dekany},
  {Smith}, {Riddle}, {Masci}, {Helou}, {Prince}, {Adams}, {Barbarino},
  {Barlow}, {Bauer}, {Beck}, {Belicki}, {Biswas}, {Blagorodnova}, {Bodewits},
  {Bolin}, {Brinnel}, {Brooke}, {Bue}, {Bulla}, {Burruss}, {Cenko}, {Chang},
  {Connolly}, {Coughlin}, {Cromer}, {Cunningham}, {De}, {Delacroix}, {Desai},
  {Duev}, {Eadie}, {Farnham}, {Feeney}, {Feindt}, {Flynn}, {Franckowiak},
  {Frederick}, {Fremling}, {Gal-Yam}, {Gezari}, {Giomi}, {Goldstein},
  {Golkhou}, {Goobar}, {Groom}, {Hacopians}, {Hale}, {Henning}, {Ho}, {Hover},
  {Howell}, {Hung}, {Huppenkothen}, {Imel}, {Ip}, {Ivezi{\'c}}, {Jackson},
  {Jones}, {Juric}, {Kasliwal}, {Kaspi}, {Kaye}, {Kelley}, {Kowalski},
  {Kramer}, {Kupfer}, {Landry}, {Laher}, {Lee}, {Lin}, {Lin}, {Lunnan},
  {Giomi}, {Mahabal}, {Mao}, {Miller}, {Monkewitz}, {Murphy}, {Ngeow},
  {Nordin}, {Nugent}, {Ofek}, {Patterson}, {Penprase}, {Porter}, {Rauch},
  {Rebbapragada}, {Reiley}, {Rigault}, {Rodriguez}, {van Roestel}, {Rusholme},
  {van Santen}, {Schulze}, {Shupe}, {Singer}, {Soumagnac}, {Stein}, {Surace},
  {Sollerman}, {Szkody}, {Taddia}, {Terek}, {Van Sistine}, {van Velzen},
  {Vestrand}, {Walters}, {Ward}, {Ye}, {Yu}, {Yan}, \& {Zolkower}}]{Bellm19}
{Bellm}, E.~C., {Kulkarni}, S.~R., {Graham}, M.~J., {et~al.} 2019, \pasp, 131,
  018002, \dodoi{10.1088/1538-3873/aaecbe}

\bibitem[{{Bolton} {et~al.}(2012){Bolton}, {Schlegel}, {Aubourg}, {Bailey},
  {Bhardwaj}, {Brownstein}, {Burles}, {Chen}, {Dawson}, {Eisenstein}, {Gunn},
  {Knapp}, {Loomis}, {Lupton}, {Maraston}, {Muna}, {Myers}, {Olmstead},
  {Padmanabhan}, {P{\^a}ris}, {Percival}, {Petitjean}, {Rockosi}, {Ross},
  {Schneider}, {Shu}, {Strauss}, {Thomas}, {Tremonti}, {Wake}, {Weaver}, \&
  {Wood-Vasey}}]{Bolton12}
{Bolton}, A.~S., {Schlegel}, D.~J., {Aubourg}, {\'E}., {et~al.} 2012, \aj, 144,
  144, \dodoi{10.1088/0004-6256/144/5/144}

\bibitem[{{Cackett} \& {Horne}(2006)}]{Cackett06}
{Cackett}, E.~M., \& {Horne}, K. 2006, \mnras, 365, 1180,
  \dodoi{10.1111/j.1365-2966.2005.09795.x}

\bibitem[{{Cai} {et~al.}(2019){Cai}, {Sun}, {Wang}, {Zhu}, {Gu}, \&
  {Yuan}}]{Cai19}
{Cai}, Z., {Sun}, Y., {Wang}, J., {et~al.} 2019, Science China Physics,
  Mechanics, and Astronomy, 62, 69511, \dodoi{10.1007/s11433-018-9330-4}

\bibitem[{{Cai} {et~al.}(2016){Cai}, {Wang}, {Gu}, {Sun}, {Wu}, {Huang}, \&
  {Chen}}]{Cai16}
{Cai}, Z.-Y., {Wang}, J.-X., {Gu}, W.-M., {et~al.} 2016, \apj, 826, 7,
  \dodoi{10.3847/0004-637X/826/1/7}

\bibitem[{{Chambers} {et~al.}(2016){Chambers}, {Magnier}, {Metcalfe},
  {Flewelling}, {Huber}, {Waters}, {Denneau}, {Draper}, {Farrow}, {Finkbeiner},
  {Holmberg}, {Koppenhoefer}, {Price}, {Rest}, {Saglia}, {Schlafly}, {Smartt},
  {Sweeney}, {Wainscoat}, {Burgett}, {Chastel}, {Grav}, {Heasley}, {Hodapp},
  {Jedicke}, {Kaiser}, {Kudritzki}, {Luppino}, {Lupton}, {Monet}, {Morgan},
  {Onaka}, {Shiao}, {Stubbs}, {Tonry}, {White}, {Ba{\~n}ados}, {Bell},
  {Bender}, {Bernard}, {Boegner}, {Boffi}, {Botticella}, {Calamida},
  {Casertano}, {Chen}, {Chen}, {Cole}, {Deacon}, {Frenk}, {Fitzsimmons},
  {Gezari}, {Gibbs}, {Goessl}, {Goggia}, {Gourgue}, {Goldman}, {Grant},
  {Grebel}, {Hambly}, {Hasinger}, {Heavens}, {Heckman}, {Henderson}, {Henning},
  {Holman}, {Hopp}, {Ip}, {Isani}, {Jackson}, {Keyes}, {Koekemoer}, {Kotak},
  {Le}, {Liska}, {Long}, {Lucey}, {Liu}, {Martin}, {Masci}, {McLean}, {Mindel},
  {Misra}, {Morganson}, {Murphy}, {Obaika}, {Narayan}, {Nieto-Santisteban},
  {Norberg}, {Peacock}, {Pier}, {Postman}, {Primak}, {Rae}, {Rai}, {Riess},
  {Riffeser}, {Rix}, {R{\"o}ser}, {Russel}, {Rutz}, {Schilbach}, {Schultz},
  {Scolnic}, {Strolger}, {Szalay}, {Seitz}, {Small}, {Smith}, {Soderblom},
  {Taylor}, {Thomson}, {Taylor}, {Thakar}, {Thiel}, {Thilker}, {Unger},
  {Urata}, {Valenti}, {Wagner}, {Walder}, {Walter}, {Watters}, {Werner},
  {Wood-Vasey}, \& {Wyse}}]{Chambers16}
{Chambers}, K.~C., {Magnier}, E.~A., {Metcalfe}, N., {et~al.} 2016, arXiv
  e-prints, arXiv:1612.05560.
\newblock \doarXiv{1612.05560}

\bibitem[{{Cohen} {et~al.}(1986){Cohen}, {Rudy}, {Puetter}, {Ake}, \&
  {Foltz}}]{Cohen86}
{Cohen}, R.~D., {Rudy}, R.~J., {Puetter}, R.~C., {Ake}, T.~B., \& {Foltz},
  C.~B. 1986, \apj, 311, 135, \dodoi{10.1086/164758}

\bibitem[{{Cui} {et~al.}(2012){Cui}, {Zhao}, {Chu}, {Li}, {Li}, {Zhang}, {Su},
  {Yao}, {Wang}, {Xing}, {Li}, {Zhu}, {Wang}, {Gu}, {Luo}, {Xu}, {Zhang},
  {Liu}, {Zhang}, {Yang}, {Cao}, {Chen}, {Chen}, {Chen}, {Chen}, {Chu}, {Feng},
  {Gong}, {Hou}, {Hu}, {Hu}, {Hu}, {Jia}, {Jiang}, {Jiang}, {Jiang}, {Jin},
  {Li}, {Li}, {Li}, {Liu}, {Liu}, {Lu}, {Mao}, {Men}, {Qi}, {Qi}, {Shi},
  {Tang}, {Tao}, {Wang}, {Wang}, {Wang}, {Wang}, {Wang}, {Wang}, {Wang},
  {Wang}, {Wang}, {Wang}, {Wang}, {Wang}, {Xu}, {Xu}, {Yang}, {Yu}, {Yuan},
  {Yuan}, {Zhai}, {Zhang}, {Zhang}, {Zhang}, {Zhao}, {Zhou}, {Zhou}, {Zhu}, \&
  {Zou}}]{Cui12}
{Cui}, X.-Q., {Zhao}, Y.-H., {Chu}, Y.-Q., {et~al.} 2012, Research in Astronomy
  and Astrophysics, 12, 1197, \dodoi{10.1088/1674-4527/12/9/003}

\bibitem[{{Dawson} {et~al.}(2016){Dawson}, {Kneib}, {Percival}, {Alam},
  {Albareti}, {Anderson}, {Armengaud}, {Aubourg}, {Bailey}, {Bautista},
  {Berlind}, {Bershady}, {Beutler}, {Bizyaev}, {Blanton}, {Blomqvist},
  {Bolton}, {Bovy}, {Brandt}, {Brinkmann}, {Brownstein}, {Burtin}, {Busca},
  {Cai}, {Chuang}, {Clerc}, {Comparat}, {Cope}, {Croft}, {Cruz-Gonzalez}, {da
  Costa}, {Cousinou}, {Darling}, {de la Macorra}, {de la Torre}, {Delubac}, {du
  Mas des Bourboux}, {Dwelly}, {Ealet}, {Eisenstein}, {Eracleous}, {Escoffier},
  {Fan}, {Finoguenov}, {Font-Ribera}, {Frinchaboy}, {Gaulme}, {Georgakakis},
  {Green}, {Guo}, {Guy}, {Ho}, {Holder}, {Huehnerhoff}, {Hutchinson}, {Jing},
  {Jullo}, {Kamble}, {Kinemuchi}, {Kirkby}, {Kitaura}, {Klaene}, {Laher},
  {Lang}, {Laurent}, {Le Goff}, {Li}, {Liang}, {Lima}, {Lin}, {Lin}, {Lin},
  {Long}, {Lundgren}, {MacDonald}, {Geimba Maia}, {Malanushenko},
  {Malanushenko}, {Mariappan}, {McBride}, {McGreer}, {M{\'e}nard}, {Merloni},
  {Meza}, {Montero-Dorta}, {Muna}, {Myers}, {Nandra}, {Naugle}, {Newman},
  {Noterdaeme}, {Nugent}, {Ogando}, {Olmstead}, {Oravetz}, {Oravetz},
  {Padmanabhan}, {Palanque-Delabrouille}, {Pan}, {Parejko}, {P{\^a}ris},
  {Peacock}, {Petitjean}, {Pieri}, {Pisani}, {Prada}, {Prakash}, {Raichoor},
  {Reid}, {Rich}, {Ridl}, {Rodriguez-Torres}, {Carnero Rosell}, {Ross},
  {Rossi}, {Ruan}, {Salvato}, {Sayres}, {Schneider}, {Schlegel}, {Seljak},
  {Seo}, {Sesar}, {Shandera}, {Shu}, {Slosar}, {Sobreira}, {Streblyanska},
  {Suzuki}, {Taylor}, {Tao}, {Tinker}, {Tojeiro}, {Vargas-Maga{\~n}a}, {Wang},
  {Weaver}, {Weinberg}, {White}, {Wood-Vasey}, {Yeche}, {Zhai}, {Zhao}, {Zhao},
  {Zheng}, {Ben Zhu}, \& {Zou}}]{Dawson16}
{Dawson}, K.~S., {Kneib}, J.-P., {Percival}, W.~J., {et~al.} 2016, \aj, 151,
  44, \dodoi{10.3847/0004-6256/151/2/44}

\bibitem[{{Denney}(2012)}]{Denney12}
{Denney}, K.~D. 2012, \apj, 759, 44, \dodoi{10.1088/0004-637X/759/1/44}

\bibitem[{{Denney} {et~al.}(2009){Denney}, {Peterson}, {Dietrich},
  {Vestergaard}, \& {Bentz}}]{Denney09}
{Denney}, K.~D., {Peterson}, B.~M., {Dietrich}, M., {Vestergaard}, M., \&
  {Bentz}, M.~C. 2009, \apj, 692, 246, \dodoi{10.1088/0004-637X/692/1/246}

\bibitem[{{Denney} {et~al.}(2014){Denney}, {De Rosa}, {Croxall}, {Gupta},
  {Bentz}, {Fausnaugh}, {Grier}, {Martini}, {Mathur}, {Peterson}, {Pogge}, \&
  {Shappee}}]{Denney14}
{Denney}, K.~D., {De Rosa}, G., {Croxall}, K., {et~al.} 2014, \apj, 796, 134,
  \dodoi{10.1088/0004-637X/796/2/134}

\bibitem[{{DESI Collaboration} {et~al.}(2016){DESI Collaboration}, {Aghamousa},
  {Aguilar}, {Ahlen}, {Alam}, {Allen}, {Allende Prieto}, {Annis}, {Bailey},
  {Balland}, {Ballester}, {Baltay}, {Beaufore}, {Bebek}, {Beers}, {Bell},
  {Bernal}, {Besuner}, {Beutler}, {Blake}, {Bleuler}, {Blomqvist}, {Blum},
  {Bolton}, {Briceno}, {Brooks}, {Brownstein}, {Buckley-Geer}, {Burden},
  {Burtin}, {Busca}, {Cahn}, {Cai}, {Cardiel-Sas}, {Carlberg}, {Carton},
  {Casas}, {Castand er}, {Cervantes-Cota}, {Claybaugh}, {Close}, {Coker},
  {Cole}, {Comparat}, {Cooper}, {Cousinou}, {Crocce}, {Cuby}, {Cunningham},
  {Davis}, {Dawson}, {de la Macorra}, {De Vicente}, {Delubac}, {Derwent},
  {Dey}, {Dhungana}, {Ding}, {Doel}, {Duan}, {Ealet}, {Edelstein},
  {Eftekharzadeh}, {Eisenstein}, {Elliott}, {Escoffier}, {Evatt}, {Fagrelius},
  {Fan}, {Fanning}, {Farahi}, {Farihi}, {Favole}, {Feng}, {Fernandez},
  {Findlay}, {Finkbeiner}, {Fitzpatrick}, {Flaugher}, {Flender}, {Font-Ribera},
  {Forero-Romero}, {Fosalba}, {Frenk}, {Fumagalli}, {Gaensicke}, {Gallo},
  {Garcia-Bellido}, {Gaztanaga}, {Pietro Gentile Fusillo}, {Gerard},
  {Gershkovich}, {Giannantonio}, {Gillet}, {Gonzalez-de-Rivera},
  {Gonzalez-Perez}, {Gott}, {Graur}, {Gutierrez}, {Guy}, {Habib}, {Heetderks},
  {Heetderks}, {Heitmann}, {Hellwing}, {Herrera}, {Ho}, {Holland}, {Honscheid},
  {Huff}, {Hutchinson}, {Huterer}, {Hwang}, {Illa Laguna}, {Ishikawa},
  {Jacobs}, {Jeffrey}, {Jelinsky}, {Jennings}, {Jiang}, {Jimenez}, {Johnson},
  {Joyce}, {Jullo}, {Juneau}, {Kama}, {Karcher}, {Karkar}, {Kehoe}, {Kennamer},
  {Kent}, {Kilbinger}, {Kim}, {Kirkby}, {Kisner}, {Kitanidis}, {Kneib},
  {Koposov}, {Kovacs}, {Koyama}, {Kremin}, {Kron}, {Kronig}, {Kueter-Young},
  {Lacey}, {Lafever}, {Lahav}, {Lambert}, {Lampton}, {Land riau}, {Lang},
  {Lauer}, {Le Goff}, {Le Guillou}, {Le Van Suu}, {Lee}, {Lee}, {Leitner},
  {Lesser}, {Levi}, {L'Huillier}, {Li}, {Liang}, {Lin}, {Linder}, {Loebman},
  {Luki{\'c}}, {Ma}, {MacCrann}, {Magneville}, {Makarem}, {Manera}, {Manser},
  {Marshall}, {Martini}, {Massey}, {Matheson}, {McCauley}, {McDonald},
  {McGreer}, {Meisner}, {Metcalfe}, {Miller}, {Miquel}, {Moustakas}, {Myers},
  {Naik}, {Newman}, {Nichol}, {Nicola}, {Nicolati da Costa}, {Nie}, {Niz},
  {Norberg}, {Nord}, {Norman}, {Nugent}, {O'Brien}, {Oh}, {Olsen}, {Padilla},
  {Padmanabhan}, {Padmanabhan}, {Palanque-Delabrouille}, {Palmese},
  {Pappalardo}, {P{\^a}ris}, {Park}, {Patej}, {Peacock}, {Peiris}, {Peng},
  {Percival}, {Perruchot}, {Pieri}, {Pogge}, {Pollack}, {Poppett}, {Prada},
  {Prakash}, {Probst}, {Rabinowitz}, {Raichoor}, {Ree}, {Refregier}, {Regal},
  {Reid}, {Reil}, {Rezaie}, {Rockosi}, {Roe}, {Ronayette}, {Roodman}, {Ross},
  {Ross}, {Rossi}, {Rozo}, {Ruhlmann-Kleider}, {Rykoff}, {Sabiu}, {Samushia},
  {Sanchez}, {Sanchez}, {Schlegel}, {Schneider}, {Schubnell}, {Secroun},
  {Seljak}, {Seo}, {Serrano}, {Shafieloo}, {Shan}, {Sharples}, {Sholl},
  {Shourt}, {Silber}, {Silva}, {Sirk}, {Slosar}, {Smith}, {Smoot}, {Som},
  {Song}, {Sprayberry}, {Staten}, {Stefanik}, {Tarle}, {Sien Tie}, {Tinker},
  {Tojeiro}, {Valdes}, {Valenzuela}, {Valluri}, {Vargas-Magana}, {Verde},
  {Walker}, {Wang}, {Wang}, {Weaver}, {Weaverdyck}, {Wechsler}, {Weinberg},
  {White}, {Yang}, {Yeche}, {Zhang}, {Zhao}, {Zheng}, {Zhou}, {Zhou}, {Zhu},
  {Zou}, \& {Zu}}]{DESICollaboration16}
{DESI Collaboration}, {Aghamousa}, A., {Aguilar}, J., {et~al.} 2016, arXiv
  e-prints, arXiv:1611.00036.
\newblock \doarXiv{1611.00036}

\bibitem[{{Dexter} \& {Agol}(2011)}]{Dexter11}
{Dexter}, J., \& {Agol}, E. 2011, \apjl, 727, L24,
  \dodoi{10.1088/2041-8205/727/1/L24}

\bibitem[{{Dexter} \& {Begelman}(2019)}]{Dexter19b}
{Dexter}, J., \& {Begelman}, M.~C. 2019, \mnras, 483, L17,
  \dodoi{10.1093/mnrasl/sly213}

\bibitem[{{Dey} {et~al.}(2019){Dey}, {Schlegel}, {Lang}, {Blum}, {Burleigh},
  {Fan}, {Findlay}, {Finkbeiner}, {Herrera}, {Juneau}, {Landriau}, {Levi},
  {McGreer}, {Meisner}, {Myers}, {Moustakas}, {Nugent}, {Patej}, {Schlafly},
  {Walker}, {Valdes}, {Weaver}, {Y{\`e}che}, {Zou}, {Zhou}, {Abareshi},
  {Abbott}, {Abolfathi}, {Aguilera}, {Alam}, {Allen}, {Alvarez}, {Annis},
  {Ansarinejad}, {Aubert}, {Beechert}, {Bell}, {BenZvi}, {Beutler}, {Bielby},
  {Bolton}, {Brice{\~n}o}, {Buckley-Geer}, {Butler}, {Calamida}, {Carlberg},
  {Carter}, {Casas}, {Castander}, {Choi}, {Comparat}, {Cukanovaite}, {Delubac},
  {DeVries}, {Dey}, {Dhungana}, {Dickinson}, {Ding}, {Donaldson}, {Duan},
  {Duckworth}, {Eftekharzadeh}, {Eisenstein}, {Etourneau}, {Fagrelius},
  {Farihi}, {Fitzpatrick}, {Font-Ribera}, {Fulmer}, {G{\"a}nsicke},
  {Gaztanaga}, {George}, {Gerdes}, {Gontcho}, {Gorgoni}, {Green}, {Guy},
  {Harmer}, {Hernand ez}, {Honscheid}, {Huang}, {James}, {Jannuzi}, {Jiang},
  {Joyce}, {Karcher}, {Karkar}, {Kehoe}, {Kneib}, {Kueter-Young}, {Lan},
  {Lauer}, {Le Guillou}, {Le Van Suu}, {Lee}, {Lesser}, {Perreault Levasseur},
  {Li}, {Mann}, {Marshall}, {Mart{\'\i}nez-V{\'a}zquez}, {Martini}, {du Mas des
  Bourboux}, {McManus}, {Meier}, {M{\'e}nard}, {Metcalfe},
  {Mu{\~n}oz-Guti{\'e}rrez}, {Najita}, {Napier}, {Narayan}, {Newman}, {Nie},
  {Nord}, {Norman}, {Olsen}, {Paat}, {Palanque-Delabrouille}, {Peng},
  {Poppett}, {Poremba}, {Prakash}, {Rabinowitz}, {Raichoor}, {Rezaie},
  {Robertson}, {Roe}, {Ross}, {Ross}, {Rudnick}, {Safonova}, {Saha},
  {S{\'a}nchez}, {Savary}, {Schweiker}, {Scott}, {Seo}, {Shan}, {Silva},
  {Slepian}, {Soto}, {Sprayberry}, {Staten}, {Stillman}, {Stupak}, {Summers},
  {Sien Tie}, {Tirado}, {Vargas-Maga{\~n}a}, {Vivas}, {Wechsler}, {Williams},
  {Yang}, {Yang}, {Yapici}, {Zaritsky}, {Zenteno}, {Zhang}, {Zhang}, {Zhou}, \&
  {Zhou}}]{Dey19}
{Dey}, A., {Schlegel}, D.~J., {Lang}, D., {et~al.} 2019, \aj, 157, 168,
  \dodoi{10.3847/1538-3881/ab089d}

\bibitem[{{Drake} {et~al.}(2009){Drake}, {Djorgovski}, {Mahabal}, {Beshore},
  {Larson}, {Graham}, {Williams}, {Christensen}, {Catelan}, {Boattini},
  {Gibbs}, {Hill}, \& {Kowalski}}]{Drake09}
{Drake}, A.~J., {Djorgovski}, S.~G., {Mahabal}, A., {et~al.} 2009, \apj, 696,
  870, \dodoi{10.1088/0004-637X/696/1/870}

\bibitem[{{Drake} {et~al.}(2013){Drake}, {Catelan}, {Djorgovski}, {Torrealba},
  {Graham}, {Belokurov}, {Koposov}, {Mahabal}, {Prieto}, {Donalek}, {Williams},
  {Larson}, {Christensen}, \& {Beshore}}]{Drake13}
{Drake}, A.~J., {Catelan}, M., {Djorgovski}, S.~G., {et~al.} 2013, \apj, 763,
  32, \dodoi{10.1088/0004-637X/763/1/32}

\bibitem[{{Eracleous} {et~al.}(2012){Eracleous}, {Boroson}, {Halpern}, \&
  {Liu}}]{Eracleous12}
{Eracleous}, M., {Boroson}, T.~A., {Halpern}, J.~P., \& {Liu}, J. 2012, \apjs,
  201, 23, \dodoi{10.1088/0067-0049/201/2/23}

\bibitem[{{Flaugher} {et~al.}(2015){Flaugher}, {Diehl}, {Honscheid}, {Abbott},
  {Alvarez}, {Angstadt}, {Annis}, {Antonik}, {Ballester}, {Beaufore},
  {Bernstein}, {Bernstein}, {Bigelow}, {Bonati}, {Boprie}, {Brooks},
  {Buckley-Geer}, {Campa}, {Cardiel-Sas}, {Castand er}, {Castilla}, {Cease},
  {Cela-Ruiz}, {Chappa}, {Chi}, {Cooper}, {da Costa}, {Dede}, {Derylo},
  {DePoy}, {de Vicente}, {Doel}, {Drlica-Wagner}, {Eiting}, {Elliott}, {Emes},
  {Estrada}, {Fausti Neto}, {Finley}, {Flores}, {Frieman}, {Gerdes},
  {Gladders}, {Gregory}, {Gutierrez}, {Hao}, {Holland}, {Holm}, {Huffman},
  {Jackson}, {James}, {Jonas}, {Karcher}, {Karliner}, {Kent}, {Kessler},
  {Kozlovsky}, {Kron}, {Kubik}, {Kuehn}, {Kuhlmann}, {Kuk}, {Lahav}, {Lathrop},
  {Lee}, {Levi}, {Lewis}, {Li}, {Mand richenko}, {Marshall}, {Martinez},
  {Merritt}, {Miquel}, {Mu{\~n}oz}, {Neilsen}, {Nichol}, {Nord}, {Ogando},
  {Olsen}, {Palaio}, {Patton}, {Peoples}, {Plazas}, {Rauch}, {Reil}, {Rheault},
  {Roe}, {Rogers}, {Roodman}, {Sanchez}, {Scarpine}, {Schindler}, {Schmidt},
  {Schmitt}, {Schubnell}, {Schultz}, {Schurter}, {Scott}, {Serrano}, {Shaw},
  {Smith}, {Soares-Santos}, {Stefanik}, {Stuermer}, {Suchyta}, {Sypniewski},
  {Tarle}, {Thaler}, {Tighe}, {Tran}, {Tucker}, {Walker}, {Wang}, {Watson},
  {Weaverdyck}, {Wester}, {Woods}, {Yanny}, \& {DES
  Collaboration}}]{Flaugher15}
{Flaugher}, B., {Diehl}, H.~T., {Honscheid}, K., {et~al.} 2015, \aj, 150, 150,
  \dodoi{10.1088/0004-6256/150/5/150}

\bibitem[{{Goad} {et~al.}(1993){Goad}, {O'Brien}, \& {Gondhalekar}}]{Goad93}
{Goad}, M.~R., {O'Brien}, P.~T., \& {Gondhalekar}, P.~M. 1993, \mnras, 263,
  149, \dodoi{10.1093/mnras/263.1.149}

\bibitem[{{Goodrich}(1989)}]{Goodrich89}
{Goodrich}, R.~W. 1989, \apj, 340, 190, \dodoi{10.1086/167384}

\bibitem[{{Graham} {et~al.}(2020){Graham}, {Ross}, {Stern}, {Drake},
  {McKernan}, {Ford}, {Djorgovski}, {Mahabal}, {Glikman}, {Larson}, \&
  {Christensen}}]{Graham20}
{Graham}, M.~J., {Ross}, N.~P., {Stern}, D., {et~al.} 2020, \mnras, 491, 4925,
  \dodoi{10.1093/mnras/stz3244}

\bibitem[{{Gravity Collaboration} {et~al.}(2018){Gravity Collaboration},
  {Sturm}, {Dexter}, {Pfuhl}, {Stock}, {Davies}, {Lutz}, {Cl{\'e}net},
  {Eckart}, {Eisenhauer}, {Genzel}, {Gratadour}, {H{\"o}nig}, {Kishimoto},
  {Lacour}, {Millour}, {Netzer}, {Perrin}, {Peterson}, {Petrucci}, {Rouan},
  {Waisberg}, {Woillez}, {Amorim}, {Brandner}, {F{\"o}rster Schreiber},
  {Garcia}, {Gillessen}, {Ott}, {Paumard}, {Perraut}, {Scheithauer},
  {Straubmeier}, {Tacconi}, \& {Widmann}}]{Gravity18}
{Gravity Collaboration}, {Sturm}, E., {Dexter}, J., {et~al.} 2018, \nat, 563,
  657, \dodoi{10.1038/s41586-018-0731-9}

\bibitem[{{Guo} \& {Gu}(2014)}]{Guo14b}
{Guo}, H., \& {Gu}, M. 2014, Journal of Astrophysics and Astronomy, 35, 477,
  \dodoi{10.1007/s12036-014-9257-1}

\bibitem[{{Guo} \& {Gu}(2016)}]{Guo16b}
---. 2016, \apj, 822, 26, \dodoi{10.3847/0004-637X/822/1/26}

\bibitem[{{Guo} {et~al.}(2018){Guo}, {Shen}, \& {Wang}}]{Guo18}
{Guo}, H., {Shen}, Y., \& {Wang}, S. 2018, {PyQSOFit: Python code to fit the
  spectrum of quasars}.
\newblock \doeprint{1809.008}

\bibitem[{{Guo} {et~al.}(2019){Guo}, {Sun}, {Liu}, {Wang}, {Kong}, {Wang},
  {Sheng}, \& {He}}]{Guo19a}
{Guo}, H., {Sun}, M., {Liu}, X., {et~al.} 2019, \apjl, 883, L44,
  \dodoi{10.3847/2041-8213/ab4138}

\bibitem[{{Guo} {et~al.}(2020){Guo}, {Shen}, {He}, {Wang}, {Liu}, {Wang},
  {Sun}, {Yang}, {Kong}, \& {Sheng}}]{Guo20}
{Guo}, H., {Shen}, Y., {He}, Z., {et~al.} 2020, \apj, 888, 58,
  \dodoi{10.3847/1538-4357/ab5db0}

\bibitem[{{Homan} {et~al.}(2020){Homan}, {MacLeod}, {Lawrence}, {Ross}, \&
  {Bruce}}]{Homan20}
{Homan}, D., {MacLeod}, C.~L., {Lawrence}, A., {Ross}, N.~P., \& {Bruce}, A.
  2020, \mnras, 496, 309, \dodoi{10.1093/mnras/staa1467}

\bibitem[{{Hunter}(2007)}]{Hunter07}
{Hunter}, J.~D. 2007, Computing in Science and Engineering, 9, 90,
  \dodoi{10.1109/MCSE.2007.55}

\bibitem[{{Hutchinson} {et~al.}(2016){Hutchinson}, {Bolton}, {Dawson}, {Allende
  Prieto}, {Bailey}, {Bautista}, {Brownstein}, {Conroy}, {Guy}, {Myers},
  {Newman}, {Prakash}, {Carnero-Rosell}, {Seo}, {Tojeiro}, {Vivek}, \& {Ben
  Zhu}}]{Hutchinson16}
{Hutchinson}, T.~A., {Bolton}, A.~S., {Dawson}, K.~S., {et~al.} 2016, \aj, 152,
  205, \dodoi{10.3847/0004-6256/152/6/205}

\bibitem[{{Hutsem{\'e}kers} {et~al.}(2019){Hutsem{\'e}kers}, {Ag{\'\i}s
  Gonz{\'a}lez}, {Marin}, {Sluse}, {Ramos Almeida}, \& {Acosta
  Pulido}}]{Hutsemekers19}
{Hutsem{\'e}kers}, D., {Ag{\'\i}s Gonz{\'a}lez}, B., {Marin}, F., {et~al.}
  2019, \aap, 625, A54, \dodoi{10.1051/0004-6361/201834633}

\bibitem[{{Hutsem{\'e}kers} {et~al.}(2017){Hutsem{\'e}kers}, {Ag{\'\i}s
  Gonz{\'a}lez}, {Sluse}, {Ramos Almeida}, \& {Acosta Pulido}}]{Hutsemekers17}
{Hutsem{\'e}kers}, D., {Ag{\'\i}s Gonz{\'a}lez}, B., {Sluse}, D., {Ramos
  Almeida}, C., \& {Acosta Pulido}, J.~A. 2017, \aap, 604, L3,
  \dodoi{10.1051/0004-6361/201731397}

\bibitem[{{Jiang} {et~al.}(2007){Jiang}, {Fan}, {Ivezi{\'c}}, {Richards},
  {Schneider}, {Strauss}, \& {Kelly}}]{Jiang07}
{Jiang}, L., {Fan}, X., {Ivezi{\'c}}, {\v{Z}}., {et~al.} 2007, \apj, 656, 680,
  \dodoi{10.1086/510831}

\bibitem[{{Jiang} {et~al.}(2008){Jiang}, {Fan}, \& {Vestergaard}}]{Jiang08}
{Jiang}, L., {Fan}, X., \& {Vestergaard}, M. 2008, \apj, 679, 962,
  \dodoi{10.1086/587868}

\bibitem[{{Jiang} {et~al.}(2013){Jiang}, {Stone}, \& {Davis}}]{Jiang13}
{Jiang}, Y.-F., {Stone}, J.~M., \& {Davis}, S.~W. 2013, \apj, 778, 65,
  \dodoi{10.1088/0004-637X/778/1/65}

\bibitem[{{Kollmeier} {et~al.}(2017){Kollmeier}, {Zasowski}, {Rix}, {Johns},
  {Anderson}, {Drory}, {Johnson}, {Pogge}, {Bird}, {Blanc}, {Brownstein},
  {Crane}, {De Lee}, {Klaene}, {Kreckel}, {MacDonald}, {Merloni}, {Ness},
  {O'Brien}, {Sanchez-Gallego}, {Sayres}, {Shen}, {Thakar}, {Tkachenko},
  {Aerts}, {Blanton}, {Eisenstein}, {Holtzman}, {Maoz}, {Nandra}, {Rockosi},
  {Weinberg}, {Bovy}, {Casey}, {Chaname}, {Clerc}, {Conroy}, {Eracleous},
  {G{\"a}nsicke}, {Hekker}, {Horne}, {Kauffmann}, {McQuinn}, {Pellegrini},
  {Schinnerer}, {Schlafly}, {Schwope}, {Seibert}, {Teske}, \& {van
  Saders}}]{Kollmeier17}
{Kollmeier}, J.~A., {Zasowski}, G., {Rix}, H.-W., {et~al.} 2017, arXiv
  e-prints, arXiv:1711.03234.
\newblock \doarXiv{1711.03234}

\bibitem[{{Korista} \& {Goad}(2000)}]{korista00}
{Korista}, K.~T., \& {Goad}, M.~R. 2000, \apj, 536, 284, \dodoi{10.1086/308930}

\bibitem[{{Krolik} {et~al.}(1991){Krolik}, {Horne}, {Kallman}, {Malkan},
  {Edelson}, \& {Kriss}}]{Krolik91}
{Krolik}, J.~H., {Horne}, K., {Kallman}, T.~R., {et~al.} 1991, \apj, 371, 541,
  \dodoi{10.1086/169918}

\bibitem[{{LaMassa} {et~al.}(2015){LaMassa}, {Cales}, {Moran}, {Myers},
  {Richards}, {Eracleous}, {Heckman}, {Gallo}, \& {Urry}}]{LaMassa15}
{LaMassa}, S.~M., {Cales}, S., {Moran}, E.~C., {et~al.} 2015, \apj, 800, 144,
  \dodoi{10.1088/0004-637X/800/2/144}

\bibitem[{{Law} {et~al.}(2009){Law}, {Kulkarni}, {Dekany}, {Ofek}, {Quimby},
  {Nugent}, {Surace}, {Grillmair}, {Bloom}, {Kasliwal}, {Bildsten}, {Brown},
  {Cenko}, {Ciardi}, {Croner}, {Djorgovski}, {van Eyken}, {Filippenko}, {Fox},
  {Gal-Yam}, {Hale}, {Hamam}, {Helou}, {Henning}, {Howell}, {Jacobsen},
  {Laher}, {Mattingly}, {McKenna}, {Pickles}, {Poznanski}, {Rahmer}, {Rau},
  {Rosing}, {Shara}, {Smith}, {Starr}, {Sullivan}, {Velur}, {Walters}, \&
  {Zolkower}}]{Law09}
{Law}, N.~M., {Kulkarni}, S.~R., {Dekany}, R.~G., {et~al.} 2009, \pasp, 121,
  1395, \dodoi{10.1086/648598}

\bibitem[{{Lawrence}(2018)}]{Lawrence18}
{Lawrence}, A. 2018, Nature Astronomy, 2, 102,
  \dodoi{10.1038/s41550-017-0372-1}

\bibitem[{{Liu} {et~al.}(2018){Liu}, {Dittmann}, {Shen}, \& {Jiang}}]{Liu18b}
{Liu}, X., {Dittmann}, A., {Shen}, Y., \& {Jiang}, L. 2018, \apj, 859, 8,
  \dodoi{10.3847/1538-4357/aabb04}

\bibitem[{{Liu} {et~al.}(2014){Liu}, {Shen}, {Bian}, {Loeb}, \&
  {Tremaine}}]{Liu14}
{Liu}, X., {Shen}, Y., {Bian}, F., {Loeb}, A., \& {Tremaine}, S. 2014, \apj,
  789, 140, \dodoi{10.1088/0004-637X/789/2/140}

\bibitem[{{Luo} {et~al.}(2020){Luo}, {Shen}, \& {Yang}}]{luo20}
{Luo}, Y., {Shen}, Y., \& {Yang}, Q. 2020, \mnras, 494, 3686,
  \dodoi{10.1093/mnras/staa972}

\bibitem[{{MacLeod} {et~al.}(2010){MacLeod}, {Ivezi{\'c}}, {Kochanek},
  {Koz{\l}owski}, {Kelly}, {Bullock}, {Kimball}, {Sesar}, {Westman}, {Brooks},
  {Gibson}, {Becker}, \& {de Vries}}]{MacLeod10}
{MacLeod}, C.~L., {Ivezi{\'c}}, {\v{Z}}., {Kochanek}, C.~S., {et~al.} 2010,
  \apj, 721, 1014, \dodoi{10.1088/0004-637X/721/2/1014}

\bibitem[{{MacLeod} {et~al.}(2016){MacLeod}, {Ross}, {Lawrence}, {Goad},
  {Horne}, {Burgett}, {Chambers}, {Flewelling}, {Hodapp}, {Kaiser}, {Magnier},
  {Wainscoat}, \& {Waters}}]{Macleod16}
{MacLeod}, C.~L., {Ross}, N.~P., {Lawrence}, A., {et~al.} 2016, \mnras, 457,
  389, \dodoi{10.1093/mnras/stv2997}

\bibitem[{{MacLeod} {et~al.}(2019){MacLeod}, {Green}, {Anderson}, {Bruce},
  {Eracleous}, {Graham}, {Homan}, {Lawrence}, {LeBleu}, {Ross}, {Ruan},
  {Runnoe}, {Stern}, {Burgett}, {Chambers}, {Kaiser}, {Magnier}, \&
  {Metcalfe}}]{MacLeod19}
{MacLeod}, C.~L., {Green}, P.~J., {Anderson}, S.~F., {et~al.} 2019, \apj, 874,
  8, \dodoi{10.3847/1538-4357/ab05e2}

\bibitem[{{Masci} {et~al.}(2019){Masci}, {Laher}, {Rusholme}, {Shupe}, {Groom},
  {Surace}, {Jackson}, {Monkewitz}, {Beck}, {Flynn}, {Terek}, {Landry},
  {Hacopians}, {Desai}, {Howell}, {Brooke}, {Imel}, {Wachter}, {Ye}, {Lin},
  {Cenko}, {Cunningham}, {Rebbapragada}, {Bue}, {Miller}, {Mahabal}, {Bellm},
  {Patterson}, {Juri{\'c}}, {Golkhou}, {Ofek}, {Walters}, {Graham}, {Kasliwal},
  {Dekany}, {Kupfer}, {Burdge}, {Cannella}, {Barlow}, {Van Sistine}, {Giomi},
  {Fremling}, {Blagorodnova}, {Levitan}, {Riddle}, {Smith}, {Helou}, {Prince},
  \& {Kulkarni}}]{Masci19}
{Masci}, F.~J., {Laher}, R.~R., {Rusholme}, B., {et~al.} 2019, \pasp, 131,
  018003, \dodoi{10.1088/1538-3873/aae8ac}

\bibitem[{{Noda} \& {Done}(2018)}]{Noda18}
{Noda}, H., \& {Done}, C. 2018, \mnras, 480, 3898,
  \dodoi{10.1093/mnras/sty2032}

\bibitem[{{Proga} {et~al.}(2000){Proga}, {Stone}, \& {Kallman}}]{Proga00}
{Proga}, D., {Stone}, J.~M., \& {Kallman}, T.~R. 2000, \apj, 543, 686,
  \dodoi{10.1086/317154}

\bibitem[{{Rau} {et~al.}(2009){Rau}, {Kulkarni}, {Law}, {Bloom}, {Ciardi},
  {Djorgovski}, {Fox}, {Gal-Yam}, {Grillmair}, {Kasliwal}, {Nugent}, {Ofek},
  {Quimby}, {Reach}, {Shara}, {Bildsten}, {Cenko}, {Drake}, {Filippenko},
  {Helfand}, {Helou}, {Howell}, {Poznanski}, \& {Sullivan}}]{Rau09}
{Rau}, A., {Kulkarni}, S.~R., {Law}, N.~M., {et~al.} 2009, \pasp, 121, 1334,
  \dodoi{10.1086/605911}

\bibitem[{{Richards} {et~al.}(2002){Richards}, {Fan}, {Newberg}, {Strauss},
  {Vanden Berk}, {Schneider}, {Yanny}, {Boucher}, {Burles}, {Frieman}, {Gunn},
  {Hall}, {Ivezi{\'c}}, {Kent}, {Loveday}, {Lupton}, {Rockosi}, {Schlegel},
  {Stoughton}, {SubbaRao}, \& {York}}]{Richards02}
{Richards}, G.~T., {Fan}, X., {Newberg}, H.~J., {et~al.} 2002, \aj, 123, 2945,
  \dodoi{10.1086/340187}

\bibitem[{{Richards} {et~al.}(2006){Richards}, {Lacy}, {Storrie-Lombardi},
  {Hall}, {Gallagher}, {Hines}, {Fan}, {Papovich}, {Vanden Berk}, {Trammell},
  {Schneider}, {Vestergaard}, {York}, {Jester}, {Anderson}, {Budav{\'a}ri}, \&
  {Szalay}}]{Richards06}
{Richards}, G.~T., {Lacy}, M., {Storrie-Lombardi}, L.~J., {et~al.} 2006, \apjs,
  166, 470, \dodoi{10.1086/506525}

\bibitem[{{Roig} {et~al.}(2014){Roig}, {Blanton}, \& {Ross}}]{Roig14}
{Roig}, B., {Blanton}, M.~R., \& {Ross}, N.~P. 2014, \apj, 781, 72,
  \dodoi{10.1088/0004-637X/781/2/72}

\bibitem[{{Ross} {et~al.}(2019){Ross}, {Graham}, {Calderone}, {Ford},
  {McKernan}, \& {Stern}}]{Ross19}
{Ross}, N.~P., {Graham}, M.~J., {Calderone}, G., {et~al.} 2019, arXiv e-prints,
  arXiv:1912.05310.
\newblock \doarXiv{1912.05310}

\bibitem[{{Ruan} {et~al.}(2014){Ruan}, {Anderson}, {Dexter}, \&
  {Agol}}]{Ruan14}
{Ruan}, J.~J., {Anderson}, S.~F., {Dexter}, J., \& {Agol}, E. 2014, \apj, 783,
  105, \dodoi{10.1088/0004-637X/783/2/105}

\bibitem[{{Ruan} {et~al.}(2019){Ruan}, {Anderson}, {Eracleous}, {Green},
  {Haggard}, {MacLeod}, {Runnoe}, \& {Sobolewska}}]{Ruan19}
{Ruan}, J.~J., {Anderson}, S.~F., {Eracleous}, M., {et~al.} 2019, \apj, 883,
  76, \dodoi{10.3847/1538-4357/ab3c1a}

\bibitem[{{Ruan} {et~al.}(2016{\natexlab{a}}){Ruan}, {Anderson}, {Green},
  {Morganson}, {Eracleous}, {Myers}, {Badenes}, {Bershady}, {Brandt},
  {Chambers}, {Davenport}, {Dawson}, {Flewelling}, {Heckman}, {Isler},
  {Kaiser}, {Kneib}, {MacLeod}, {Paris}, {Ross}, {Runnoe}, {Schlafly},
  {Schmidt}, {Schneider}, {Schwope}, {Shen}, {Stassun}, {Szkody}, {Waters}, \&
  {York}}]{Ruan16b}
{Ruan}, J.~J., {Anderson}, S.~F., {Green}, P.~J., {et~al.} 2016{\natexlab{a}},
  \apj, 825, 137, \dodoi{10.3847/0004-637X/825/2/137}

\bibitem[{{Ruan} {et~al.}(2016{\natexlab{b}}){Ruan}, {Anderson}, {Cales},
  {Eracleous}, {Green}, {Morganson}, {Runnoe}, {Shen}, {Wilkinson}, {Blanton},
  {Dwelly}, {Georgakakis}, {Greene}, {LaMassa}, {Merloni}, \&
  {Schneider}}]{Ruan16a}
{Ruan}, J.~J., {Anderson}, S.~F., {Cales}, S.~L., {et~al.} 2016{\natexlab{b}},
  \apj, 826, 188, \dodoi{10.3847/0004-637X/826/2/188}

\bibitem[{{Rumbaugh} {et~al.}(2018){Rumbaugh}, {Shen}, {Morganson}, {Liu},
  {Banerji}, {McMahon}, {Abdalla}, {Benoit-L{\'e}vy}, {Bertin}, {Brooks},
  {Buckley-Geer}, {Capozzi}, {Carnero Rosell}, {Carrasco Kind}, {Carretero},
  {Cunha}, {D'Andrea}, {da Costa}, {DePoy}, {Desai}, {Doel}, {Frieman},
  {Garc{\'\i}a-Bellido}, {Gruen}, {Gruendl}, {Gschwend}, {Gutierrez},
  {Honscheid}, {James}, {Kuehn}, {Kuhlmann}, {Kuropatkin}, {Lima}, {Maia},
  {Marshall}, {Martini}, {Menanteau}, {Plazas}, {Reil}, {Roodman}, {Sanchez},
  {Scarpine}, {Schindler}, {Schubnell}, {Sheldon}, {Smith}, {Soares-Santos},
  {Sobreira}, {Suchyta}, {Swanson}, {Walker}, {Wester}, \& {DES
  Collaboration}}]{Rumbaugh18}
{Rumbaugh}, N., {Shen}, Y., {Morganson}, E., {et~al.} 2018, \apj, 854, 160,
  \dodoi{10.3847/1538-4357/aaa9b6}

\bibitem[{{Runco} {et~al.}(2016){Runco}, {Cosens}, {Bennert}, {Scott},
  {Komossa}, {Malkan}, {Lazarova}, {Auger}, {Treu}, \& {Park}}]{Runco16}
{Runco}, J.~N., {Cosens}, M., {Bennert}, V.~N., {et~al.} 2016, \apj, 821, 33,
  \dodoi{10.3847/0004-637X/821/1/33}

\bibitem[{{Runnoe} {et~al.}(2013){Runnoe}, {Brotherton}, {Shang}, \&
  {DiPompeo}}]{Runnoe13}
{Runnoe}, J.~C., {Brotherton}, M.~S., {Shang}, Z., \& {DiPompeo}, M.~A. 2013,
  \mnras, 434, 848, \dodoi{10.1093/mnras/stt1077}

\bibitem[{{Runnoe} {et~al.}(2016){Runnoe}, {Cales}, {Ruan}, {Eracleous},
  {Anderson}, {Shen}, {Green}, {Morganson}, {LaMassa}, {Greene}, {Dwelly},
  {Schneider}, {Merloni}, {Georgakakis}, \& {Roman-Lopes}}]{Runnoe16}
{Runnoe}, J.~C., {Cales}, S., {Ruan}, J.~J., {et~al.} 2016, \mnras, 455, 1691,
  \dodoi{10.1093/mnras/stv2385}

\bibitem[{{Runnoe} {et~al.}(2017){Runnoe}, {Eracleous}, {Pennell}, {Mathes},
  {Boroson}, {Sigur{\dh}sson}, {Bogdanovi{\'c}}, {Halpern}, {Liu}, \&
  {Brown}}]{Runnoe17}
{Runnoe}, J.~C., {Eracleous}, M., {Pennell}, A., {et~al.} 2017, \mnras, 468,
  1683, \dodoi{10.1093/mnras/stx452}

\bibitem[{{Shakura} \& {Sunyaev}(1973)}]{Shakura73}
{Shakura}, N.~I., \& {Sunyaev}, R.~A. 1973, \aap, 500, 33

\bibitem[{{Shappee} {et~al.}(2014){Shappee}, {Prieto}, {Grupe}, {Kochanek},
  {Stanek}, {De Rosa}, {Mathur}, {Zu}, {Peterson}, {Pogge}, {Komossa}, {Im},
  {Jencson}, {Holoien}, {Basu}, {Beacom}, {Szczygie{\l}}, {Brimacombe},
  {Adams}, {Campillay}, {Choi}, {Contreras}, {Dietrich}, {Dubberley},
  {Elphick}, {Foale}, {Giustini}, {Gonzalez}, {Hawkins}, {Howell}, {Hsiao},
  {Koss}, {Leighly}, {Morrell}, {Mudd}, {Mullins}, {Nugent}, {Parrent},
  {Phillips}, {Pojmanski}, {Rosing}, {Ross}, {Sand}, {Terndrup}, {Valenti},
  {Walker}, \& {Yoon}}]{Shappee14}
{Shappee}, B.~J., {Prieto}, J.~L., {Grupe}, D., {et~al.} 2014, \apj, 788, 48,
  \dodoi{10.1088/0004-637X/788/1/48}

\bibitem[{{Shen}(2013)}]{Shen13}
{Shen}, Y. 2013, Bulletin of the Astronomical Society of India, 41, 61.
\newblock \doarXiv{1302.2643}

\bibitem[{{Shen} {et~al.}(2008){Shen}, {Greene}, {Strauss}, {Richards}, \&
  {Schneider}}]{Shen08}
{Shen}, Y., {Greene}, J.~E., {Strauss}, M.~A., {Richards}, G.~T., \&
  {Schneider}, D.~P. 2008, \apj, 680, 169, \dodoi{10.1086/587475}

\bibitem[{{Shen} \& {Liu}(2012)}]{Shen12}
{Shen}, Y., \& {Liu}, X. 2012, \apj, 753, 125,
  \dodoi{10.1088/0004-637X/753/2/125}

\bibitem[{{Shen} {et~al.}(2011){Shen}, {Richards}, {Strauss}, {Hall},
  {Schneider}, {Snedden}, {Bizyaev}, {Brewington}, {Malanushenko},
  {Malanushenko}, {Oravetz}, {Pan}, \& {Simmons}}]{Shen11}
{Shen}, Y., {Richards}, G.~T., {Strauss}, M.~A., {et~al.} 2011, \apjs, 194, 45,
  \dodoi{10.1088/0067-0049/194/2/45}

\bibitem[{{Shen} {et~al.}(2015){Shen}, {Brandt}, {Dawson}, {Hall}, {McGreer},
  {Anderson}, {Chen}, {Denney}, {Eftekharzadeh}, {Fan}, {Gao}, {Green},
  {Greene}, {Ho}, {Horne}, {Jiang}, {Kelly}, {Kinemuchi}, {Kochanek},
  {P{\^a}ris}, {Peters}, {Peterson}, {Petitjean}, {Ponder}, {Richards},
  {Schneider}, {Seth}, {Smith}, {Strauss}, {Tao}, {Trump}, {Wood-Vasey}, {Zu},
  {Eisenstein}, {Pan}, {Bizyaev}, {Malanushenko}, {Malanushenko}, \&
  {Oravetz}}]{Shen15}
{Shen}, Y., {Brandt}, W.~N., {Dawson}, K.~S., {et~al.} 2015, \apjs, 216, 4,
  \dodoi{10.1088/0067-0049/216/1/4}

\bibitem[{{Shen} {et~al.}(2019){Shen}, {Hall}, {Horne}, {Zhu}, {McGreer},
  {Simm}, {Trump}, {Kinemuchi}, {Brandt}, {Green}, {Grier}, {Guo}, {Ho},
  {Homayouni}, {Jiang}, {I-Hsiu Li}, {Morganson}, {Petitjean}, {Richards},
  {Schneider}, {Starkey}, {Wang}, {Chambers}, {Kaiser}, {Kudritzki}, {Magnier},
  \& {Waters}}]{Shen19}
{Shen}, Y., {Hall}, P.~B., {Horne}, K., {et~al.} 2019, \apjs, 241, 34,
  \dodoi{10.3847/1538-4365/ab074f}

\bibitem[{{Sheng} {et~al.}(2017){Sheng}, {Wang}, {Jiang}, {Yang}, {Yan}, {Dou},
  \& {Peng}}]{Sheng17}
{Sheng}, Z., {Wang}, T., {Jiang}, N., {et~al.} 2017, \apjl, 846, L7,
  \dodoi{10.3847/2041-8213/aa85de}

\bibitem[{{Sheng} {et~al.}(2020){Sheng}, {Wang}, {Jiang}, {Ding}, {Cai}, {Guo},
  {Sun}, {Dou}, \& {Yang}}]{Sheng20}
---. 2020, \apj, 889, 46, \dodoi{10.3847/1538-4357/ab5af9}

\bibitem[{{Stoughton} {et~al.}(2002){Stoughton}, {Lupton}, {Bernardi},
  {Blanton}, {Burles}, {Castand er}, {Connolly}, {Eisenstein}, {Frieman},
  {Hennessy}, {Hindsley}, {Ivezi{\'c}}, {Kent}, {Kunszt}, {Lee}, {Meiksin},
  {Munn}, {Newberg}, {Nichol}, {Nicinski}, {Pier}, {Richards}, {Richmond},
  {Schlegel}, {Smith}, {Strauss}, {SubbaRao}, {Szalay}, {Thakar}, {Tucker},
  {Vand en Berk}, {Yanny}, {Adelman}, {Anderson}, {Anderson}, {Annis},
  {Bahcall}, {Bakken}, {Bartelmann}, {Bastian}, {Bauer}, {Berman},
  {B{\"o}hringer}, {Boroski}, {Bracker}, {Briegel}, {Briggs}, {Brinkmann},
  {Brunner}, {Carey}, {Carr}, {Chen}, {Christian}, {Colestock}, {Crocker},
  {Csabai}, {Czarapata}, {Dalcanton}, {Davidsen}, {Davis}, {Dehnen},
  {Dodelson}, {Doi}, {Dombeck}, {Donahue}, {Ellman}, {Elms}, {Evans}, {Eyer},
  {Fan}, {Federwitz}, {Friedman}, {Fukugita}, {Gal}, {Gillespie}, {Glazebrook},
  {Gray}, {Grebel}, {Greenawalt}, {Greene}, {Gunn}, {de Haas}, {Haiman},
  {Haldeman}, {Hall}, {Hamabe}, {Hansen}, {Harris}, {Harris}, {Harvanek},
  {Hawley}, {Hayes}, {Heckman}, {Helmi}, {Henden}, {Hogan}, {Hogg}, {Holmgren},
  {Holtzman}, {Huang}, {Hull}, {Ichikawa}, {Ichikawa}, {Johnston}, {Kauffmann},
  {Kim}, {Kimball}, {Kinney}, {Klaene}, {Kleinman}, {Klypin}, {Knapp},
  {Korienek}, {Krolik}, {Kron}, {Krzesi{\'n}ski}, {Lamb}, {Leger},
  {Limmongkol}, {Lindenmeyer}, {Long}, {Loomis}, {Loveday}, {MacKinnon},
  {Mannery}, {Mantsch}, {Margon}, {McGehee}, {McKay}, {McLean}, {Menou},
  {Merelli}, {Mo}, {Monet}, {Nakamura}, {Narayanan}, {Nash}, {Neilsen},
  {Newman}, {Nitta}, {Odenkirchen}, {Okada}, {Okamura}, {Ostriker}, {Owen},
  {Pauls}, {Peoples}, {Peterson}, {Petravick}, {Pope}, {Pordes}, {Postman},
  {Prosapio}, {Quinn}, {Rechenmacher}, {Rivetta}, {Rix}, {Rockosi}, {Rosner},
  {Ruthmansdorfer}, {Sandford}, {Schneider}, {Scranton}, {Sekiguchi}, {Sergey},
  {Sheth}, {Shimasaku}, {Smee}, {Snedden}, {Stebbins}, {Stubbs}, {Szapudi},
  {Szkody}, {Szokoly}, {Tabachnik}, {Tsvetanov}, {Uomoto}, {Vogeley}, {Voges},
  {Waddell}, {Walterbos}, {Wang}, {Watanabe}, {Weinberg}, {White}, {White},
  {Wilhite}, {Wolfe}, {Yasuda}, {York}, {Zehavi}, \& {Zheng}}]{Stoughton02}
{Stoughton}, C., {Lupton}, R.~H., {Bernardi}, M., {et~al.} 2002, \aj, 123, 485,
  \dodoi{10.1086/324741}

\bibitem[{{STScI Development Team}(2013)}]{STScIDevelopmentTeam13}
{STScI Development Team}. 2013, {pysynphot: Synthetic photometry software
  package}.
\newblock \doeprint{1303.023}

\bibitem[{{Sun} {et~al.}(2018){Sun}, {Xue}, {Cai}, \& {Guo}}]{Sun18b}
{Sun}, M., {Xue}, Y., {Cai}, Z., \& {Guo}, H. 2018, \apj, 857, 86,
  \dodoi{10.3847/1538-4357/aab786}

\bibitem[{{Sun} {et~al.}(2015){Sun}, {Trump}, {Shen}, {Brand t}, {Dawson},
  {Denney}, {Hall}, {Ho}, {Horne}, {Jiang}, {Richards}, {Schneider}, {Bizyaev},
  {Kinemuchi}, {Oravetz}, {Pan}, \& {Simmons}}]{Sun15}
{Sun}, M., {Trump}, J.~R., {Shen}, Y., {et~al.} 2015, \apj, 811, 42,
  \dodoi{10.1088/0004-637X/811/1/42}

\bibitem[{{Sun} {et~al.}(2020){Sun}, {Xue}, {Brandt}, {Gu}, {Trump}, {Cai},
  {He}, {Lin}, {Liu}, \& {Wang}}]{Sun20}
{Sun}, M., {Xue}, Y., {Brandt}, W.~N., {et~al.} 2020, \apj, 891, 178,
  \dodoi{10.3847/1538-4357/ab789e}

\bibitem[{{Sun} {et~al.}(2014){Sun}, {Wang}, {Chen}, \& {Zheng}}]{Sun14}
{Sun}, Y.-H., {Wang}, J.-X., {Chen}, X.-Y., \& {Zheng}, Z.-Y. 2014, \apj, 792,
  54, \dodoi{10.1088/0004-637X/792/1/54}

\bibitem[{{Tohline} \& {Osterbrock}(1976)}]{Tohline76}
{Tohline}, J.~E., \& {Osterbrock}, D.~E. 1976, \apjl, 210, L117,
  \dodoi{10.1086/182317}

\bibitem[{{Trakhtenbrot} {et~al.}(2019){Trakhtenbrot}, {Arcavi}, {MacLeod},
  {Ricci}, {Kara}, {Graham}, {Stern}, {Harrison}, {Burke}, {Hiramatsu},
  {Hosseinzadeh}, {Howell}, {Smartt}, {Rest}, {Prieto}, {Shappee}, {Holoien},
  {Bersier}, {Filippenko}, {Brink}, {Zheng}, {Li}, {Remillard}, \&
  {Loewenstein}}]{Trakhtenbrot19}
{Trakhtenbrot}, B., {Arcavi}, I., {MacLeod}, C.~L., {et~al.} 2019, \apj, 883,
  94, \dodoi{10.3847/1538-4357/ab39e4}

\bibitem[{{Tran} {et~al.}(1992){Tran}, {Osterbrock}, \& {Martel}}]{Tran92}
{Tran}, H.~D., {Osterbrock}, D.~E., \& {Martel}, A. 1992, \aj, 104, 2072,
  \dodoi{10.1086/116382}

\bibitem[{{Urry} \& {Padovani}(1995)}]{Urry95}
{Urry}, C.~M., \& {Padovani}, P. 1995, \pasp, 107, 803, \dodoi{10.1086/133630}

\bibitem[{{van der Walt} {et~al.}(2011){van der Walt}, {Colbert}, \&
  {Varoquaux}}]{Numpy}
{van der Walt}, S., {Colbert}, S.~C., \& {Varoquaux}, G. 2011, Computing in
  Science and Engineering, 13, 22, \dodoi{10.1109/MCSE.2011.37}

\bibitem[{{Vanden Berk} {et~al.}(2001){Vanden Berk}, {Richards}, {Bauer},
  {Strauss}, {Schneider}, {Heckman}, {York}, {Hall}, {Fan}, {Knapp},
  {Anderson}, {Annis}, {Bahcall}, {Bernardi}, {Briggs}, {Brinkmann}, {Brunner},
  {Burles}, {Carey}, {Castander}, {Connolly}, {Crocker}, {Csabai}, {Doi},
  {Finkbeiner}, {Friedman}, {Frieman}, {Fukugita}, {Gunn}, {Hennessy},
  {Ivezi{\'c}}, {Kent}, {Kunszt}, {Lamb}, {Leger}, {Long}, {Loveday}, {Lupton},
  {Meiksin}, {Merelli}, {Munn}, {Newberg}, {Newcomb}, {Nichol}, {Owen}, {Pier},
  {Pope}, {Rockosi}, {Schlegel}, {Siegmund}, {Smee}, {Snir}, {Stoughton},
  {Stubbs}, {SubbaRao}, {Szalay}, {Szokoly}, {Tremonti}, {Uomoto}, {Waddell},
  {Yanny}, \& {Zheng}}]{VandenBerk01}
{Vanden Berk}, D.~E., {Richards}, G.~T., {Bauer}, A., {et~al.} 2001, \aj, 122,
  549, \dodoi{10.1086/321167}

\bibitem[{{Vestergaard} \& {Peterson}(2006)}]{Vestergaard06}
{Vestergaard}, M., \& {Peterson}, B.~M. 2006, \apj, 641, 689,
  \dodoi{10.1086/500572}

\bibitem[{{Wang} {et~al.}(2018){Wang}, {Xu}, \& {Wei}}]{Wang18}
{Wang}, J., {Xu}, D.~W., \& {Wei}, J.~Y. 2018, \apj, 858, 49,
  \dodoi{10.3847/1538-4357/aab88b}

\bibitem[{{Wang} {et~al.}(2020){Wang}, {Shen}, {Jiang}, {Grier}, {Horne},
  {Homayouni}, {Peterson}, {Trump}, {Brandt}, {Hall}, {Ho}, {Li}, {Kinemuchi},
  {McGreer}, \& {Schneider}}]{Wang20}
{Wang}, S., {Shen}, Y., {Jiang}, L., {et~al.} 2020, arXiv e-prints,
  arXiv:2006.06178.
\newblock \doarXiv{2006.06178}

\bibitem[{{Wilhite} {et~al.}(2008){Wilhite}, {Brunner}, {Grier}, {Schneider},
  \& {vanden Berk}}]{Wilhite08}
{Wilhite}, B.~C., {Brunner}, R.~J., {Grier}, C.~J., {Schneider}, D.~P., \&
  {vanden Berk}, D.~E. 2008, \mnras, 383, 1232,
  \dodoi{10.1111/j.1365-2966.2007.12655.x}

\bibitem[{{Wilhite} {et~al.}(2006){Wilhite}, {Vanden Berk}, {Brunner}, \&
  {Brinkmann}}]{Wilhite06}
{Wilhite}, B.~C., {Vanden Berk}, D.~E., {Brunner}, R.~J., \& {Brinkmann}, J.~V.
  2006, \apj, 641, 78, \dodoi{10.1086/500421}

\bibitem[{{Wilhite} {et~al.}(2005){Wilhite}, {Vanden Berk}, {Kron},
  {Schneider}, {Pereyra}, {Brunner}, {Richards}, \& {Brinkmann}}]{Wilhite05}
{Wilhite}, B.~C., {Vanden Berk}, D.~E., {Kron}, R.~G., {et~al.} 2005, \apj,
  633, 638, \dodoi{10.1086/430821}

\bibitem[{{Wills} {et~al.}(1993){Wills}, {Brotherton}, {Fang}, {Steidel}, \&
  {Sargent}}]{Wills93}
{Wills}, B.~J., {Brotherton}, M.~S., {Fang}, D., {Steidel}, C.~C., \&
  {Sargent}, W. L.~W. 1993, \apj, 415, 563, \dodoi{10.1086/173186}

\bibitem[{{Xie} {et~al.}(2016){Xie}, {Shao}, {Shen}, {Liu}, \& {Li}}]{Xie16}
{Xie}, X., {Shao}, Z., {Shen}, S., {Liu}, H., \& {Li}, L. 2016, \apj, 824, 38,
  \dodoi{10.3847/0004-637X/824/1/38}

\bibitem[{{Yang} {et~al.}(2018){Yang}, {Wu}, {Fan}, {Jiang}, {McGreer},
  {Shangguan}, {Yao}, {Wang}, {Joshi}, {Green}, {Wang}, {Feng}, {Fu}, {Yang},
  \& {Liu}}]{Yang18}
{Yang}, Q., {Wu}, X.-B., {Fan}, X., {et~al.} 2018, \apj, 862, 109,
  \dodoi{10.3847/1538-4357/aaca3a}

\bibitem[{{Yang} {et~al.}(2020){Yang}, {Shen}, {Chen}, {Liu}, {Annis}, {Avila},
  {Bertin}, {Brooks}, {Buckley-Geer}, {Carnero Rosell}, {Carrasco Kind},
  {Carretero}, {da Costa}, {Desai}, {Thomas Diehl}, {Doel}, {Frieman},
  {Garcia-Bellido}, {Gaztanaga}, {Gerdes}, {Gruen}, {Gruendl}, {Gschwend},
  {Gutierrez}, {Hollowood}, {Honscheid}, {Hoyle}, {James}, {Krause}, {Kuehn},
  {Lidman}, {Lima}, {Maia}, {Marshall}, {Martini}, {Menanteau}, {Miquel},
  {Plazas Malag{\'o}n}, {Sanchez}, {Scarpine}, {Schindler}, {Schubnell},
  {Serrano}, {Sevilla}, {Smith}, {Soares-Santos}, {Sobreira}, {Suchyta},
  {Swanson}, {Tarle}, {Vikram}, \& {Walker}}]{Yang20}
{Yang}, Q., {Shen}, Y., {Chen}, Y.-C., {et~al.} 2020, \mnras, 493, 5773,
  \dodoi{10.1093/mnras/staa645}

\end{thebibliography}

\begin{longtable*}{lcl}
\caption{FITS Catalog Format}\label{tab:format}\\
\hline \hline \\[-2ex]
   \multicolumn{1}{c}{\textbf{Column}} &
   \multicolumn{1}{c}{\textbf{Format}} &
   \multicolumn{1}{c}{\textbf{Description}} \\[0.5ex] \hline
   \\[-1.8ex]
\endfirsthead
1\dotfill  & LONG       & Object ID in repeat DR16 catalog \\
2\dotfill  & LONG       & Spectroscopic plate number \\
3\dotfill  & LONG       & MJD of spectroscopic observation\\
4\dotfill  & LONG       & Spectroscopic fiber number \\
5\dotfill  & DOUBLE   & Right ascension in decimal degrees (J2000.0) \\
6\dotfill  & DOUBLE   & Declination in decimal degrees (J2000.0) \\
7\dotfill  & DOUBLE   & Redshift \\
8\dotfill  & LONG     & Number of spectroscopic observations\\
9\dotfill  & LONG   & First detection: detected = 1; not detected = 0; out of FIRST footprint = $-1$\\
10\dotfill & DOUBLE   & Observed FIRST flux density at 20 cm [mJy]\\
11\dotfill & LONG   & Flag represents the brightest (1) and faintest epochs ($-1$) in an object\\
12\dotfill & DOUBLE   & Rest frame continuum variability at 1450\AA, $\delta$ V $\equiv$ $\rm \frac{Max-Min}{0.5(Max+Min)}$ \\
13\dotfill & DOUBLE   & Observed optical flux density at rest frame 1450\AA\ [\ergcmsA]\\
14\dotfill & DOUBLE   & Signal-to-noise ratio of continuum around 1450\AA\\
15\dotfill & DOUBLE   & Signal-to-noise ratio of \civ\ line \\
16\dotfill & DOUBLE   & Monochromatic luminosity at 1350\AA\ [$\log (L_{\rm 1350}/\rm ergs^{-1})$]\\
17\dotfill & DOUBLE   & Uncertainty in $\log L_{\rm 1350}$\\
18\dotfill & DOUBLE   & FWHM of the whole \civ\ [\kms]\\
19\dotfill & DOUBLE   & Uncertainty in \civ\ FWHM\\
20\dotfill & DOUBLE   & Restframe equivalent width of the broad \civ\ [\AA]\\
21\dotfill & DOUBLE   & Uncertainty in ${\rm EW_{CIV}}$\\
22\dotfill & DOUBLE   & Virial BH mass based on \civ\ [VP06, $\log (M_{\rm BH,vir}/M_\odot$)]\\
23\dotfill & DOUBLE   & Measurement uncertainty in $\log M_{\rm BH,vir}$ (\civ, VP06)\\
24\dotfill & DOUBLE   & The adopted fiducial virial BH mass: $M_{\rm BH}=\frac{1}{2}(M_{\rm bright}+M_{\rm faint})$ [$\log (M_{\rm BH,vir}/M_\odot$)]\\
25\dotfill & DOUBLE   & Uncertainty in the fiducial virial BH mass (measurement uncertainty only)\\
26\dotfill & DOUBLE   & Eddington ratio based on the fiducial virial BH mass\\
27\dotfill & DOUBLE   & Reduced $\chi^2$ for the \civ\ line fitting\\
28\dotfill & LONG     & Type = 0 if CLQs, otherwise type = 1 for EVQs \\
\hline
\end{longtable*}

\appendix 

\renewcommand\thefigure{\thesection.\arabic{figure}} 
\setcounter{figure}{0}
\section{Spectral Flux Calibration Problem} \label{sec:calibration}
Each SDSS spectrum is flux-calibrated by matching the counts in the mean of the high SN spectra of the spectrophotometric and reddening standards on the plate and equating this to the synthetic composite F8 sub-dwarf spectrum. This is placed on an absolute scale by matching the synthesized magnitudes of these stars to the SDSS photometry. Based on previous studies, two important aspects should be considered regarding to the SDSS spectral variability: 1) the original uncertainties of SDSS spectral flux calibration (Figure \ref{fig:calibration}); 2) the fiber-drop issue (i.e., the fiber is partially dropped from a plate resulting in a significant drop in spectral flux, see Figure \ref{fig:fake_CLQ}). 

\subsection{Intrinsic scatter}\label{sec:scatter}
\begin{figure}
\centering
\includegraphics[width=9.cm]{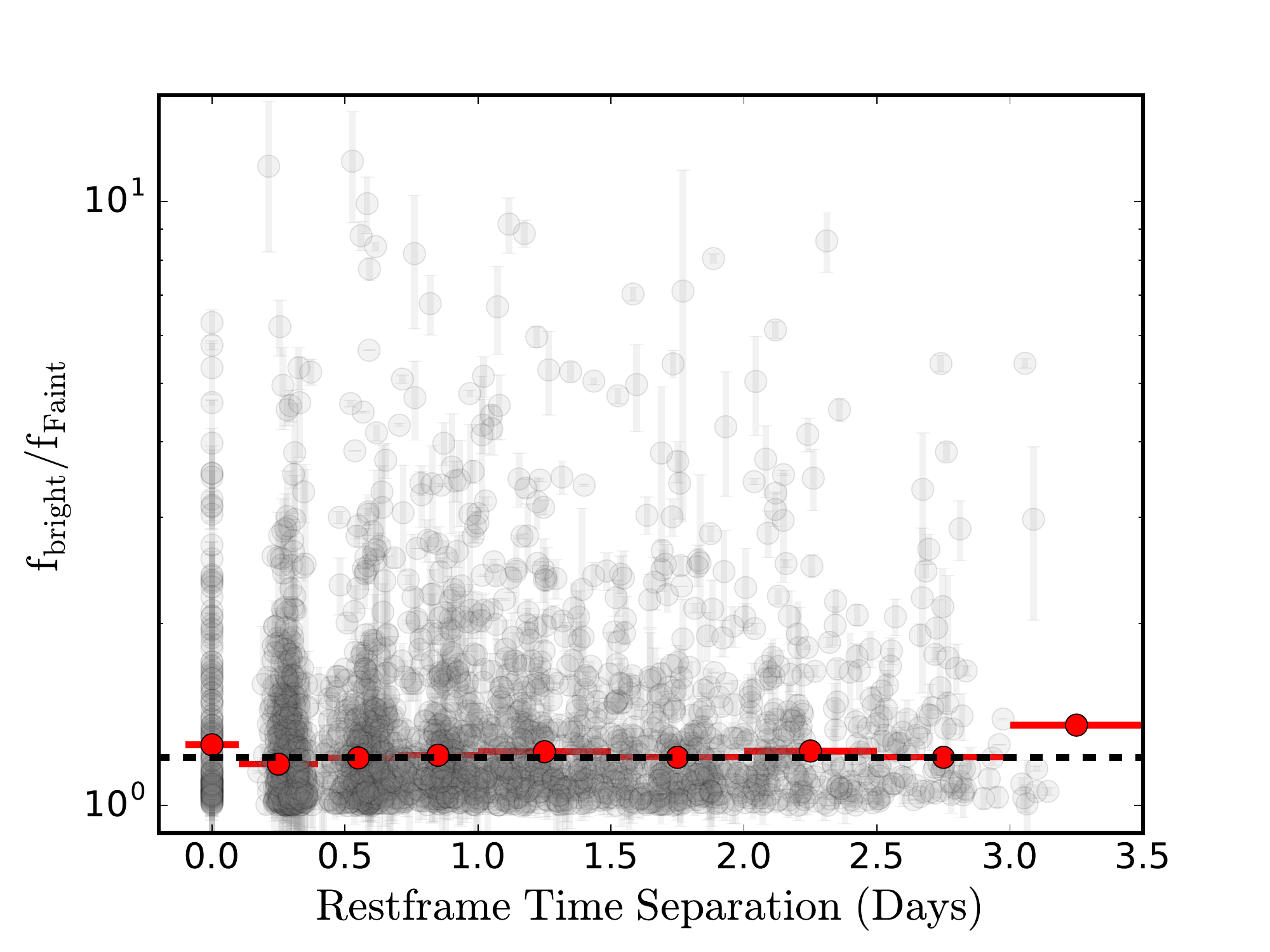}
\caption{Rest frame time separation versus continuum flux ratio at 1450\AA. The grey dots are 2400 high-redshift (z $>$1.5), two-epoch radio-quiet quasars with observed time separation less than 10 days. The red dots are the median values in different bins (horizontal error bars), and the black dashed line is the uncertainty level of 20\%.}
\label{fig:calibration}
\end{figure}

\begin{figure}
\hspace{-0.5cm}
\centering
\includegraphics[width=9.cm]{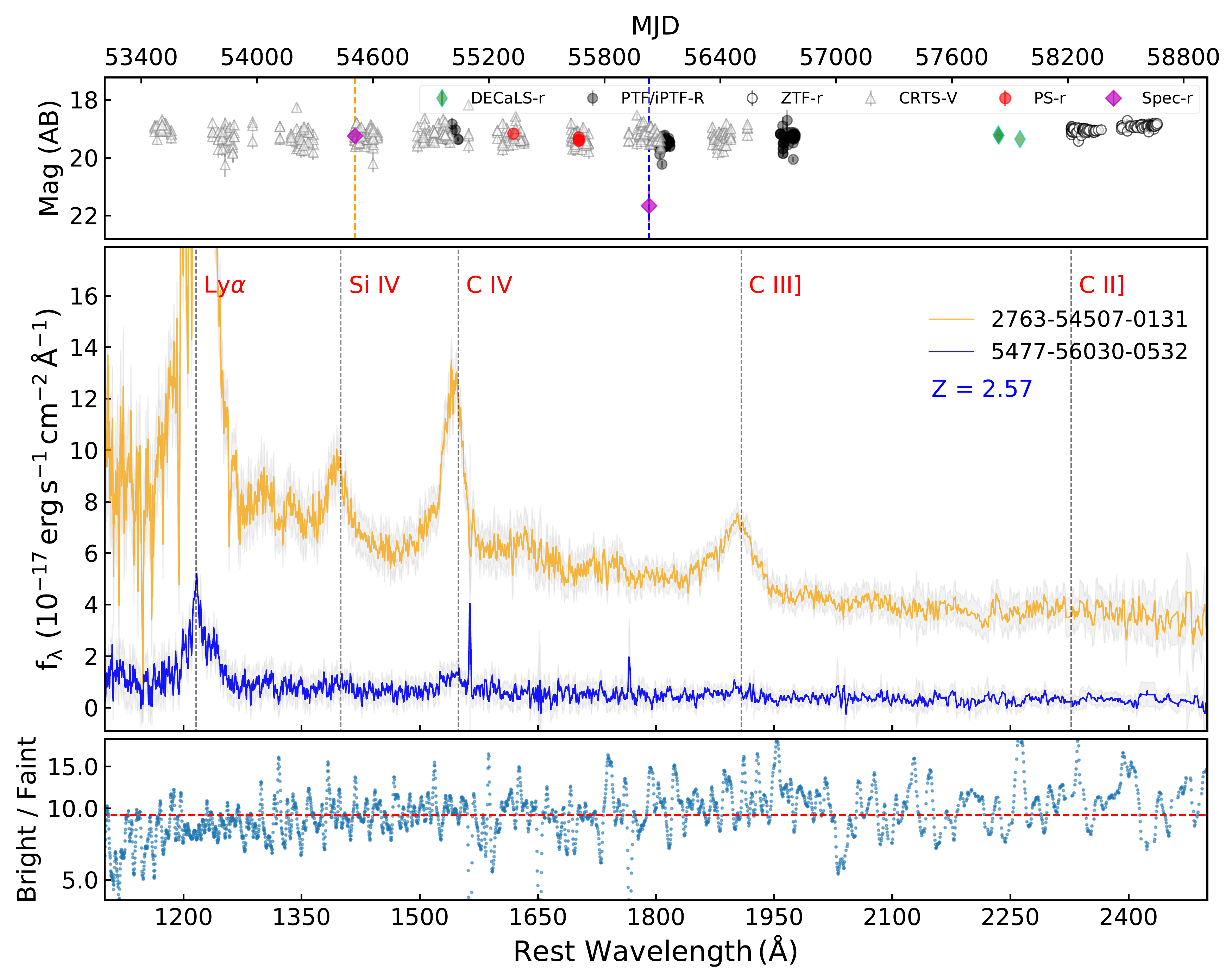}
\caption{Example of a fiber-drop quasar which mimics a \ciii\ and/or \civ\ CLQ. The synthetic magnitude from the blue spectrum (MJD = 56030) is $>$ 2 mag fainter than concurrent multi-survey photometries, indicating an problematic epoch with dropped fiber rather than a true variability within such a short timescale. The spectral ratio of the bright and faint states is also shown in the bottom panel.}
\label{fig:fake_CLQ}
\end{figure}

The asserted accuracy of spectrophotometric calibration of stars is about 6\% in SDSS. However, we find that the accuracy of the spectrophotometry for high-redshift quasars (point source) based on repeat spectra is around 20\% in Figure \ref{fig:calibration}, consistent with \cite{Stoughton02}. To evaluate the flux-calibration accuracy of SDSS spectra at high redshift, we compile a sample of 2400 two-epoch radio-quiet normal quasars ($z>$ 1.5) with time separation $\Delta T$ $<$ 10 days in the observed frame and presents the time separation between the bright and faint epochs in rest frame versus continuum flux ratio at 1450\AA\ in Figure \ref{fig:calibration}. The normal quasar variability on timescale of 10 days is around 0.02 mag (5\% in flux) according to the structure function of Stripe 82 quasars \citep{MacLeod10}. However, we surprisingly find that about 1\% of objects show continuum variation ($f_{\rm bright}/f_{\rm faint}$) larger than 500\%, even for the same night observations. The median flux ratios in different bins are about 20\%, consistent with the believed accuracy in \cite{Stoughton02}. In addition, we speculate that objects with very large flux ratios (e.g., $>$500\%) are subject to the fiber-drop issue (see \S \ref{sec:fiberdrop}) to different degrees. The apparent anti-correlation between flux ratio and time separation is not real since fewer objects are distributed around 3 days.

\subsection{Fiber-drop Epochs} \label{sec:fiberdrop}
The scenario becomes even worse when fiber is inadvertently loose or partially dropped for high-redsfhit objects without any caution or flag from SDSS pipeline. This very rare phenomenon was also confirmed in repeatedly monitored TDSS and SDSS-RM programs \citep{Shen15,Sun15}. As shown in Figure \ref{fig:fake_CLQ}, a fiber-partially-dropped epoch (faint blue epoch, middle panel) mimics a \ciii\ and/or \civ\ CLQ with a significant drop of continuum flux. The synthetic magnitude for the faint spectrum totally disobeys the concurrent photometries and typical short-term variability magnitude. In addition to simultaneous photometries, a fiber-drop spectrum usually has an identical shape (i.e., continuum slope and line profile) to the bright state, rather than showing a BWB trend as a typical quasar, which is further demonstrated by the flat spectral ratio of bright and faint states (bottom panel). Moreover, with a typical exposure time (45 min) for SDSS objects, the SNR for the faint state is usually much lower than that of bright state with such a huge variation, rather than similar SNRs in both states.

\section{The Rest CLQs}\label{app:rest}

\begin{figure*}
\centering
\hspace*{-2cm}
\includegraphics[width=\paperwidth]{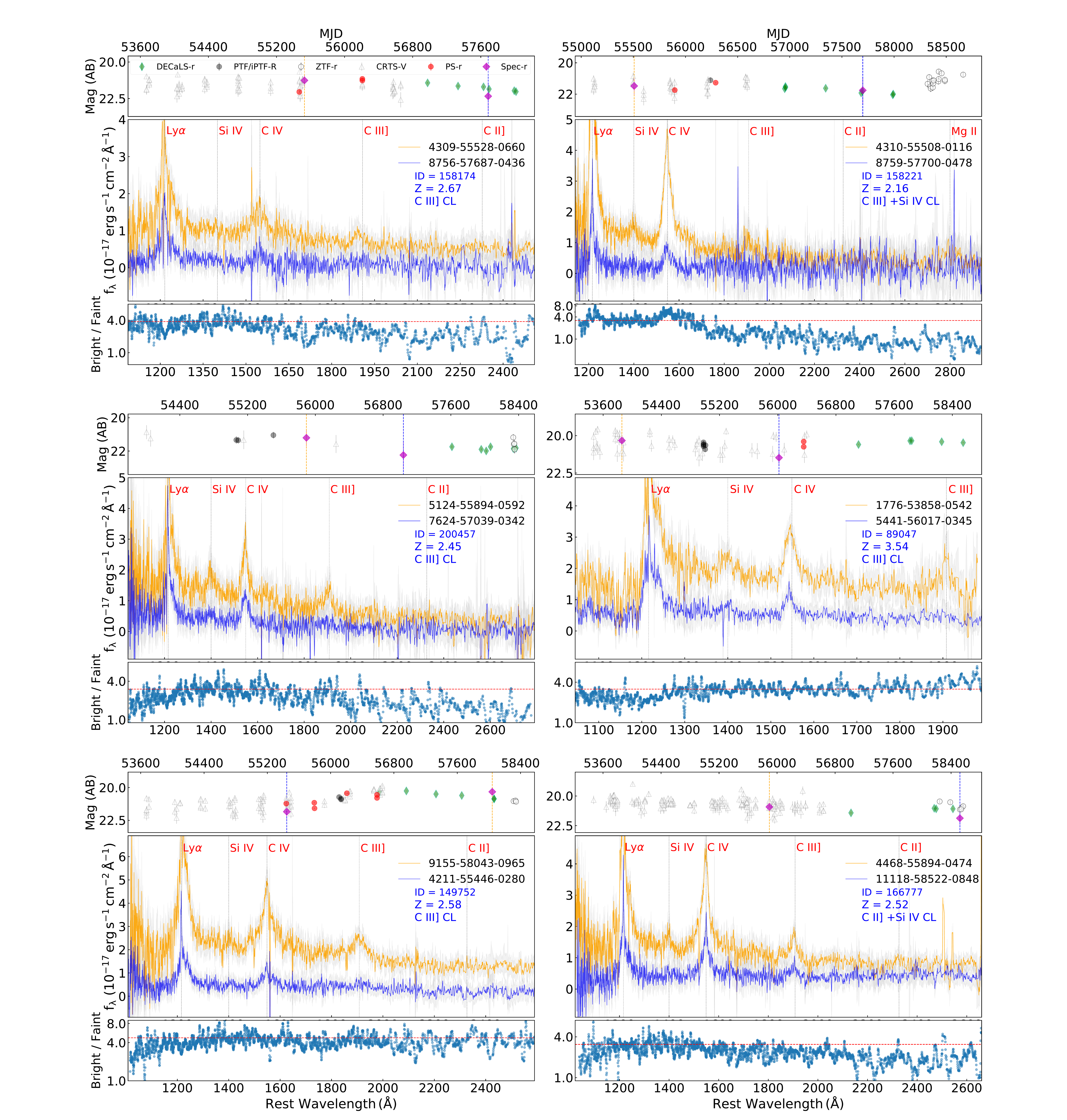}
\caption{The same as Figure \ref{fig:example}}
\label{fig:example2}
\end{figure*}

\begin{figure*}
\centering
\hspace*{-2cm}
\includegraphics[width=\paperwidth]{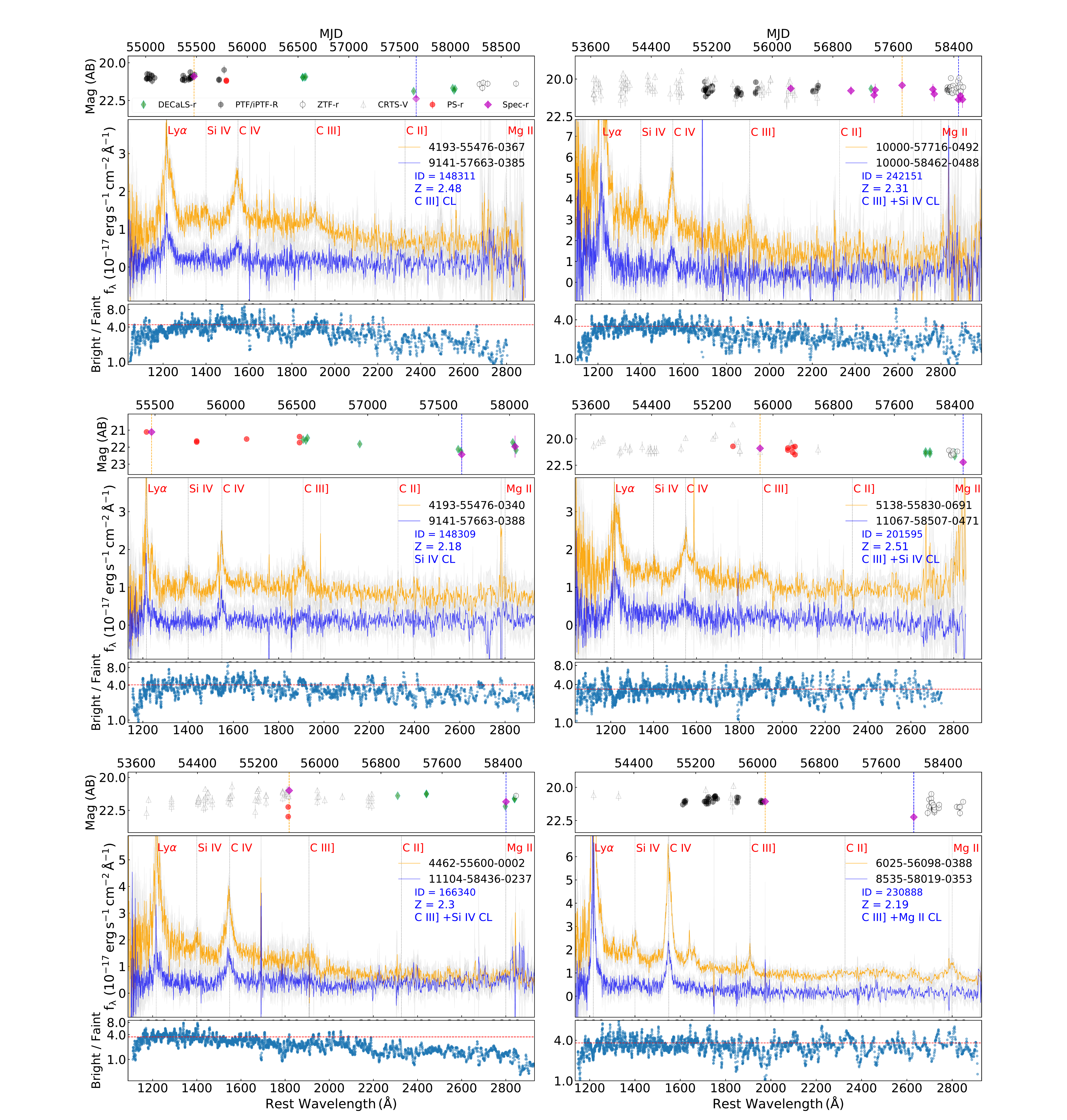}
\caption{The same as Figure \ref{fig:example}}
\label{fig:example3}
\end{figure*}

\begin{figure*}
\centering
\hspace*{-2cm}
\includegraphics[width=\paperwidth]{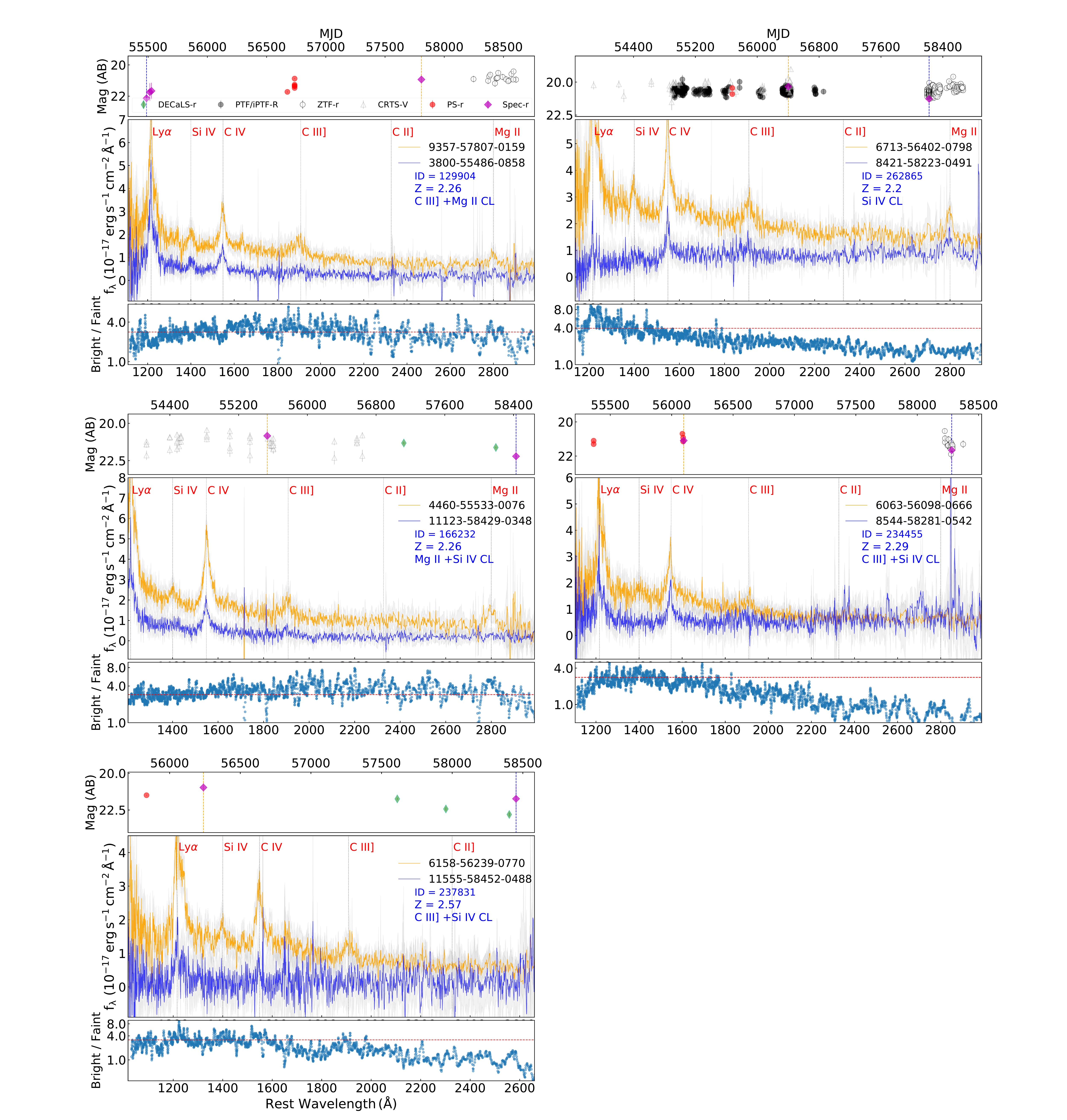}
\caption{The same as Figure \ref{fig:example}}
\label{fig:example1}
\end{figure*}

\end{CJK}

\end{document}